\documentclass[journal,hidelinks]{new-aiaa}
\usepackage[utf8]{inputenc}
\usepackage{textcomp}
\usepackage{subcaption}

\usepackage{graphicx}
\usepackage{amsmath}
\usepackage{mathrsfs}  
\usepackage[version=4]{mhchem}
\usepackage{siunitx}
\usepackage{longtable,tabularx}
\usepackage[table]{xcolor}
\usepackage{mathtools}
\usepackage[normalem]{ulem}
\usepackage{xr}

\title{Modified Marrone-Treanor dissociation model: formulation and benchmarking for diatom/atom mixtures\footnote{Portions of this work were presented as papers 2019-0789 and 2020-2191 at AIAA SciTech forums held in San Diego, CA. and Orlando, FL}}

\author{Ross S.~Chaudhry\footnote{Research Associate, Smead Aerospace Engineering Sciences, AIAA Member}}
\affil{University of Colorado, Boulder, CO 80303}
\author{Erik Torres\footnote{Research Associate, Aerospace Engineering \& Mechanics; Now at Aero-Thermo-Mechanical Laboratory, Universit\'e Libre de Bruxelles, Avenue F.D. Roosevelt 50, B-1050 Brussels, Belgium; erik.matthias.torres@ulb.be (Corresponding author)},
Thomas E.~Schwartzentruber\footnote{Professor, Aerospace Engineering \& Mechanics, AIAA Associate Fellow} and 
Graham V.~Candler\footnote{McKnight Presidential Endowed Chair, Aerospace Engineering \& Mechanics, AIAA Fellow}}
\affil{University of Minnesota, Minneapolis, MN 55455}

\begin{document}

\maketitle

\begin{abstract}
 We present a modified Marrone-Treanor model for dissociation with rate parameters derived exclusively from quasiclassical trajectory calculations on ab initio potential energy surfaces. Analysis of the trajectory dataset for reactant $\mathrm{O_2}$ and $\mathrm{N_2}$ diatoms sampled from Boltzmann internal energy distributions over a wide $T,T_\mathrm{v}$ range indicates that a modified version of the classical Marrone-Treanor two-temperature model captures the most relevant physics of shock-heated dissociating diatomic species very well. We find that simple correction factors account for non-Boltzmann depletion effects observed in direct molecular simulations employing the same potentials. The concentration-dependent functional form proposed for these correction factors ensures that depletion effects vanish at chemical equilibrium. Based on comparisons in isothermal and adiabatic heat baths we verify that the resulting two-temperature dissociation model accurately reproduces all major features observed in the direct molecular simulations, while remaining computationally inexpensive enough for large-scale computational fluid dynamics simulations.
\end{abstract}

%-------------------------------------------------------------------------------
\section*{Nomenclature} \label{sec:nomenclature}

{\renewcommand\arraystretch{1.0}
\noindent\begin{longtable*}{@{}l @{\quad=\quad} l@{} l@{}}
$a_U$ & fitting constant in MMT model $\, [-]$ \\
$\mathcal{A}_s$ & species $s$ chemical symbol in reaction equation $\, [-]$ \\
$C$ & modified Arrhenius pre-exponential factor coefficient $\, [L^3 \, T^{-1} \, N^{-1} \, \Theta^{-n}]$ \\
$c_0$ & Proportionality factor $= 1 \, \mathrm{atm \cdot s}$ \\
$D$ & mixture set of diatomic species $\, [-]$ \\
$D_{0,s}$ & molecular dissociation energy of species $s$ $\, [M \, L^2 \, T^{-2} \, N^{-1}]$ \\
$e_{\mathrm{v},s}$ & species $s$ vibrational energy per unit mass $\, [L^2 \, T^{-2}]$ \\
$E_\mathrm{v}$ & mixture vibrational energy per unit volume $\, [M \, L^{-1} \, T^{-2}]$ \\
$f_k^\mathrm{NB}$ & rate coefficient non-Boltzmann factor $[-]$ \\
$f_\varepsilon^\mathrm{NB}$ & vibrational energy change non-Boltzmann factor $[-]$ \\
$j$  & rotational quantum number $[-]$ \\
$k^\mathrm{Arr}$ & modified Arrhenius rate coefficient $[L^3 \, T^{-1} \, N^{-1}]$ \\
$\mathrm{k_B}$  & Boltzmann constant $= 1.38065 \times 10^{-23} \, \mathrm{J \cdot K^{-1}}$ \\
$k_{\mathrm{diss/rec}}$ & effective dissociation/recombination rate coefficient $\, [L^3 \, T^{-1} \, N^{-1}]$ \\
$k_r^{f/b}$ & forward / backward rate coefficient for generic reaction $r$ $\, [(L^{-3} \, N)^{1 - \nu_r^{f/b}} \, T^{-1} \, ]$ \\
$K_\mathrm{c}^\mathrm{eq}$ & molar concentration-based equilibrium constant $\, [(L^{-3} \, N)^{\nu_r^\mathrm{T}}]$ \\
$L$ & physical dimension of length \\
$M$ & physical dimension of mass \\
$\mathrm{M}$ & arbitrary collision partner species $\, [-]$ \\
$M_s$ & species $s$ molar mass $\, [M \, N^{-1}]$ \\
$m_{s,q}^\mathrm{high/low}$ & high/low-temperature limit vibrational relaxation time fit slope parameter for species pair $s,q$ $\, [\Theta^{1/3}]$ \\
$n_{s,q}^\mathrm{high/low}$ & high/low-temperature limit vibrational relaxation time fit offset parameter for species pair $s,q$ $\, [-]$ \\
$N$ & physical dimension of amount of substance \\
$n$ & modified Arrhenius pre-exponential factor temperature exponent $\, [-]$ \\
$n_{\mathrm{v},s}$ & number of discrete vibrational levels of diatomic species $s$ $\, [-]$ \\
$p$ & mixture static pressure $\, [M \, L^{-1} \, T^{-1}]$ \\
$P_d$ & probability of dissociation $\, [-]$ \\
$P_d \, (v)$ & vibration-specific probability of dissociation $\, [-]$ \\
$R$ & set of chemical reactions $\, [-]$ \\
$r_\mathrm{A-A}$ & inter-atomic distance between atom pair A-A $\, [L]$ \\
$\mathcal{R}_r^f, \mathcal{R}_r^b, \mathcal{R}_r$ & forward, backward and net rate per unit volume of reaction $r$ $\, [L^{-3} \, T^{-1} \, N]$ \\
$\vec{q}_\mathrm{v}$ & vibrational energy heat flux $\, [M \, T^{-3}]$ \\
$Q_{\mathrm{v},s}$ & vibrational partition function of species $s$ $\, [-]$ \\
$\tilde{Q}_{\mathrm{v},s}$ & approximate vibrational partition function of species $s$ $\, [-]$ \\
$S$ & mixture set of chemical species $\, [-]$ \\
$\mathcal{S}_d$ & statistical support for dissociation $\, [-]$ \\
$t$ & time $\, [T]$ \\
$T$ & physical dimension of time \\
$T$ & mixture translational-rotational temperature $\, [\Theta]$ \\
$T_{\mathrm{D},s}$ & species $s$ characteristic dissociation temperature $\, [\Theta]$ \\
$T_F$ & pseudotemperature in MMT model $\, [\Theta]$ \\
$T_{\mathrm{r},s}$ & species $s$ rotational temperature in DMS calculation $\, [\Theta]$ \\
$T_\mathrm{t}$ & mixture translational temperature in DMS calculation $\, [\Theta]$ \\
$T_\mathrm{t-r}$ & mixture translational-rotational temperature in DMS calculation $\, [\Theta]$ \\
$T_{\mathrm{v},s}$ & species $s$ vibrational temperature in DMS calculation $\, [\Theta]$ \\
$T_\mathrm{v}$ & mixture vibrational temperature $\, [\Theta]$ \\
$\vec{u}, \vec{v}_s$ & bulk and species $s$ diffusion velocities $\, [M \, L^{-1}]$ \\
$U, U^*$ & pseudotemperatures in MMT model $\, [\Theta]$ \\
$v$ & vibrational quantum number $\, \mathrm{[-]}$ \\
$V_D$ & potential energy of diatom $\, [M \, L^2 \, T^{-2}]$ \\
$w_s$ & rate of formation of species $s$ mass per unit volume $\, [M \, L^{-3} \, T^{-1}]$ \\
$w_{s,r}$ & rate of formation of species $s$ mass per unit volume in reaction $r$ $\, [M \, L^{-3} \, T^{-1}]$ \\
$w_{\mathrm{v}}^\mathrm{relax}$ & vibration-translational energy relaxation source term $\, [M \, L^{-1} \, T^{-3}]$ \\
$w_{\mathrm{v}}^\mathrm{chem}$ & vibrational energy-chemistry coupling source term $\, [M \, L^{-1} \, T^{-3}]$ \\
$w_{\mathrm{v},r,s}^\mathrm{chem}$ & rate of species $s$ vibrational energy change due to reaction $r$ $\, [M \, L^{-1} \, T^{-3}]$ \\
$x_s$ & mole fraction of species $s$ $\, [-]$ \\
$Z$ & vibrational nonequilibrium rate factor $\, [-]$ \\
$\varepsilon_\mathrm{int/rot/vib}$ & per-molecule internal/rotational/vibrational energy $\, [M \, L^2 \, T^{-2}]$ \\
$\varepsilon_\mathrm{rem}$ & per-molecule remaining energy for dissociation $\, [M \, L^2 \, T^{-2}]$ \\
$\varepsilon_s (v)$ & species $s$ per-molecule quantized level $v$ vibrational energy $\, [M \, L^2 \, T^{-2}]$ \\
$\langle \varepsilon_{\mathrm{v},s} \rangle_\mathrm{diss}$ & per-molecule vibrational energy removed for species $s$ due to dissociation $\, [M \, L^2 \, T^{-2}]$ \\
$\langle \Delta \varepsilon_{\mathrm{v},s} \rangle_\mathrm{diss}$ & per-molecule vibrational energy \emph{change} for species $s$ due to dissociation $\, [M \, L^2 \, T^{-2}]$ \\
$\zeta_r$ & nonequilibrium concentration ratio of reaction $r$ $\, [-]$ \\
$\Theta_{\mathrm{v},s}$ & species $s$ characteristic temperature of vibration $\, [\Theta]$\\
$\Theta$ & physical dimension of temperature \\
$\nu_{s,r}^{f/b}$ & species $s$ stoichiometric coefficient in forward / backward sense of reaction $r$ $\, [-]$ \\
$\nu_r^{f/b}$ & stoichiometric coefficient sum in forward / backward sense of reaction $r$: $\nu_r^{f/b} = \sum_{s \in S} \{ \nu_{s,r}^{f/b}\}$ \, $\mathrm{[-]}$ \\
$\nu_r^\mathrm{T}$ & total difference of stoichiometric coefficients in reaction $r$: $\nu_r^\mathrm{T} = \nu_r^{b} - \nu_r^{f}$ \, $\mathrm{[-]}$ \\
$\rho_s$ & species $s$ density $\, [M \, L^{-3}]$ \\
$\langle \tau_s^\mathrm{v} \rangle$ & species $s$ mixture-averaged vibrational relaxation time $\, [T]$ \\
$\tau_{s,q}^\mathrm{v}$ & species $s$ vibrational relaxation time with collision partner species $q$ $\, [T]$ \\
CFD & Computational Fluid Dynamics \\
DMS & Direct Molecular Simulation \\
MMT & Modified Marrone-Treanor \\
M\&W & Millikan and White \\
PES & potential energy surface \\
QCT & quasi-classical trajectory \\
QSS & quasi-steady state \\
SHO & simple harmonic oscillator
\end{longtable*}}

%-------------------------------------------------------------------------------
\section{Introduction} \label{sec:introduction}

\lettrine{T}{he} dissociation of air has been studied for decades because it plays an important role in hypersonic flows~\cite{Gnoffo1999, Candler2019}. The dissociation process is complicated because its rate depends strongly on the vibrational state of the gas. Molecules that are vibrationally excited dissociate more easily because the vibrational energy reduces the energy barrier to dissociation. Conversely, when the gas has low vibrational energy, dissociation is suppressed. In addition, dissociation affects the vibrational energy distribution, because when excited molecules dissociate, the high-energy tail of the vibrational energy distribution is depopulated. A complete dissociation model must account for these vibrational excitation and reaction dynamics.

There are many air dissociation kinetics models in the literature, based on differing assumptions about the dominant mechanisms in dissociating air. Most of these models assume that the translational and rotational modes of the gas are close to being in equilibrium with one another, and thus the mixture translational-rotational energy is modeled by a single temperature, $T$. The vibrational energy modes relax relatively slowly, and therefore a vibrational energy conservation equation is solved, with the vibrational energy represented by a single vibrational temperature $T_\mathrm{v}$. This approach was originally conceived over 60 years ago~\cite{HammerlingTK1959, TreanorM1962, MarroneT1963}. 

For a two-temperature framework, three inputs are required from a dissociation model: reaction rate coefficients, characteristic vibrational relaxation times, and the change in vibrational energy due to dissociation~\cite{GnoffoGS1989}. Most dissociation models in the literature prescribe the rate coefficients and change in vibrational energy \cite{LosevG1962, MarroneT1963, Park1988, KnabFM1995, LuoAM2018} and rely on vibrational relaxation times from elsewhere, such as the fits of Millikan and White~\cite{MillikanW1963} (M\&W). However, all of these processes are tightly coupled and an accurate model requires full consistency between all three aspects of the dissociation model.

Ab initio potential energy surfaces (PESs) have recently been developed for many of the three- and four-atom systems of interest in high-temperature air chemistry. Such PESs describe the force on each individual atom during any arbitrary spatial arrangement of the multi-atom system. Collisions of air reactants can be simulated by integrating in time on these surfaces from an initial condition to a final post-collision molecular configuration. These quasiclassical trajectories~\cite{truhlar79a} (QCTs) can be used to quantify reaction dynamics, including the role of vibrational excitation on dissociation. QCT analysis can be used to sample pre-collision states from a gas at a specified translational, rotational and vibrational temperature, and obtain reaction rate coefficients by averaging over many trajectories. Calculations for the species pairings considered in this work were conducted entirely on a set of potentials generated by the University of Minnesota Computational Chemistry group. Trajectory calculations for $\mathrm{N_2-N_2}$ pairs were performed on the electronically adiabatic $\mathrm{N_4}(^1A)$ PES~\cite{PaukkuYVT2013, PaukkuYVT2014, BenderVNPVTSC2015}. The same PES was originally used to simulate $\mathrm{N_2-N}$ trajectories by keeping the fourth $\mathrm{N}$ atom at effectively infinite distance from the interacting $\mathrm{N_3}$ sub-system. A recently released, dedicated $\mathrm{N_3}(^4A^{\prime \prime})$ PES for ground-electronic-state $\mathrm{N_2 -N}$ collisions~\cite{varga21b} now supersedes this approach. Trajectory calculations for $\mathrm{O_2-O_2}$ pairs were carried out on three $\mathrm{O_4}$ PESs by Paukku et al.~\cite{PaukkuYVSBT2017, PaukkuVT2018}, whereas a total of 9 ground-electronic-state $\mathrm{O_3}$ PESs were necessary to fully simulate $\mathrm{O_2-O}$ trajectories~\cite{VargaPT2017}. For $\mathrm{N_2-O_2}$ pairs the triplet surface by Varga et al.~\cite{VargaMSPT2016} was employed.

In prior works, QCT calculations on the Minnesota potentials have been used to study reaction dynamics by sampling initial conditions from fixed Boltzmann or approximately-Boltzmann distributions characterized by one or more temperatures (i.e. sweeps over $T$, $T_\mathrm{v}$ ranges). This approach allowed for focused analysis of a particular condition, or set of conditions, although the effect of non-Boltzmann distributions was not accounted for. For example, QCT analysis was employed by Bender et al.~\cite{BenderVNPVTSC2015} to study $\mathrm{N_2 + N_2}$ dissociation, producing estimates of thermal and multi-temperature dissociation rate coefficients, $k_\mathrm{diss} (T)$ and $k_\mathrm{diss} (T, T_\mathrm{v})$, as well as average vibrational energy removed per dissociation $\langle \varepsilon_\mathrm{v, N_2} \rangle_\mathrm{diss}$. Similar analyses were conducted for nitrogen dissociation with partners $\mathrm{N}$ and $\mathrm{O_2}$~\cite{ValentiniSBC2016, ChaudhryBSC2018}. Chaudhry et al.~\cite{ChaudhryBVSC2016, ChaudhryGBSC2018, Chaudhry2018} studied oxygen dissociation with partners $\mathrm{O_2}$, $\mathrm{O}$, and $\mathrm{N_2}$, producing equivalent estimates of $k_\mathrm{diss} (T, T_\mathrm{v})$ and $\langle \varepsilon_\mathrm{v, O_2} \rangle_\mathrm{diss}$ for these reactions. Most of these data have been collected and summarized in Ref.~\cite{Chaudhry2018}. 

Beyond this approach, relying on ab initio PESs in rovibrational state-resolved master equation calculations~\cite{PanesiJSM2013, KimB2013, JaffeSP2015} and Direct Molecular Simulations (DMS)~\cite{SchwartzentruberGV2018} makes it possible to predict the dynamics of the vibrational energy distribution in a dynamically relaxing and dissociating gas mixture. Over multiple DMS studies~\cite{ValentiniSBNC2015, ValentiniSBC2016, grover19a, grover19b, torres20b, torres24a} it has consistently been observed that the distribution of internal energies is non-Boltzmann, resulting in a reduction of the dissociation rate by a factor of $2$ to $5$ relative to an assumed Boltzmann distribution at $T_\mathrm{v}$. Thus, non-Boltzmann effects must be accounted for in some capacity in any practical multi-temperature model for air dissociation. Recent examples of both kinetic and continuum models that include non-Boltzmann effects can be found in Refs.~\cite{singh20b, singh20c}, where it is shown that reasonable model assumptions could lead to more simplified two-temperature models for Computational Fluid Dynamics (CFD). Recent DMS calculations of air dissociation in the quasi-steady-state (QSS) regime~\cite{torres24a} in particular, have yielded estimates for non-Boltzmann rate coefficients and vibrational energy removed by dissociation for the most relevant reactions in high-temperature air. A subset of these results is further analyzed in Sec.~\ref{sec:nb_effects} to calibrate the non-Boltzmann correction factors of the two-temperature model proposed in this paper. Furthermore, DMS calculations have yielded estimates of characteristic vibrational relaxation times for the major diatomic species in air~\cite{ValentiniSBC2016, grover19a, grover19b, torres24b}. Thus, with accurate PESs it is possible to compute data for all three critical components of a dissociation model. 

In this work, we analyze a database of QCT results. We investigate the mechanics of dissociation and identify key trends that exist for the dissociation reactions considered. We find that the classical Marrone-Treanor model~\cite{MarroneT1963} captures these trends quite well and we propose several modifications, resulting in a new multi-temperature model for air dissociation. This paper is organized as follows. In Sec.~\ref{sec:mmt_model} we review portions of the governing fluid equations relevant to modeling finite-rate chemistry of partially dissociated air within the context of a two-temperature $(T,T_\mathrm{v})$ description and introduce the Modified Marrone-Treanor (MMT) model. Section~\ref{sec:qct_analysis} summarizes the main findings of our QCT database analysis. We identify the most important functional dependencies for probability of dissociation and vibrational energy change per dissociation revealed by the aggregate trajectory data. In Sec.~\ref{sec:fitting_qct_data} we introduce the analytical functional forms used to curve-fit the two-temperature dissociation rate coefficient and vibrational energy change per dissociation to the QCT database. In Sec.~\ref{sec:nb_effects} we discuss the need for and implementation of correction factors to account for non-Boltzmann depletion effects during quasi-steady-state dissociation in a shock-heated gas. In Sec.~\ref{sec:approach_to_equilibrium} we propose a functional dependence on mixture composition for these non-Boltzmann factors to ensure the MMT model will reproduce the correct reaction and vibrational energy removal rates as the mixture approaches thermo-chemical equilibrium. In Sec.~\ref{sec:verif_vs_dms} we benchmark our CFD implementation of the MMT model against DMS reference solutions of nonequilibrium dissociation in $\mathrm{N_2/N}$ and $\mathrm{O_2/O}$ mixtures. All model parameters for thermodynamics, vibrational relaxation and dissociation chemistry are directly derived from, or fully consistent with the ab initio PESs which DMS also relies upon. Finally, in Sec.~\ref{sec:conclusion} we formulate the conclusions of this work. 

A full parametrization of the MMT model for 5-species air, including vibrational relaxation times derived from the ab initio PESs, as well as benchmarking against DMS results for dissociating air will be discussed in an upcoming follow-up article.

%-------------------------------------------------------------------------------
\section{The Modified Marrone-Treanor model} \label{sec:mmt_model}

%-------------------------------------------------------------------------------
\subsection{Conservation equations for a reacting gas mixture} \label{sec:conservation_eqns}

We consider a model for a reacting gas mixture in thermal and chemical nonequilibrium, where the generic species set $S$ comprises diatomic and atomic species (for 5-species air $S = \{ \mathrm{N_2}, \mathrm{O_2}, \mathrm{NO}, \mathrm{N}, \mathrm{O} \}$, but for the calculations in Sec.~\ref{sec:verif_vs_dms}, either $S = \{ \mathrm{N_2}, \mathrm{N} \}$, or $S = \{ \mathrm{O_2}, \mathrm{O} \}$). The mass conservation equation for each species is:
\begin{equation}
 \frac{\partial\rho_s}{\partial t} + \nabla \cdot (\rho_s \vec{u} + \rho_s \vec{v}_s) = w_s, \qquad s \in S, \label{eq:species_continuity}
\end{equation}
where $\rho_s$ is the density of species $s$, $\vec{u}$ is the bulk velocity, $\vec{v}_s$ is the mass diffusion velocity and $w_s$ is the rate of species $s$ formation. The chemical kinetics model must provide an expression for $w_s$ for each species as a function of the gas mixture state. This source term may be written in a generalized form:
\begin{equation}
 w_s = \sum_{r \in R} w_{s,r} = M_s \sum_{r \in R} ( \nu_{s,r}^b - \nu_{s,r}^f ) \left( \mathcal{R}_r^f - \mathcal{R}_r^b \right), \qquad s \in S, \label{eq:species_source_term}
\end{equation}
where $M_s$ is the species molar mass and the sum extends over all \emph{reversible} reactions $r$ in set $R$. Coefficients $\nu_{s,r}^f$ and $\nu_{s,r}^b$ represent pre- and post-reaction stoichiometric coefficients for species $s$ in reaction $r$ written in generic form, when read left-to-right:
\begin{equation}
 \sum_{q \in S} \nu_{q,r}^f \, \mathcal{A}_{q} \xrightleftharpoons[k_r^b]{k_r^f} \sum_{s \in S} \nu_{s,r}^b \, \mathcal{A}_{s}, \qquad r \in R. \label{eq:reaction_equation}
\end{equation}

Here $\mathcal{A}_{q}$ and $\mathcal{A}_{s}$ represent the chemical symbol of a particular species on the reactant, or product side of the equation. If a given species does not participate as reactant or product, its corresponding pre- or post-reaction stoichiometric coefficient in Eq.~(\ref{eq:reaction_equation}) is zero. The terms $\mathcal{R}_r^f$ and $\mathcal{R}_r^b$ in Eq.~(\ref{eq:species_source_term}) represent the \emph{forward} and \emph{backward} rates of reaction $r$ per unit volume respectively:
\begin{equation}
 \mathcal{R}_r^f = k_r^f \prod_{q \in S} \left( \frac{\rho_q}{M_q} \right)^{\nu_{q,r}^f}, \qquad \mathcal{R}_r^b = k_r^b \prod_{s \in S} \left( \frac{\rho_s}{M_s} \right)^{\nu_{s,r}^b}, \qquad r \in R. \label{eq:net_rate}
\end{equation}
with $k_r^f$ and $k_r^b$ as the corresponding forward and backward rate coefficients.

In this work we assume for simplicity that the vibrational energy density of the gas mixture $E_\mathrm{v}$ can be represented with a single conservation equation, characterized by a mixture vibrational temperature $T_\mathrm{v}$:
\begin{equation}
 \frac{\partial E_\mathrm{v}}{\partial t}+\nabla\cdot (E_\mathrm{v} \vec{u})+\nabla\cdot\sum_{s \in D}(\rho_s \, e_{\mathrm{v},s} \vec{v}_s) + \nabla \cdot \vec{q}_\mathrm{v} = w_{\mathrm{v}}^\mathrm{relax} + w_{\mathrm{v}}^\mathrm{chem}, \label{eq:vibconservation}
\end{equation}
where $D$ represents the set of diatomic species contributing to the vibrational energy mode. In a mixture containing multiple diatomic species, such as five-species air, this would be $D = \{ \mathrm{N_2}, \mathrm{O_2}, \mathrm{NO} \}$ and $T_\mathrm{v}$ would be implicitly defined through the relation $E_\mathrm{v} = \sum\limits_{s \in D} \rho_s \, e_{\mathrm{v},s} (T_\mathrm{v})$. In the simpler diatom/atom ($\mathrm{A_2/A}$) mixtures considered in Sec.~\ref{sec:verif_vs_dms}, this automatically reduces to $D = \{ \mathrm{A_2} \}$ and implies that $T_\mathrm{v} = T_\mathrm{v, A_2}$.

Terms on the left hand side of Eq.~(\ref{eq:vibconservation}) account for the local rate of change in $E_\mathrm{v}$ and vibrational energy transport by advection and diffusion respectively, while the two vibrational energy source terms sit on the right hand side. The first, $w_{\mathrm{v}}^\mathrm{relax}$, represents the vibration-translation relaxation source term. It is typically modeled as a sum of Landau-Teller terms:
\begin{equation}
 w_{\mathrm{v}}^\mathrm{relax} = \sum_{s \in D} \rho_s\frac{e_{\mathrm{v},s}(T)-e_{\mathrm{v},s}(T_\mathrm{v})}{\langle \tau_s^\mathrm{v} \rangle}, \label{eq:vib_relax_equation}
\end{equation}
where $\langle \tau_s^\mathrm{v} \rangle$ is an appropriately averaged vibrational relaxation time for species $s$. Refer to Sec.~\ref{sec:tau_vt_fits} for further details on how we compute vibrational relaxation times within the context of the current multi-temperature model.

The chemistry-vibrational energy coupling term $w_{\mathrm{v}}^\mathrm{chem} = \sum\limits_{s \in D} \sum\limits_{r \in R} w_{\mathrm{v},r,s}^\mathrm{chem}$ in Eq.~(\ref{eq:vibconservation}) accounts for net removal, or replenishment of vibrational energy during chemical reactions. When written in generic form, the individual terms may be expressed as:
\begin{equation}
 w_{\mathrm{v},r,s}^\mathrm{chem} = ( \nu_{s,r}^b - \nu_{s,r}^f ) \left( \langle \varepsilon_{\mathrm{v},s} \rangle_r^f \, \mathcal{R}_r^f - \langle \varepsilon_{\mathrm{v},s} \rangle_r^b \, \mathcal{R}_r^b    \right)  \qquad r \in R, s \in D \label{eq:evib_chem_source}
\end{equation}
where $\langle \varepsilon_{\mathrm{v},s} \rangle_r^f$ represents the average amount of vibrational energy per unit mole of species $s$ being \emph{removed} in the forward sense of reaction $r$, whereas $\langle \varepsilon_{\mathrm{v},s} \rangle_r^b$ is the corresponding amount being \emph{replenished} by the reverse process. In principle, $\langle \varepsilon_{\mathrm{v},s} \rangle_r^f$ and $\langle \varepsilon_{\mathrm{v},s} \rangle_r^b$ may take on different values, for instance when the local gas state exhibits significant departure from thermal equilibrium. However, when $\langle \varepsilon_{\mathrm{v},s} \rangle_r^f \approx \langle \varepsilon_{\mathrm{v},s} \rangle_r^b = \langle \varepsilon_{\mathrm{v},s} \rangle_r$ holds, Eq.~(\ref{eq:evib_chem_source}) can be simplified to:
\begin{equation}
 w_{\mathrm{v},r,s}^\mathrm{chem} = ( \nu_{s,r}^b - \nu_{s,r}^f ) \left( \mathcal{R}_r^f - \mathcal{R}_r^b \right) \langle \varepsilon_{\mathrm{v},s} \rangle_r \qquad r \in R, s \in D \label{eq:evib_chem_source_eff}
\end{equation}

Closure of Eq.~(\ref{eq:vibconservation}) requires all relevant $\langle \varepsilon_{\mathrm{v},s} \rangle_r$ also be provided by the chemical kinetics model. One common option is to assume \emph{non-preferential} vibrational energy-chemistry coupling. This assumes that the amount of vibrational energy of species $s$ removed/replenished per reaction $r$ will be equal to the average of that species in the mixture at the local thermodynamic state, i.e. $\langle \varepsilon_{\mathrm{v},s} \rangle_r^\mathrm{non-pref.} = e_{\mathrm{v},s} (T_\mathrm{v}) / M_s$. When substituted into Eq.~(\ref{eq:evib_chem_source_eff}) for every species and combined with Eq.~(\ref{eq:species_source_term}), this yields the non-preferential (i.e. vibrationally unbiased) form of the overall source term:
\begin{equation}
 w_{\mathrm{v}}^\mathrm{chem, (non-pref.)} = \sum_{s \in D} w_s \, e_{\mathrm{v},s} (T_\mathrm{v}). \label{eq:evib_chem_source_nonpref}
\end{equation}

It is revealed by the QCT calculations in Secs.~\ref{sec:qct_analysis} and \ref{sec:fitting_qct_data} that assuming non-preferential vibrational energy removal is a poor fit for dissociation reactions. Such reactions exhibit significant vibrational bias in their reaction rate coefficients, as well as for the average vibrational energy removed by dissociation. Thus, a major feature of the chemical-kinetics model described in Sec.~\ref{sec:mmt_model_explanation} is that it accounts for this preferential treatment by providing a more accurate expression for $\langle \varepsilon_{\mathrm{v},s} \rangle_r$ in dissociation reactions.

%-------------------------------------------------------------------------------
\subsection{The MMT model formulation} \label{sec:mmt_model_explanation}

In this work we propose modifications to the classic Marrone-Treanor model to represent air dissociation dynamics. The Modified Marrone-Treanor model is motivated by the analysis of a large quasiclassical trajectory database created using accurate potential energy surfaces developed by researchers at the University of Minnesota~\cite{potlib21}. At its core the MMT model consists of three main parts: (1) an analytical expression for the two-temperature nonequilibrium dissociation rate coefficient, (2) a corresponding expression for the vibrational energy change per dissociation and (3) a set of QCT- and DMS-derived kinetic parameters necessary to compute both these quantities. The MMT model's non-Boltzmann, two-temperature dissociation rate coefficient $k_\mathrm{diss}^\mathrm{MMT-NB}$ is defined as:
\begin{equation}
 k_\mathrm{diss}^\mathrm{MMT-NB} (T,T_\mathrm{v}) = k^\mathrm{Arr} (T) \, Z (T,T_\mathrm{v}) \, f_k^\mathrm{NB}, \label{eq:mmt_rate_coefficient}
\end{equation}
where $k^\mathrm{Arr}$ represents the \emph{thermal equilibrium} dissociation rate coefficient in modified Arrhenius form:
\begin{equation}
 k^\mathrm{Arr} (T) = C T^n\exp(-T_{\mathrm{D},s}/T), \label{eq:karr}
\end{equation}
while $Z (T,T_\mathrm{v})$ is the MMT model's vibrational nonequilibrium factor, exactly as derived by Marrone and Treanor~\cite{MarroneT1963}:
\begin{equation}
 Z(T,T_\mathrm{v}) = \frac{\tilde{Q}_{\mathrm{v},s}(T)\,\tilde{Q}_{\mathrm{v},s}(T_F)}{\tilde{Q}_{\mathrm{v},s}(T_\mathrm{v})\,\tilde{Q}_{\mathrm{v},s}(-U)}, \label{eq:Zfact}
\end{equation}
and $f_k^\mathrm{NB}$ is a non-Boltzmann correction discussed in more detail in Secs.~\ref{sec:nb_effects} and \ref{sec:approach_to_equilibrium}. The $\tilde{Q}$'s appearing in Eq.~(\ref{eq:Zfact}) are approximate vibrational partition functions of the form:
\begin{equation}
 \tilde{Q}_{\mathrm{v},s}(T) = \frac{1 - \exp\,(-T_{\mathrm{D},s}/T) }{1 - \exp\,(-\Theta_{\mathrm{v},s}/T)} \label{eq:approx_partfunc}
\end{equation}
and $T_F$ and $U$ are pseudotemperatures defined as
\begin{equation}
 T_F (T,T_\mathrm{v}) = \left( \frac{1}{T_\mathrm{v}}-\frac{1}{T}-\frac{1}{U(T)} \right)^{-1} \label{eq:TF}
\end{equation}
and
\begin{equation}
 U(T) = \left( \frac{a_U}{T}+\frac{1}{U^*} \right)^{-1} \label{eq:NewFit_U}
\end{equation}
respectively. Thus, in addition to the local $(T,T_\mathrm{v})$ values, Eqs.~(\ref{eq:mmt_rate_coefficient})-(\ref{eq:NewFit_U}) depend on the reaction-specific Arrhenius parameters $C$ and $n$, characteristic vibrational and dissociation temperatures $\Theta_{\mathrm{v},s}$ and $T_{\mathrm{D},s}$ of the dissociating diatom, as well as the MMT reaction-specific parameters $U^*$ and $a_U$. 

The average vibrational energy removed per dissociation in the MMT model is calculated using Knab's formula~\cite{KnabFM1995}:
\begin{equation}
 \langle \varepsilon_{\mathrm{v},s} \rangle_\mathrm{diss}^\mathrm{Knab} (T, T_\mathrm{v}) = \frac{\mathrm{k_B} \, \Theta_{\mathrm{v},s}}{\exp\,(\Theta_{\mathrm{v},s}/T_F) - 1} - \frac{\mathrm{k_B} \, T_{\mathrm{D},s}}{\exp\,(T_{\mathrm{D},s}/T_F)-1}, \label{eq:Knab}
\end{equation}
where, by convention, $\langle \varepsilon_{\mathrm{v},s} \rangle_\mathrm{diss}^\mathrm{Knab} > 0$ when vibrational energy is being removed during dissociation of species $s$. Apart from being parameterized by the dissociating species' characteristic vibrational and dissociation temperatures, this expression shares with the MMT rate coefficient an indirect dependence on vibrational temperature through the pseudotemperature $T_F (T,T_\mathrm{v})$. As discussed in Sec.~\ref{sec:mmt_pseudotemp}, we take advantage of this commonality to determine the MMT parameters $a_U$ and $U^{*}$ applicable to both the vibrational energy removed and the rate coefficient from QCT data for $\langle \varepsilon_{\mathrm{v},s} \rangle_\mathrm{diss}$ under thermal-equilibrium conditions, i.e. where $T = T_\mathrm{v}$ and $f_k^\mathrm{NB} = f_\varepsilon^\mathrm{NB} = 1$.

Analogous to the MMT dissociation rate coefficient, non-Boltzmann depletion effects modify the average vibrational energy removed by dissociation. They are accounted for through a correction factor $f_\varepsilon^\mathrm{NB}$ applied to Eq.~(\ref{eq:Knab}).
\begin{equation}
 \langle \varepsilon_{\mathrm{v},s} \rangle_\mathrm{diss}^\mathrm{MMT-NB} =  \langle \varepsilon_{\mathrm{v},s} \rangle_\mathrm{diss}^\mathrm{Knab} f_\varepsilon^\mathrm{NB}. \label{eq:devib_nb_factor}
\end{equation}

In Table~\ref{tab:mmt_params} we list model parameters for $\mathrm{N_2}$ dissociation with collision partners $\mathrm{M} = \{ \mathrm{N_2}, \mathrm{N}, \mathrm{O_2} \}$ and for $\mathrm{O_2}$ dissociation with partners $\mathrm{M} = \{ \mathrm{O_2}, \mathrm{O}, \mathrm{N_2} \}$ respectively. We present two distinct, but nearly equivalent parameter sets. In sub-table~\ref{tab:td_variable} we list values obtained when the QCT-derived rate coefficients are curve-fit to Eq.~(\ref{eq:karr}) with the dissociation temperature treated as the third unknown in addition to $C$ and $n$. Thus, $T_{\mathrm{D},\mathrm{N_2}}$ and $T_{\mathrm{D},\mathrm{O_2}}$ differ for every reaction listed in Table~\ref{tab:td_variable}, even for reactions where the dissociating diatomic species $\mathrm{A_2}$ is the same. In sub-table~\ref{tab:td_fixed} we list values obtained from alternative two-parameter curve fits with dissociation temperatures held fixed at the commonly accepted values $T_{\mathrm{D},\mathrm{N_2}} = 113\,200 \mathrm{K}$ and $T_{\mathrm{D},\mathrm{O_2}} = 59\,380 \mathrm{K}$ respectively. The values for $C$, $n$ and $T_{\mathrm{D},\mathrm{A_2}}$ (columns 3-5) appeared originally in Tables 3.1 and 3.2 of Ref.~\cite{Chaudhry2018} on rows marked as ``effective dissociation'' (i.e. sum of simple + swap dissociation\footnotemark[1] rates). Differences in MMT parameters $a_U$ and $U^*$ (columns 7-8) between sub-tables~\ref{tab:td_variable} and \ref{tab:td_fixed} for the same reaction are a consequence of curve-fitting the QCT data for vibrational energy change per dissociation to Eq.~(\ref{eq:Knab}) with different values of $T_{\mathrm{D},\mathrm{A_2}}$ (see Sec.~\ref{sec:mmt_pseudotemp}). These parameters were originally listed in Tables 3.3 and 3.4 of Ref.~\cite{Chaudhry2018}.

\footnotetext[1]{\emph{Simple}, or \emph{single} dissociation refers to quasiclassical trajectories of type $\mathrm{AB} + \mathrm{CD} \rightarrow \mathrm{A} + \mathrm{B} + \mathrm{CD}$, as opposed to those involving simultaneous swap-dissociation $\mathrm{AB} + \mathrm{CD} \rightarrow \mathrm{A} + \mathrm{C} + \mathrm{BD}$, or double-dissociation $\mathrm{AB} + \mathrm{CD} \rightarrow \mathrm{A} + \mathrm{B} + \mathrm{C} + \mathrm{D}$.}

Further note minor differences in the listed values for characteristic vibrational temperature of the dissociating diatomic species (column 6) both in Tables~\ref{tab:td_variable} and \ref{tab:td_fixed}. Whereas $\Theta_{\mathrm{v},\mathrm{N_2}} = 3\,411 \, \mathrm{K}$ for $\mathrm{N_2}$ dissociation with partners $\mathrm{N_2}$ and $\mathrm{N}$ (in rows 1-2), it becomes $\Theta_{\mathrm{v},\mathrm{N_2}} = 3\,415 \, \mathrm{K}$ for dissociation with collision partner $\mathrm{O_2}$ (3rd row). The analog happens for $\mathrm{O_2}$ dissociation with collision partners $\mathrm{O_2}$ and $\mathrm{O}$ (rows 4-5), where $\Theta_{\mathrm{v},\mathrm{O_2}} = 2\,280 \, \mathrm{K}$ vs. $2\,263 \, \mathrm{K}$ with partner $\mathrm{N_2}$ (row 6). Such discrepancies can be traced back to the slightly different shapes of diatomic potential curves $V_D (r_\mathrm{A-A})$ associated with different PES versions. The characteristic vibrational temperatures listed in Table~\ref{tab:mmt_params} are defined as $\Theta_{\mathrm{v},\mathrm{A_2}} = [ \varepsilon_\mathrm{A_2} (v\!=\!1) - \varepsilon_\mathrm{A_2} (v\!=\!0) ]/\mathrm{k_B}$ and thus depend on the precise vibrational energies of the lowest-energy level and its immediate higher-lying neighbor. The values of all $\varepsilon_\mathrm{A_2} (v)$, in turn, are found as approximate solutions to an eigenvalue problem taking $V_D (r_\mathrm{A-A})$ as an input (see Sec. 3.2 of Bender~\cite{Bender2016} for more details). Whereas all trajectory calculations for reactions in rows 1-2 were performed on the same $\mathrm{N_4}(^1A)$ PES~\cite{PaukkuYVT2013, PaukkuYVT2014, Bender2016}, QCT calculations for reaction 3 relied on the $\mathrm{N_2O_2}(^3A)$ PES~\cite{VargaMSPT2016} instead. The latter PES employed slightly different shape parameters in defining the nitrogen diatomic potential $V_D (r_\mathrm{N-N})$. In a similar fashion, reactions 4-5 employed the combined set of $\mathrm{O_4}$~\cite{PaukkuYVSBT2017, PaukkuVT2018} and $\mathrm{O_3}$~\cite{VargaPT2017} PESs all sharing a common diatomic potential curve for $\mathrm{O_2}$. Reaction in row 6 again relied on the triplet $\mathrm{N_2O_2}$ PES, whose shape parameters for defining $V_D (r_\mathrm{O-O})$ differed slightly from those for the pure oxygen PESs. 

At the fluid scale of the two-temperature model proposed in this work, the slight discrepancies in vibrational energy ladders generated with different PESs have a negligible influence on the overall predictions and can be mostly ignored. By contrast, in multi-species DMS calculations~\cite{torres24a, torres24b} that resolve individual molecular collisions, a common set of diatomic potential curves across all PESs involved becomes crucial to maintaining energy conservation over successive time steps. This consideration prompted the Minnesota chemists to recently re-fit all their currently available PESs with a consistent set of $V_D (r_\mathrm{N-N})$, $V_D (r_\mathrm{O-O})$ and $V_D (r_\mathrm{N-O})$ curves. In an upcoming article we rely on this updated set of PESs to generate new MMT model parameters applicable to a five-species air mixture ($S = \{ \mathrm{N_2}, \mathrm{O_2}, \mathrm{NO}, \mathrm{N}, \mathrm{O} \}$). Regarding the present work, the parameters of Table~\ref{tab:mmt_params} are sufficient for benchmarking the MMT model against DMS reference calculations of simpler $\mathrm{N_2/N}$ and $\mathrm{O_2/O}$ mixtures unaffected by the aforementioned inconsistencies (see Sec.~\ref{sec:verif_vs_dms}).

Columns 9-10 of Table~\ref{tab:mmt_params} list the non-Boltzmann correction factors applied in Eqs.~(\ref{eq:mmt_rate_coefficient}) and (\ref{eq:devib_nb_factor}) respectively. Refer to Sec.~\ref{sec:nb_effects} for further details and the justification for employing common values of $f_k^\mathrm{NB} = 0.5$ and $f_\varepsilon^\mathrm{NB} = 0.85$ for all six reactions. In summary, Table~\ref{tab:mmt_params} contains all parameters to completely evaluate the MMT dissociation rate coefficient and vibrational energy loss terms, regardless of whether one prefers those derived from variable, or fixed-$T_{\mathrm{D}}$ Arrhenius fits. 

Finally, recall that Eqs.~(\ref{eq:mmt_rate_coefficient})-(\ref{eq:devib_nb_factor}) have been implicitly formulated within the framework of a fluid model with only a single vibrational temperature. Specifically, wherever a vibrational temperature is required as input to the MMT expressions, the mixture-averaged value $T_\mathrm{v}$ is used. However, if one was to implement the model within a multi-$T_\mathrm{v}$ fluid description, one should replace $T_\mathrm{v}$ everywhere in Eqs.~(\ref{eq:mmt_rate_coefficient})-(\ref{eq:devib_nb_factor}) with the vibrational temperature of the particular dissociating species.

%-------------------------------------------------------------------------------

\begin{table}%[hbt!]
  
  \caption{Coefficients for the MMT model; $\boldsymbol{\mathrm{A_2}}$ is the dissociating molecular species and $\boldsymbol{\mathrm{M}}$ is the collision partner}
  \label{tab:mmt_params}
  \centering
  
  \begin{subtable}[h]{\textwidth}
   \subcaption{MMT parameters with variable $T_{\mathrm{D}, \mathrm{A_2}}$} \label{tab:td_variable}
   \centering
   
   \begin{tabular}{cccccccccc}
    %\hline
    $\mathrm{A_2}$ & $\mathrm{M}$   & $C \, \mathrm{[cm^3\, s^{-1}\, mol^{-1}\, K^{-n}]}$ & $n$ & $T_{\mathrm{D},\mathrm{A_2}} \, \mathrm{[K]}$ & $\Theta_{\mathrm{v},\mathrm{A_2}} \, \mathrm{[K]}$ & $a_U$ & $U^* \, \mathrm{[K]}$ & $f_k^\mathrm{NB}$ & $f_\varepsilon^\mathrm{NB}$ \\ \hline \hline
    $\mathrm{N_2}$ & $\mathrm{N_2}$ & $3.5967\times 10^{18}$ & $-0.7017$ & $117\,529$ & $3\,411$ & $0.3868$ & $254\,556$ & $0.5$ & $0.85$ \\
    $\mathrm{N_2}$ & $\mathrm{N}$   & $7.9920\times 10^{17}$ & $-0.5625$ & $113\,957$ & $3\,411$ & $0.3668$ & $478\,708$ & $0.5$ & $0.85$ \\
    $\mathrm{N_2}$ & $\mathrm{O_2}$ & $5.0420\times 10^{19}$ & $-0.9991$ & $116\,892$ & $3\,415$ & $0.3001$ & $210\,253$ & $0.5$ & $0.85$ \\
    \hline
    $\mathrm{O_2}$ & $\mathrm{O_2}$ & $3.6932\times 10^{18}$ & $-0.7695$ &  $60\,540$ & $2\,280$ & $0.3965$ &  $57\,343$ & $0.5$ & $0.85$ \\
    $\mathrm{O_2}$ & $\mathrm{O}$   & $9.2109\times 10^{17}$ & $-0.6541$ &  $60\,552$ & $2\,280$ & $0.3537$ & $237\,290$ & $0.5$ & $0.85$ \\
    $\mathrm{O_2}$ & $\mathrm{N_2}$ & $3.1942\times 10^{17}$ & $-0.5430$ &  $62\,949$ & $2\,263$ & $0.3620$ & $385\,466$ & $0.5$ & $0.85$ \\
    \\
   \end{tabular}
  \end{subtable}
  
  \begin{subtable}[h]{\textwidth}
   \subcaption{MMT parameters with $T_{\mathrm{D}, \mathrm{A_2}}$ held fixed} \label{tab:td_fixed}
   \centering
   
   \begin{tabular}{cccccccccc}
    $\mathrm{A_2}$ & $\mathrm{M}$   & $C \, \mathrm{[cm^3\, s^{-1}\, mol^{-1}\, K^{-n}]}$ & $n$ & $T_{\mathrm{D},\mathrm{A_2}} \, \mathrm{[K]}$ & $\Theta_{\mathrm{v},\mathrm{A_2}} \, \mathrm{[K]}$ & $a_U$ & $U^* \, \mathrm{[K]}$ & $f_k^\mathrm{NB}$ & $f_\varepsilon^\mathrm{NB}$ \\ \hline \hline
    $\mathrm{N_2}$ & $\mathrm{N_2}$ & $1.5210\times 10^{17}$ & $-0.4062$ & $113\,200$ & $3\,411$ & $0.5166$ & $365\,046$ & $0.5$ & $0.85$ \\
    $\mathrm{N_2}$ & $\mathrm{N}$   & $4.5968\times 10^{17}$ & $-0.5108$ & $113\,200$ & $3\,411$ & $0.3836$ & $503\,817$ & $0.5$ & $0.85$ \\
    $\mathrm{N_2}$ & $\mathrm{O_2}$ & $3.3947\times 10^{18}$ & $-0.7470$ & $113\,200$ & $3\,415$ & $0.3664$ & $192\,401$ & $0.5$ & $0.85$ \\
    \hline
    $\mathrm{O_2}$ & $\mathrm{O_2}$ & $6.9032\times 10^{17}$ & $-0.6003$ &  $59\,380$ & $2\,280$ & $0.4670$ &  $59\,814$ & $0.5$ & $0.85$ \\
    $\mathrm{O_2}$ & $\mathrm{O}$   & $1.6913\times 10^{17}$ & $-0.4831$ &  $59\,380$ & $2\,280$ & $0.3761$ & $137\,160$ & $0.5$ & $0.85$ \\
    $\mathrm{O_2}$ & $\mathrm{N_2}$ & $1.8313\times 10^{15}$ & $-0.0223$ &  $59\,380$ & $2\,263$ & $0.5801$ &$-174\,299$ & $0.5$ & $0.85$ \\
  \end{tabular}
  \end{subtable}
  
\end{table}

%-------------------------------------------------------------------------------
%\clearpage
\subsection{Vibrational relaxation times} \label{sec:tau_vt_fits}

In our proposed two-temperature nonequilibrium model the relaxation of the combined trans-rotational and vibrational modes toward thermal equilibrium is described by Eq.~(\ref{eq:vib_relax_equation}), the source term $w_\mathrm{v}^\mathrm{relax}$  in the vibrational energy balance equation. Each diatomic species' contribution is modeled as a Landau-Teller relaxation term, where the numerator represents departure of its vibrational energy $e_{\mathrm{v}, s} (T_\mathrm{v})$ at local vibrational temperature $T_\mathrm{v}$ from the corresponding equilibrium value $e_{\mathrm{v}, s} (T)$ at the local gas trans-rotational temperature $T$. The denominators in Eq.~(\ref{eq:vib_relax_equation}) contain every diatomic species' average vibrational relaxation time, itself defined as a weighted average over the inverse of pair-wise relaxation times $\tau_{s,q}^\mathrm{v}$, with $x_q$ as the mole fraction of every collision partner species in the mixture: 
\begin{equation}
 \langle \tau_{s}^\mathrm{v} \rangle = \left( \sum_{q \in S} \frac{x_{q}}{\tau_{s,q}^\mathrm{v}} \right)^{-1}, \qquad s \in D. \label{eq:species_relaxation_time_mw}
\end{equation}

For the species in partially dissociated air one would traditionally rely on the Millikan and White correlation~\cite{MillikanW1963} to determine these pair-wise relaxation times. However, experimental evidence~\cite{IbraguimovaSLSTZ2013, streicher20c, streicher21a, streicher22a, streicher22b}, as well as recent DMS studies~\cite{grover19a, grover19b, torres24b} point to significant departures from the M\&W predictions for some species pairings, especially in the high-temperature limit. In this section we present an alternative expression together with the necessary parameters to calculate characteristic vibrational relaxation times for $\mathrm{N_2- N_2}$, $\mathrm{N_2 - N}$, $\mathrm{O_2 - O_2}$ and $\mathrm{O_2 - O}$ pairs. In a subsequent article, we will present additional parameters for a 5-species air mixture. We have derived all these parameters from curve fits to the DMS results of Ref.~\cite{torres24b} which relied on the most recent ab initio PESs released by the Minnesota chemists. The analytical expression chosen to curve-fit the DMS-derived relaxation times has the form:
\begin{equation}
 \tau_{s,q}^\mathrm{v} = \frac{c_0}{p} \left( \exp \left( m_{s,q}^\mathrm{low} \, T^{-1/3} + n_{s,q}^\mathrm{low} \right) + \exp \left( m_{s,q}^\mathrm{high} \, T^{-1/3} + n_{s,q}^\mathrm{high} \right) \right), \qquad s \in D, q \in S, \label{eq:two_slope_fit}
\end{equation}
where $s$ and $q$ are species indices, $p$ is the static pressure, factor $c_0 = 1 \, \mathrm{atm \cdot s}$ and $T$ is the trans-rotational temperature. This expression assumes that there are two distinct ``low-temperature''  and ``high-temperature'' limiting behaviors for each $\tau_{s,q}^\mathrm{v}$ which can be parameterized by separate straight-line segments on a $\log ( \tau_{s,q}^\mathrm{v} \cdot p )$ vs. $T^{-1/3}$ plot. 

\begin{figure}%[htb]
 \centering
 
 \includegraphics[width=0.75\textwidth]{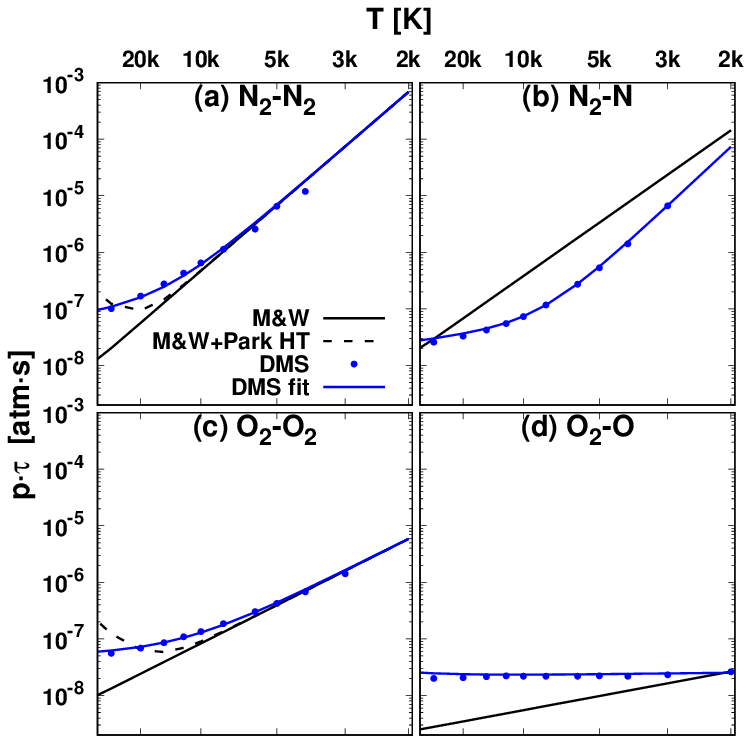}
 
 \caption{Vibrational relaxation times for nitrogen and oxygen. DMS data~\cite{torres24b} and curve fits (blue), M\&W correlation~\cite{MillikanW1963} plus Park's high-temperature correction~\cite{Park1993} (black).}
 \label{fig:ptau_vib_plots}
\end{figure}

The DMS data taken from Ref.~\cite{torres24b} span a wide temperature range, from an extreme upper limit of $T = 100\,000 \, \mathrm{K}$ down to values between $2\,000 - 4\,000 \, \mathrm{K}$, depending on the species pair. The precise temperature points we took into account for fitting the high- and low-temperature portions of Eq.~(\ref{eq:two_slope_fit}) respectively were chosen by trial and error and vary slightly for each species pair. Our goal was to obtain an overall curve fit for every individual pair that would reproduce DMS predictions at both ends of the temperature scale, while still adequately bridging between the two limiting straight-line behaviors. We list the resulting numerical parameters for the four relaxation pairs in Table~\ref{tab:tau_v_fit_reduced} and plot the resulting curve fits for the four collision pairs (blue lines) in Figs.~\ref{fig:ptau_vib_plots}(a)-(d), together with the original DMS data points (blue symbols) from Ref.~\cite{torres24b}. Each sub-figure also shows the Millikan and White correlation (continuous black lines) and for $\mathrm{N_2 - N_2}$ and $\mathrm{O_2 - O_2}$ pairs the added Park high-temperature correction~\cite{Park1993} (dashed black lines).

For the $\mathrm{N_2 - N_2}$ and $\mathrm{O_2 - O_2}$ curve fits in Figs.~\ref{fig:ptau_vib_plots}(a) and (c) in particular, the low-temperature behavior is identical to the one predicted by the respective Millikan and White correlation, as we deliberately set their low-temperature parameters in Eq.~(\ref{eq:two_slope_fit}) to be equivalent to the M\&W ones. We made this choice, because the M\&W correlation for these two relaxing pairs is particularly well-supported by experimental data in the low-temperature limit~\cite{MillikanW1963}. At the same time, it is precisely at temperatures below $4\,000$ and $3\,000 \, \mathrm{K}$ respectively, where the DMS-derived $\tau^\mathrm{v}$ data for $\mathrm{N_2 - N_2}$  and $\mathrm{O_2 - O_2}$ is missing. At the upper end of the temperature scale, both species' DMS curve fits gradually depart from the straight-line Millikan and White trends to predict comparatively longer vibrational relaxation times. This behavior is what motivated Park's high-temperature correction~\cite{Park1993} in the first place. In Figs.~\ref{fig:ptau_vib_plots}(a) and (c) we observe qualitative agreement between the DMS curve fits and Park's correction. However, the correction kicks in at slightly higher temperatures and is more aggressive than the DMS predictions in raising the vibrational relaxation times above those of the M\&W correlation.

For the $\mathrm{N_2 - N}$ pair, across most of the temperature range shown in Fig.~\ref{fig:ptau_vib_plots}(b), the DMS calculations predict vibrational relaxation times somewhat shorter than those obtained with Millikan and White. However, as already reported in Ref.~\cite{grover19b}, we observe the widest gap between Millikan and White and DMS for the $\mathrm{O_2 - O}$-pair. We show its DMS curve fit in Fig.~\ref{fig:ptau_vib_plots}(d) and see it predicting a nearly constant value on a $\log p \cdot \tau^\mathrm{v}$ vs. $T^{-1/3}$ plot. For these diatom-atom pairs the vibrational relaxation times are particularly difficult to access experimentally, as they only become significant in partially dissociated mixtures. Thus, shock-tube experiments attempting to measure them have to operate at higher post-shock temperatures, precisely where the M\&W correlation is unlikely to provide a good fit to the data. Therefore, we relied exclusively on the available DMS data and made no attempt to force any part of either pair's curve fit to agree with prior M\&W correlation predictions.

%-------------------------------------------------------------------------------

\begin{table}%[htb]
 \centering
 \caption{Subset of DMS-derived~\cite{torres24b} fit parameters for pressure-weighted characteristic vibrational relaxation times $\mathrm{[atm \cdot s]}$ to be evaluated according to Eq.~(\ref{eq:two_slope_fit})}
 \label{tab:tau_v_fit_reduced}
 
 \begin{tabular}{l c c c c}
  Species               & $m^\mathrm{low}$ & $n^\mathrm{low}$ & $m^\mathrm{high}$ & $n^\mathrm{high}$ \\
  pair $s-q$            & $\mathrm{[K^{1/3}]}$ & $\mathrm{[-]}$ & $\mathrm{[K^{1/3}]}$ & $\mathrm{[-]}$ \\ \hline \hline
  $\mathrm{N_2-N_2}$    & 221.0 & -24.83 &  33.30 & -17.31 \\
  $\mathrm{N_2-N}$      & 239.4 & -28.52 &  32.34 & -18.39 \\
  \hline
  $\mathrm{O_2-O_2}$    & 129.0 & -22.29 & -4.384  & -16.70 \\
  $\mathrm{O_2-O}$      & 3.018 & -17.74 & -121.6  & -15.82 \\
 \end{tabular} 
 
\end{table}

%-------------------------------------------------------------------------------

%-------------------------------------------------------------------------------
%\clearpage
\section{Quasi-classical trajectory analysis} \label{sec:qct_analysis}

In this section we summarize how we used an extensive QCT database to analyze air dissociation dynamics and correctly scale the reaction rate coefficients as a function of the translational-rotational and vibrational temperatures. This summary constitutes a highly condensed version of the analysis carried out in chapter~3 of Ref.~\cite{Chaudhry2018}. Because we constructed the trajectory data with the assumption of a Boltzmann vibrational distribution, we do not consider non-Boltzmann effects until later, in Sec.~\ref{sec:nb_effects}. The methods and procedures for actually generating the QCT data are fully described in Refs.~\cite{BenderVNPVTSC2015, Bender2016}.

%-------------------------------------------------------------------------------
\subsection{Vibrational energy change per dissociation} \label{sec:devib}
%Sec. 3.4 of Chaudhry2018

\begin{figure}[p]
   \centering
   \begin{subfigure}[t]{0.48\textwidth}
      \centering
      \includegraphics[width=\textwidth]{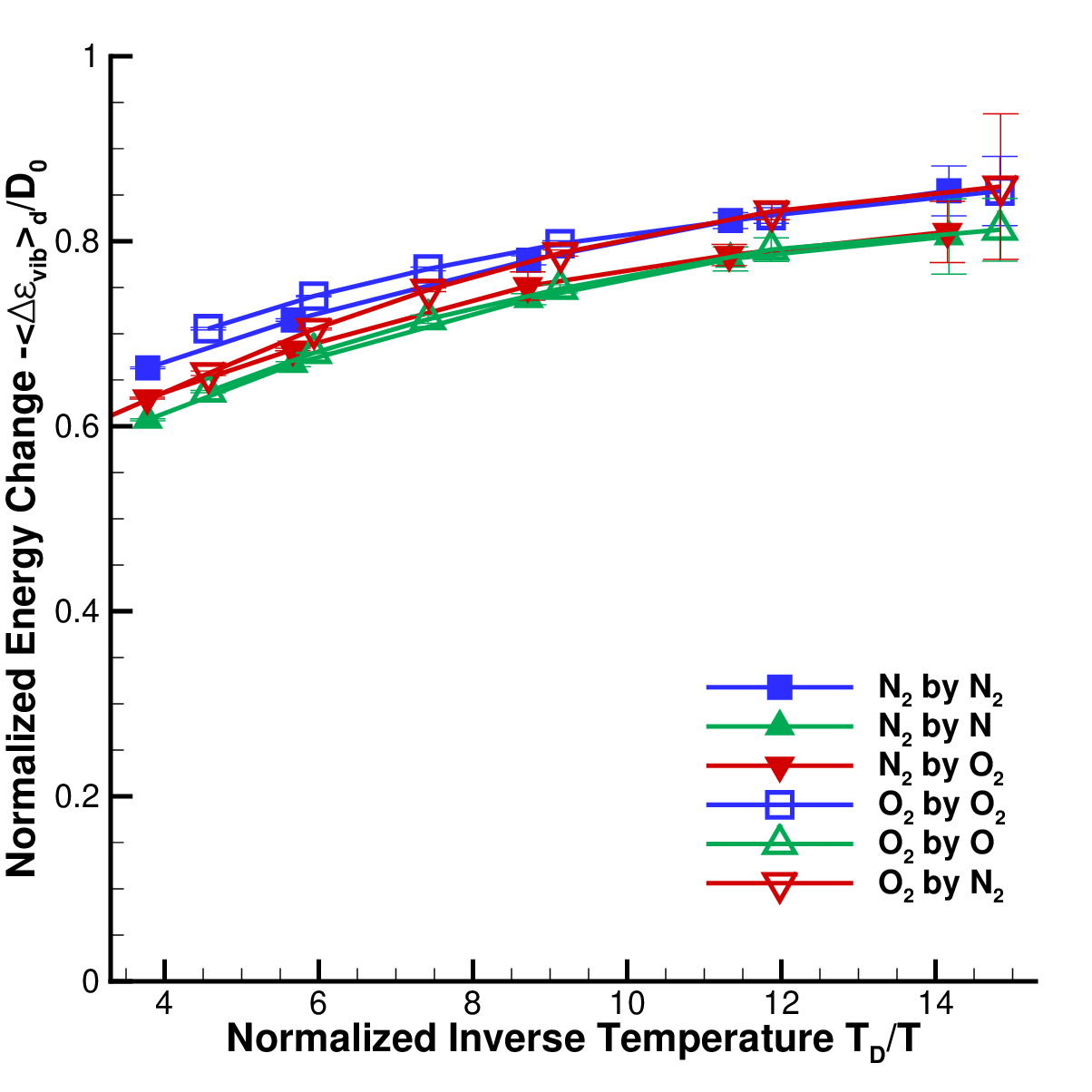}
      \caption{Equilibrium test set}
   \end{subfigure}
   \begin{subfigure}[t]{0.48\textwidth}
      \centering
      \includegraphics[width=\textwidth]{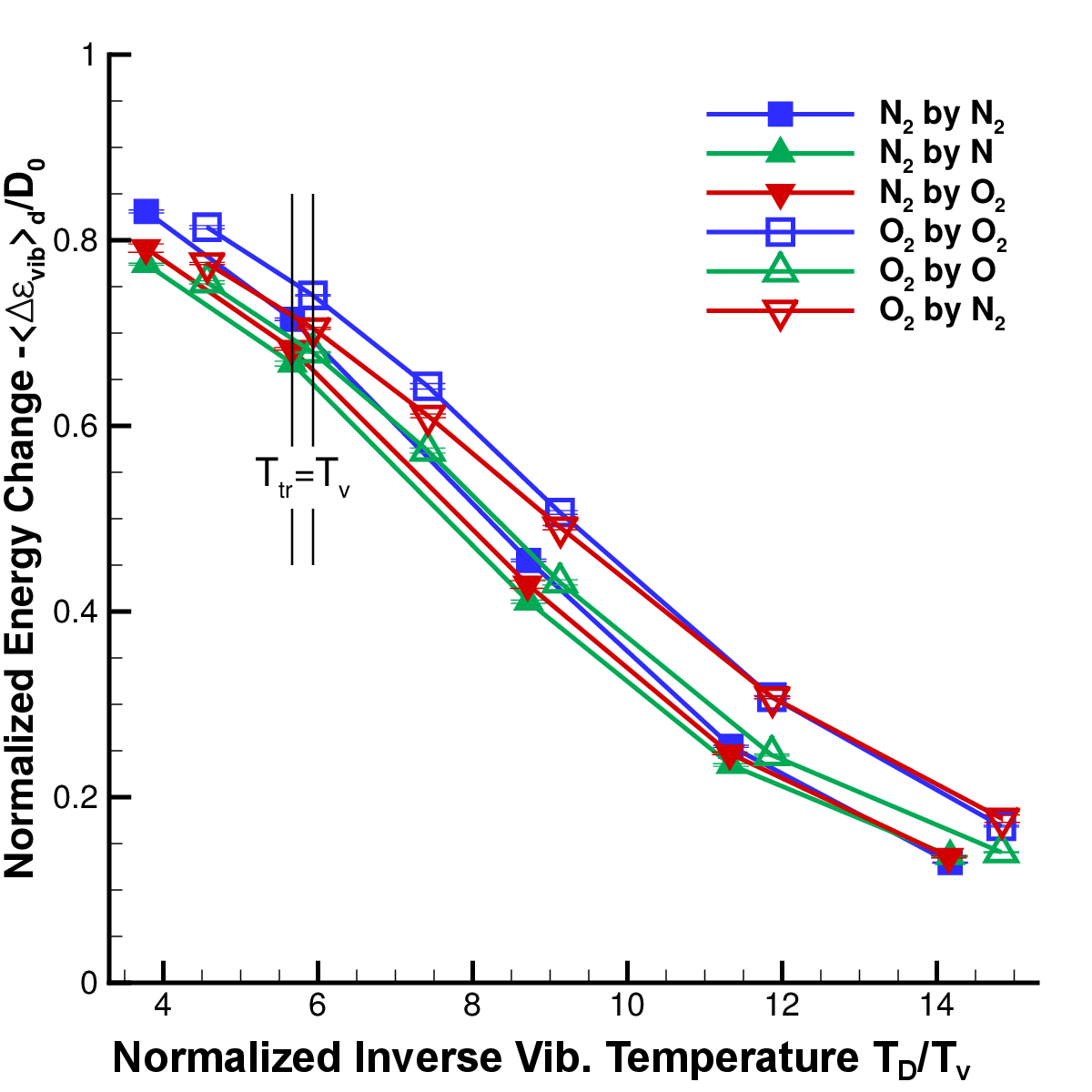}
      \caption{Nonequilibrium test set ($T = 20\,000 \, \mathrm{K}$ for $\mathrm{N_2 \, by \, M}$ and $T = 10\,000 \, \mathrm{K}$ for $\mathrm{O_2 \, by \, M}$).}
   \end{subfigure}
   
   \caption{Average change in vibrational energy per dissociation normalized by $D_{0,s}$ for all 6 dissociation reactions, temperatures normalized by $T_{\mathrm{D},s}$.}
   \label{fig:All_Devib_Norm}
\end{figure}

We first present QCT-derived data for the average vibrational energy removed, or alternatively the negative energy change per dissociation\footnote{For all dissociation reactions considered in this work, $\nu_{s,\mathrm{diss}}^b - \nu_{s,\mathrm{diss}}^f = -1$, which means that vibrational energy \emph{change} and energy \emph{removed} are always related as: $- \langle \Delta \varepsilon_{\mathrm{v},s} \rangle_\mathrm{diss} = \langle \varepsilon_{\mathrm{v},s} \rangle_\mathrm{diss}$}. As discussed at the end of Sec.~\ref{sec:conservation_eqns}, this variable quantifies the effect that dissociation has on the vibrational energy content of the gas. The raw QCT data for $\mathrm{O_2 \, by \, M}$ dissociation with collision partners $\mathrm{M} = \{ \mathrm{O_2, O, N_2} \}$ and $\mathrm{N_2 \, by \, M}$ dissociation with partners $\mathrm{M} = \{ \mathrm{N_2, N, O_2} \}$ is shown in Fig.~3.22(a) of Ref.~\cite{Chaudhry2018} for equilibrium conditions ($T\!=\!T_\mathrm{v}$). Oxygen dissociation results were gathered within the range $4\,000 \, \mathrm{K} \le T \le 13\,000 \, \mathrm{K}$, whereas nitrogen results are available for the interval $8\,000 \, \mathrm{K} \le T \le 30\,000 \, \mathrm{K}$. For the same collision pairs Fig.~3.22(b) of Ref.~\cite{Chaudhry2018} presents QCT-derived vibrational energy change per dissociation at nonequilibrium conditions characterized by a range of $(T,T_\mathrm{v})$ combinations. For oxygen dissociation the sweep over $T_\mathrm{v}$ is shown for constant $T = 10\,000 \, \mathrm{K}$, whereas the nitrogen dissociation sweep is shown for a trans-rotational temperature of $20\, 000 \, \mathrm{K}$. We normalize this change in vibrational energy per dissociation by the corresponding dissociation energy $D_{0,s}$, and the temperature by the corresponding dissociation temperature, $T_{\mathrm{D},s}$. We plot this normalized vibrational energy change per collision in Fig.~\ref{fig:All_Devib_Norm}, where we can see that the normalized vibrational energy change collapses to within about 10\% when normalized by the dissociation energy for both vibrational equilibrium and nonequilibrium conditions.

These results indicate the character of molecules that dominate the dissociation process. As described by Bender~\cite{BenderVNPVTSC2015} for the case of $\mathrm{N_2 + N_2}$ dissociation, when conditions are sampled from thermal equilibrium, molecules that dissociate are primarily vibrationally excited. This specificity weakens as temperature increases, because other modes of energy are more able to make up for a lack of vibrational energy. When conditions are sampled with $T_\mathrm{v}\!<\!T$, rotation compensates for vibration and the molecules which tend to dissociate more easily are rotationally excited instead. The present results demonstrate that those findings apply in general for both molecular nitrogen and oxygen dissociation with all the collision partners studied. Refer to Sec.~3.4 of Ref.~\cite{Chaudhry2018} for further details.

%-------------------------------------------------------------------------------
\subsection{Effect of reactant states} \label{sec:reactant_states}
%Sec. 3.5 of Chaudhry2018

We next consider the effect that each reactant state has on whether dissociation will occur. As described in Ref.~\cite{ChaudhryBSC2018}, the range in probability of dissociation as a function of a specific state quantifies its effect. For example, a low-$v$ molecule is much less likely to dissociate than a high-$v$ molecule and thus the vibrational quantum number has a large effect on dissociation. However, the probability of dissociation varies substantially for different reactions and conditions, which is problematic when aggregating data in the QCT database. 

In order to facilitate the analysis, for every reactant pair of interest (either $\mathrm{O_2}(v) \, \mathrm{by} \, \mathrm{M}$, or $\mathrm{N_2}(v) \, \mathrm{by} \, \mathrm{M}$) we may compute the vibration-specific, thermally averaged dissociation probability $P_d \, (v) = P_d \, (\varepsilon_s(v), T)$ as the ratio of all quasiclassical trajectories where the dissociating molecule initially possesses vibrational energy $\varepsilon_s(v)$ relative to the total number of trajectories for the reactant pair, regardless of outcome. We further calculate the Boltzmann-weighted, thermally averaged dissociation probability $P_d = P_d (T, T_\mathrm{v}) = \sum_{v=0}^{n_{v,s}-1} \{ P_d \, (v) \exp \left( - \varepsilon_s (v) / \mathrm{k_B} T_\mathrm{v} \right) \} / Q_{\mathrm{v}, s} (T_\mathrm{v})$. The ratio of both quantities yields the support factor:
\begin{equation}
 \mathcal{S}_d \, (v)=\frac{P_d \, (v)}{P_d}. \label{eq:Support}
\end{equation}

If a reactant state exhibits substantial variation in $\mathcal{S}_d$, then that state has a large effect on whether dissociation will occur. This is illustrated by Fig.~3.25(a) of Ref.~\cite{Chaudhry2018} for the special case of $\mathrm{O_2}(v) + \mathrm{N_2}$ dissociation at $T\!=\!T_\mathrm{v} = 10\,000 \, \mathrm{K}$. At these conditions the support factor is seen to increase over nearly 4 orders of magnitude from the lowest to the highest-lying $v$-states of oxygen. When the support factor is then plotted for a range of temperatures, as in Fig.~3.26 of Ref.~\cite{Chaudhry2018}, it also becomes clear that the selectivity for dissociation on $v$-level is highest at low $T$ and gradually becomes less pronounced at high temperatures.

\begin{figure}%[p]
   \centering
   \begin{subfigure}[t]{0.48\textwidth}
      \centering
      \includegraphics[width=\textwidth]{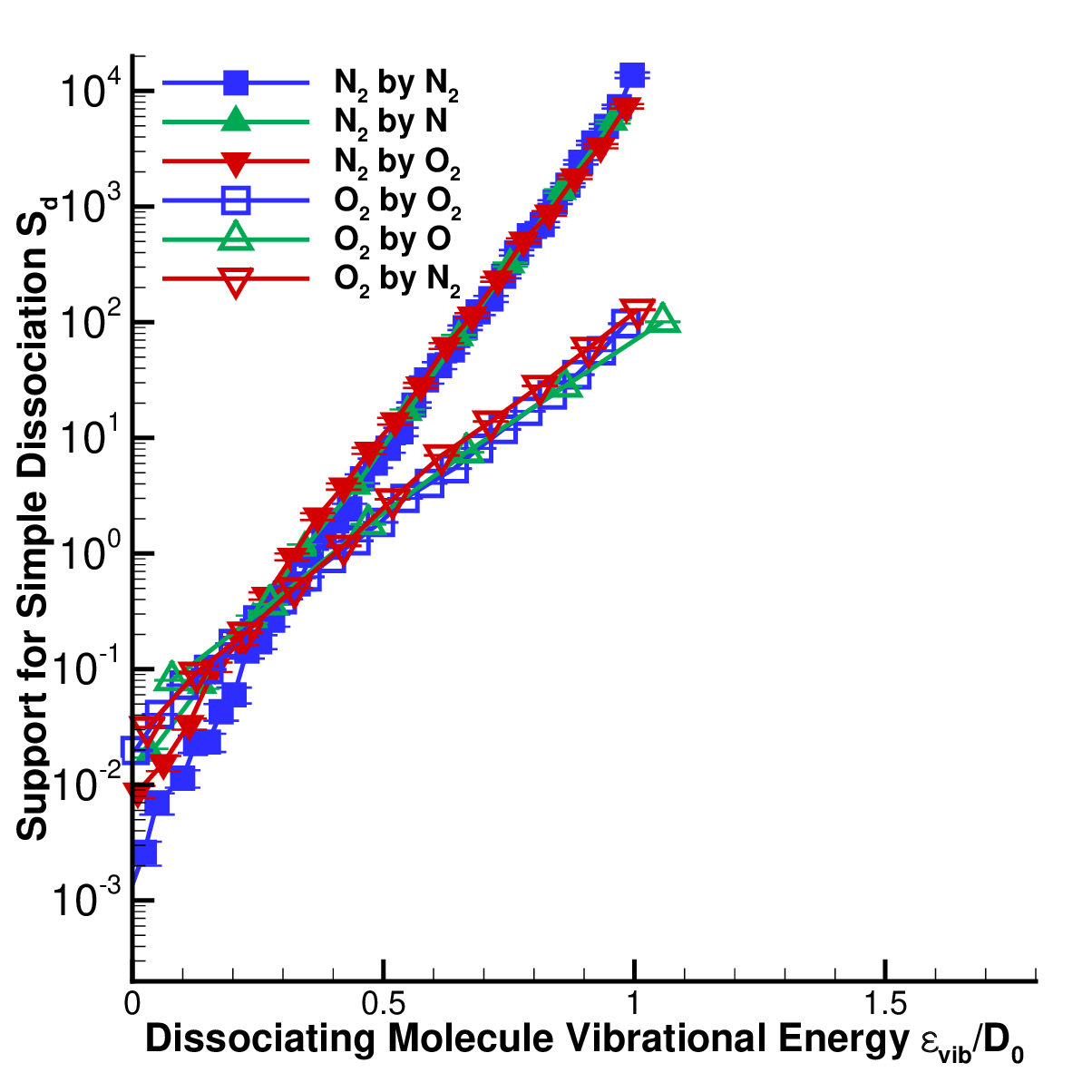}
      \caption{Dependence on vibrational energy}
   \end{subfigure}
   \begin{subfigure}[t]{0.48\textwidth}
      \centering
      \includegraphics[width=\textwidth]{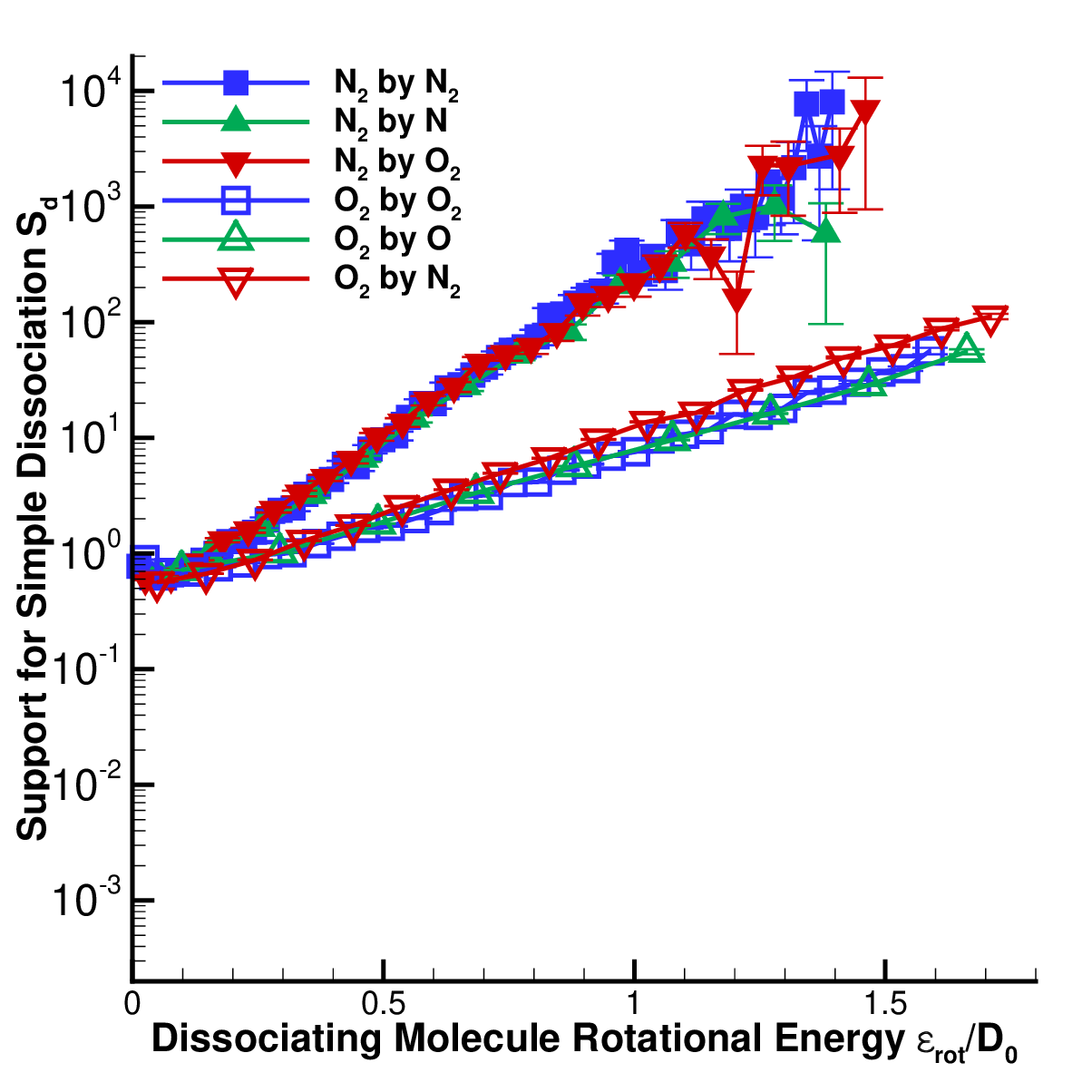}
      \caption{Dependence on rotational energy}
   \end{subfigure}
   
   \caption{Support for single/simple dissociation for all dissociation reactions, at $T = T_\mathrm{v} = 10\,000 \, \mathrm{K}$.}
   \label{fig:Support_All}
\end{figure}

The concept of the support for dissociation may be generalized. Instead of plotting against discrete $v$-levels, the support factor may alternatively be parameterized in terms of the rotational, vibrational or overall internal energies, or the relative translational energy of the collision pair. This analysis is done in Sec.~3.5.2 of Ref.~\cite{Chaudhry2018} for all six oxygen and nitrogen dissociation reactions at $T\!=\!T_\mathrm{v} = 10\,000 \, \mathrm{K}$. Although the support is seen to depend on all these quantities to some extent, the strongest and most direct dependence is observed for vibrational and rotational energies of the dissociating diatom. Therefore, in Fig.~\ref{fig:Support_All} we compare the effect that the vibrational and rotational components of the dissociating molecule's internal energy have on $\mathcal{S}_d$. These plots show that the rotational and vibrational energy both affect dissociation, and that the effect of vibrational energy is larger than that of rotational energy. Importantly, these normalized results also collapse, simplifying the development of a model to capture vibrational favoring of dissociation.

%-------------------------------------------------------------------------------

%-------------------------------------------------------------------------------

%-------------------------------------------------------------------------------
\subsection{Role of vibration} \label{sec:role_of_vibration}
%Sec. 3.5.5 of Chaudhry2018

In Secs.~\ref{sec:devib} and \ref{sec:reactant_states} we presented aggregate QCT-derived data for vibrational energy change and support factor for all six dissociation reactions in Table~\ref{tab:mmt_params} over a wide range of temperatures. Perhaps unsurprisingly, the overall message of this analysis is that the vibrational energy of the dissociating diatom has the strongest effect on dissociation probability. In this subsection, we attempt to explain \emph{why} vibrational energy has a stronger effect on dissociation than rotational energy. For this discussion we use the $\mathrm{O_2 + O \rightarrow 3O}$ reaction, averaged over all ground-electronic-state interactions at $T = T_\mathrm{v} = 10\,000 \, \mathrm{K}$. 

Figure~3.33 of Ref.~\cite{Chaudhry2018} shows a contour plot of dissociation probability as a function of rovibrational level $(v,j)$ for this reaction. The molecules most likely to dissociate (orange to red shaded area) populate levels with either very large $v$ (bottom right corner), very large $j$ (top left corner) or $(v,j)$ combinations running along the upper-right edge of the plot. These molecules all have in common that they possess internal (i.e. vibrational+rotational) energies close to, or above the dissociation threshold $D_0$ for $\mathrm{O_2}$. Furthermore, as shown in greater detail by Figs.~3.34(a) and (b) of Ref.~\cite{Chaudhry2018}, we find that the probability of dissociation from each rovibrational level is not fully predicted by either vibrational or rotational energy alone.

\begin{figure}
   \centering
   \begin{subfigure}[t]{0.48\textwidth}
      \centering
      \includegraphics[width=\textwidth]{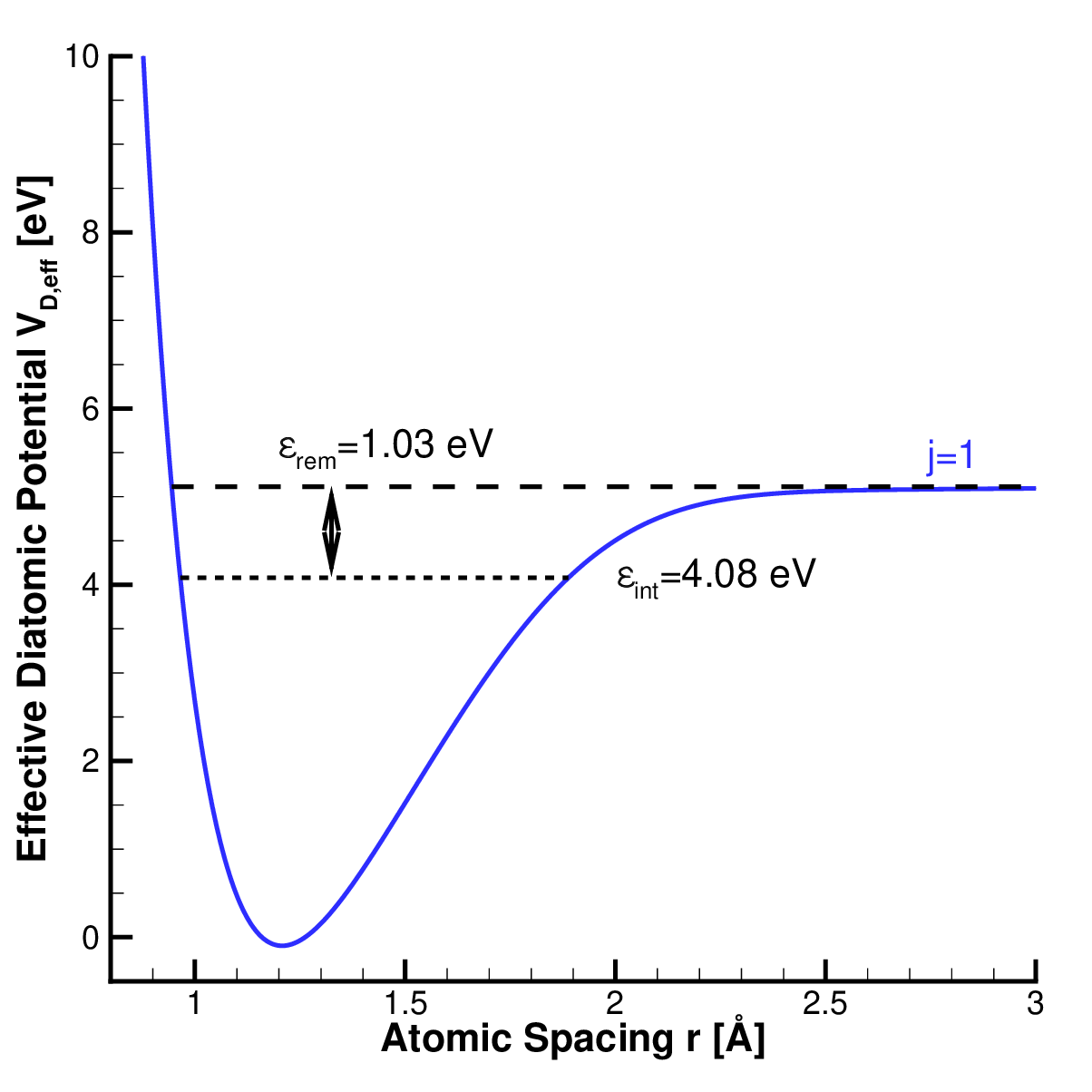}
      \caption{$v=26$, $j=1$; high $P_d$}
   \end{subfigure}
   \begin{subfigure}[t]{0.48\textwidth}
      \centering
      \includegraphics[width=\textwidth]{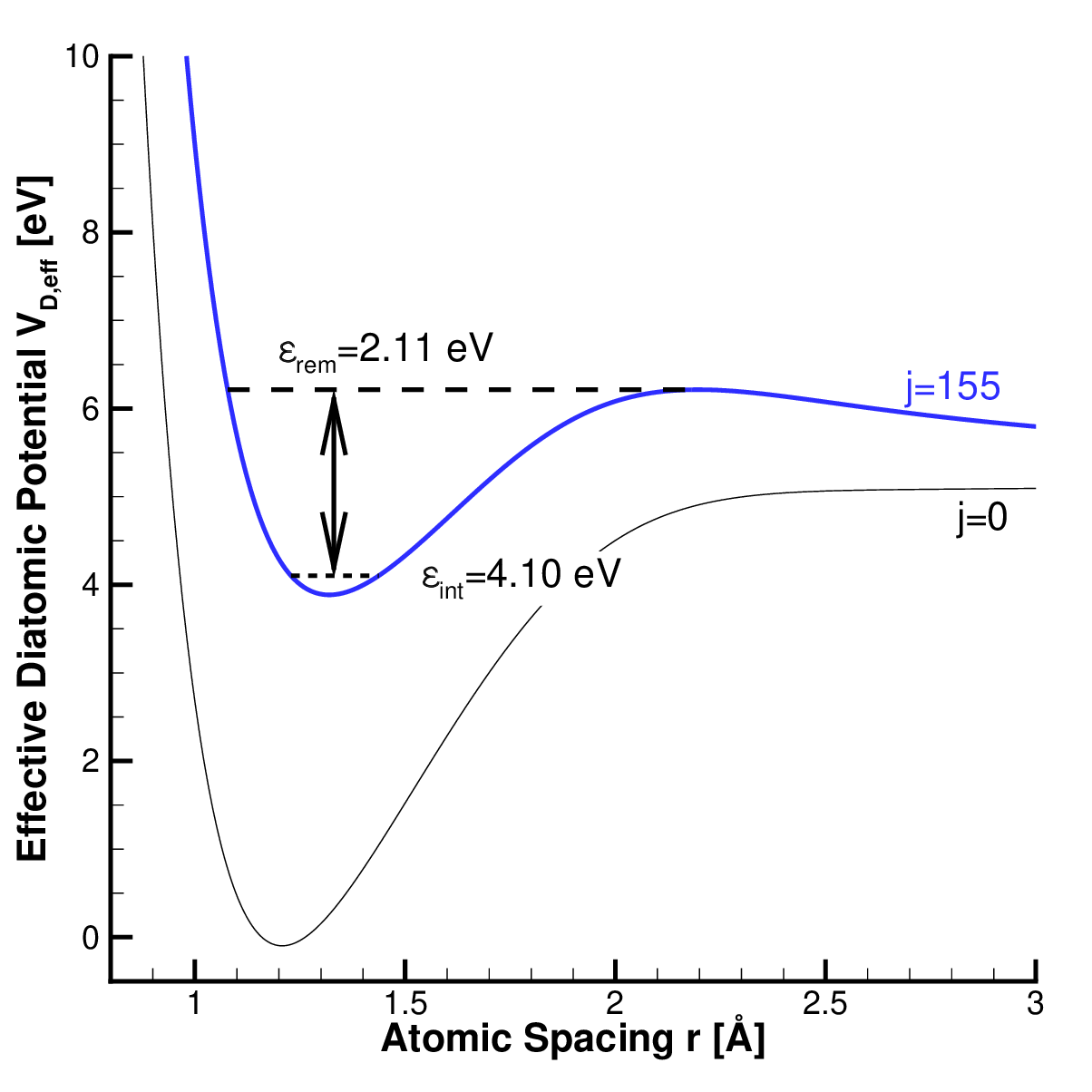}
      \caption{$v=1$, $j=155$; low $P_d$}
   \end{subfigure}
   
   \caption{Effective diatomic potential for two rovibrational states with similar internal energies}
   \label{fig:jpdf_diapots}
\end{figure}

For a given internal energy, we find molecules that are more vibrationally excited to be more likely to dissociate; this conclusion is consistent with our findings using the support factor. For example, we identify two sample rovibrational levels with similar internal energy, one with small $\varepsilon_\mathrm{vib}$ and the other with large $\varepsilon_\mathrm{vib}$ (see Fig.~3.35(b) in Ref.~\cite{Chaudhry2018}). The state which has most of its internal energy in rotation is approximately $8.5$ times less likely to dissociate than the state which has its internal energy in vibration. We show the effective diatomic potentials for these two states in Fig.~\ref{fig:jpdf_diapots}. Although neither of these molecules is quasibound, the molecule in Fig.~\ref{fig:jpdf_diapots}(b), with mostly rotational energy requires more energy to dissociate due to the centrifugal barrier, assuming its angular momentum remains constant. We quantify this by defining the \emph{remainder energy}, $\varepsilon_\mathrm{rem}$, which is the difference between the centrifugal barrier height and the internal energy $\varepsilon_\mathrm{int} = \varepsilon_\mathrm{vib} + \varepsilon_\mathrm{rot}$,
\begin{equation}
   \varepsilon_\mathrm{rem}(v,j)= \max\left[ V_\mathrm{D,eff}(j) \right] - \varepsilon_\mathrm{int}(v,j) \label{eq:erem}
\end{equation}
in which function $\max[V_\mathrm{D, eff}(j)]$ refers to the local maximum of the effective diatomic potential curve; see Figure \ref{fig:jpdf_diapots}.

\begin{figure}
   \centering
   \includegraphics[width=0.48\textwidth]{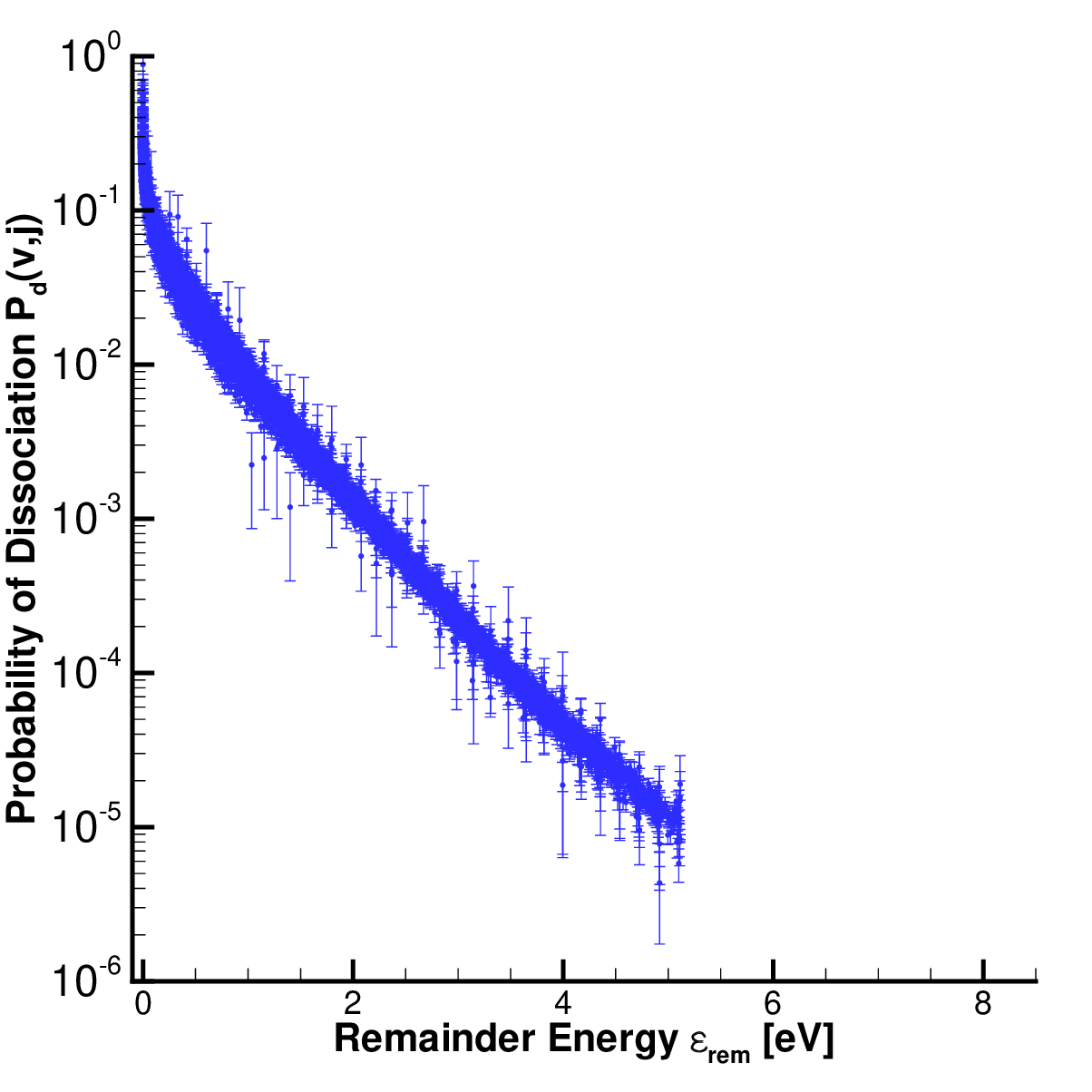}
   
   \caption{Probability of dissociation from a specific remainder energy $\varepsilon_\mathrm{rem}$}
   \label{fig:jpdf_erem}
\end{figure}

Several analyses in the literature have used the barrier height in theoretical arguments, often as an effective dissociation energy~\cite{Nikitin1974, Jaffe1986, LuoAM2018} and our interpretation is similar. Here we directly evaluate its validity as a predictor for dissociation using the QCT data. We show the probability of dissociation compared to the remainder energy in Fig.~\ref{fig:jpdf_erem}. We find that the remainder energy is a relatively good predictor of dissociation; the scatter in probability is approximately a factor of 2, a substantial portion of which is due to statistical uncertainty. We made similar observations for the other dissociation reactions we have considered. Thus, the remainder energy appears to be a good predictor for single dissociation (see footnote\footnotemark[1] in Sec.~\ref{sec:mmt_model_explanation}). This result also indicates an important physical mechanism: the reason why rotational energy is less effective at promoting dissociation is because rotational energy also increases the centrifugal barrier. Thus, adding one unit of rotational energy reduces $\varepsilon_\mathrm{rem}$ by an amount less than the energy added. This reduction does not occur for vibrational energy.

Admittedly, the preceding explanation has its limitations. Whether any specific molecule splits apart will not only be a function of its effective diatomic potential curve and associated remainder energy, but will depend in a non-trivial manner on the precise collision dynamics, i.e. the collision pair's relative translational energy, the spatial orientation of the collision partner's incoming trajectory relative to the molecule, as well as the timing of closest approach. Even after individual trajectories have been aggregated to average out these dynamic effects, Fig.~\ref{fig:jpdf_erem} represents only a single equilibrium temperature condition. When $T$ is changed, it shifts the relative translational energy distribution of colliding pairs. This will undoubtedly change the aforementioned dissociation probability ratio of $8.5$ between the two chosen levels, as well as affect the overall $P_d$ vs. $\varepsilon_\mathrm{rem}$ curve. Thus, the height of the centrifugal barrier is likely not the only factor for explaining dissociation probability. Regardless of the precise reasons, the aggregate QCT results presented in this section confirm the strong vibrational bias in dissociation reactions. This ultimately justifies the major modeling choices presented in Sec.~\ref{sec:mmt_model_explanation} for the MMT model.

%-------------------------------------------------------------------------------
\section{Fitting to QCT data} \label{sec:fitting_qct_data}
%Sec. 3.6 of Chaudhry2018

In this section, we continue summarizing parts of chapter~3 (in particular Sec.~3.6) of Ref.~\cite{Chaudhry2018}. We describe the methodology and results for fitting aggregate quantities of interest from QCT to relatively simple analytic forms. When possible, we use functional forms that exist in the literature. We consider fits for both the dissociation rate and the change in vibrational energy per dissociation for the equilibrium and nonequilibrium test sets. But, as we will show, we can extract the model parameters appearing in Table~\ref{tab:mmt_params} from just the equilibrium test sets for dissociation rate coefficient and energy removed by dissociation. We consider each dissociation reaction with a specific collision partner separately.

%-------------------------------------------------------------------------------
\subsection{Equilibrium dissociation rate coefficient} \label{sec:equil_diss_rate_coeff}

We first fit dissociation rate data for the equilibrium test set ($T\!=\!T_\mathrm{v}$) to the modified Arrhenius form of Eq.~(\ref{eq:karr}). We write the equation in terms of inverse temperature and the 
natural logarithm of the dissociation rate coefficient,
\begin{equation}
   \ln k^\mathrm{Arr} = \ln C - n \ln \frac{1}{T} - \frac{T_{\mathrm{D},s}}{T}.
\end{equation}

We choose to use this form for several reasons. First, the equation is now linear in the coefficients we wish to determine $\ln C$, $-n$, and $-T_{\mathrm{D},s}$. We can therefore use a standard least squares estimator to determine the coefficients (e.g., Riley et al.~\cite{RileyHB2006} [p.~1271–1277]). Second, a least squares minimization in $\ln k^\mathrm{Arr}$ corresponds to minimizing the relative error in $k^\mathrm{Arr}$, which is desirable. For the temperature ranges of interest, $k^\mathrm{Arr}$ varies by many orders of magnitude, so minimizing the absolute error would essentially neglect relative error in the small but important $k^\mathrm{Arr}$ values at low temperatures.

We show results for all dissociation reactions we have considered in Table \ref{tab:mmt_params}. We find the data to be well described by the modified Arrhenius form; the goodness-of-fit metric $R^2$ (computed in $\ln k^\mathrm{Arr}$ space) is better than $0.9998$ for all results shown. For all effective dissociation rate coefficients, the maximum deviation is less than 5\%, and it is often better than the estimated statistical QCT error for low-temperature conditions. The fitted values of $n$ are typically between $0$ and $-1.5$, which is consistent with theoretical results (e.g. Ref.~\cite{Byron1959}) and data available in the literature (e.g., Ref.~\cite{Park1996}). Note that the fitted dissociation temperature in Table~\ref{tab:td_variable} is within 5\% of the dissociation temperature based on the diatomic PES's dissociation energy. In the experimental literature, it is common to fit Arrhenius coefficients with a fixed $T_{\mathrm{D},s}$. We have done this with accepted values of $T_{\mathrm{D},s}$ in Table~\ref{tab:td_fixed}, and find that the fits are only slightly changed relative to those shown in Table \ref{tab:td_variable}. A more detailed listing containing all modified Arrhenius fit parameters obtained with and without a fixed $T_{\mathrm{D},s}$ can be found in Tables 3.1 and 3.2 of Ref.~\cite{Chaudhry2018}.

%-------------------------------------------------------------------------------
\subsection{Nonequilibrium dissociation rate coefficient} \label{sec:neq_diss_rate_coeff}

Next, we consider the dissociation rate coefficient for the nonequilibrium QCT dataset. The rate coefficient at thermal nonequilibrium is often considered relative to the one at thermal equilibrium by defining and fitting a nonequilibrium correction factor, Eq.~(\ref{eq:Zfact}). We define the dissociation rate coefficients at equilibrium by the Arrhenius form, with values taken from the previous section's fits. Similar to the equilibrium rate coefficient, we fit in terms of the natural logarithm (i.e. $\ln Z$). A variety of nonequilibrium correction factors are defined in the literature; the book chapter by Losev et al.~\cite{LosevKMPS1996} provides a good summary. Here we focus on the Marrone-Treanor model, because we found it to be consistent with the QCT data. Prior work~\cite{Chaudhry2018} compared the Losev $\beta$ model~\cite{LosevG1962}, the Park $T-T_\mathrm{v}$ model~\cite{Park1988_Two-Temperature, Park1990}, and the more recent Luo et al.~\cite{LuoAM2018} model.

The Marrone-Treanor preferential dissociation model~\cite{MarroneT1963} assumes that vibrationally-excited molecules are more likely to dissociate, with an exponential dependence determined by the model parameter $U$. Rewriting Eq.~(\ref{eq:Zfact}) in logarithmic form, we have:
\begin{equation}
   \ln Z(T,T_\mathrm{v}) = \ln \tilde{Q}_{\mathrm{v},s} (T) + \ln \tilde{Q}_{\mathrm{v},s} (T_F) - \ln \tilde{Q}_{\mathrm{v},s} (T_\mathrm{v}) - \ln \tilde{Q}_{\mathrm{v},s} (-U)
\end{equation}
with $T_F$ defined in Eq.~(\ref{eq:TF}). We use a nonlinear least squares fit to determine the $U$ model parameter for the nonequilibrium dissociation reactions. 

\begin{figure}
   \begin{subfigure}[t]{0.48\linewidth}
      \centering
      \includegraphics[width=\textwidth]{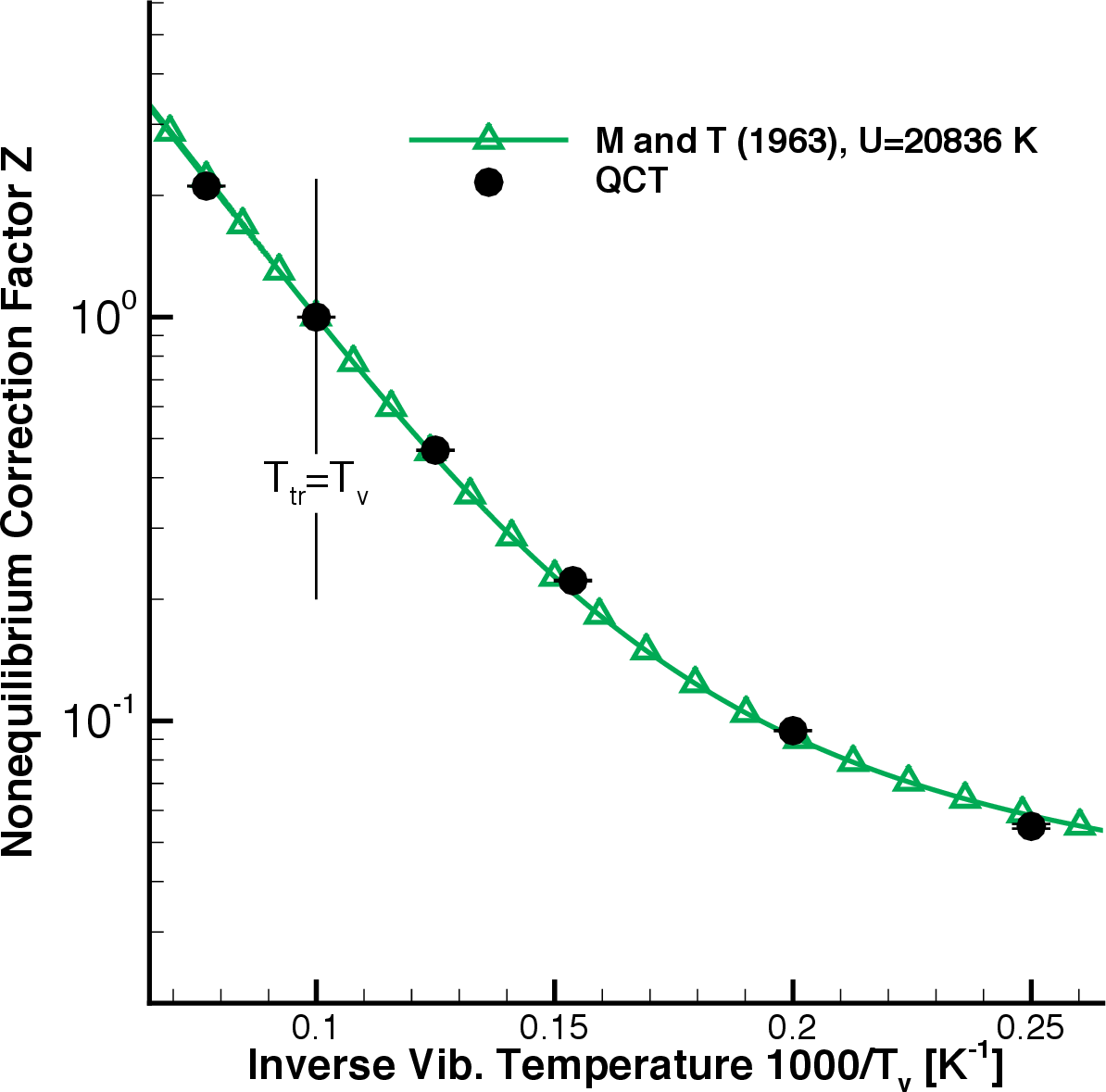}
      \caption{Partner $\mathrm{N_2}$ $Z$}
   \end{subfigure}
   \begin{subfigure}[t]{0.48\linewidth}
      \centering
      \includegraphics[width=\textwidth]{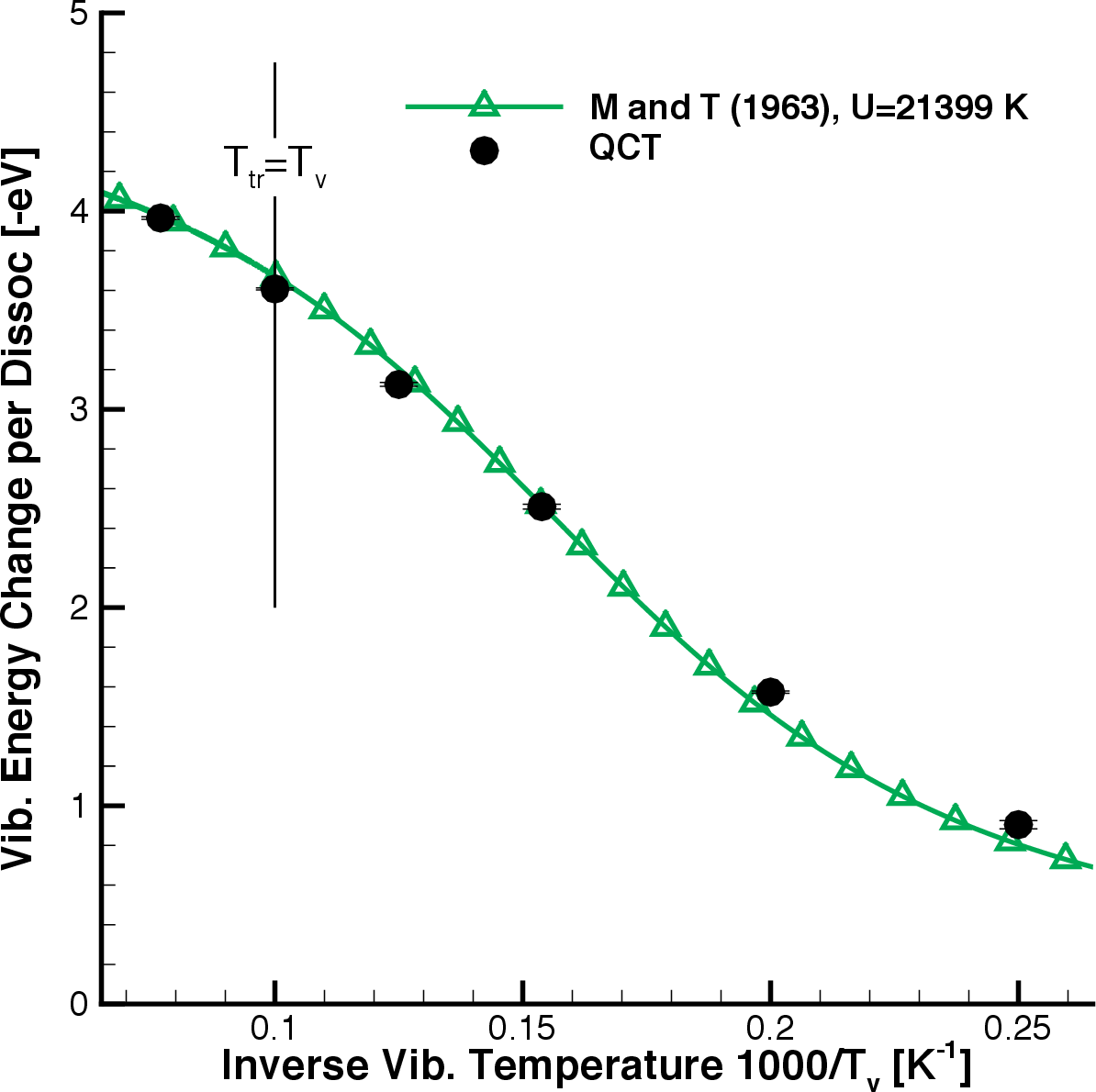}
      \caption{Partner $\mathrm{N_2}$ $\langle\varepsilon_v\rangle$}
   \end{subfigure}
   
   \caption{Nonequilibrium factor $Z$ and average change in vibrational energy per dissociation $\langle \varepsilon_{\mathrm{v},s} \rangle_\mathrm{diss}$ for effective oxygen dissociation in nonequilibrium test set.}
   \label{fig:O2_Noneq_Fits}
\end{figure}

In Fig.~\ref{fig:O2_Noneq_Fits}(a) we show the curve fit to $Z(T,T_\mathrm{v})$ for effective oxygen dissociation with collision partner $\mathrm{N_2}$ at $T=10\,000\,\mathrm{K}$. The best fit to the QCT data is achieved for $U = 20\,836 \, \mathrm{K}$. We have chosen to show just this one reaction at a single trans-rotational temperature for brevity, but equivalent plots for $Z$ in $\mathrm{O_2}$ dissociation with partners $\mathrm{O_2}$ and $\mathrm{O}$ can be found in Fig.~3.38(a) and 3.38(b) of Ref.~\cite{Chaudhry2018}. We have further analyzed nitrogen dissociation with the collision partners listed in Table~\ref{tab:mmt_params} in the same fashion (not shown) and found the Marrone-Treanor model to fit the data well; the maximum relative deviation is within 5\% for nitrogen dissociation with partners $\mathrm{N_2}$ and $\mathrm{N}$ and oxygen dissociation with partners $\mathrm{O_2}$ and $\mathrm{O}$, and within 7.5\% for both dissociation reactions of $\mathrm{N_2 + O_2}$ pairs.

%-------------------------------------------------------------------------------
\subsection{Nonequilibrium vibrational energy change per dissociation} \label{sec:neq_devib}

We now consider the vibrational energy change per dissociation for the nonequilibrium QCT dataset. As we have previously described, this term represents the effect of dissociation on the nonequilibrium thermodynamic state. In the past, this term's behavior in the limit of extreme thermal nonequilibrium has prevented several preferential dissociation models from being successfully used in CFD (see Ref.~\cite{ChaudhryBSC2018}).

The Marrone-Treanor model yields for the vibrational energy removed per dissociation (see Eq.~(8) in Ref.\cite{MarroneT1963}):
\begin{equation}
  \langle \varepsilon_{\mathrm{v},s} \rangle_\mathrm{diss} \left( T_F \right) = \frac{1}{Q_{\mathrm{v},s} (T_F)} \sum_{v=0}^{n_{v,s}-1} \varepsilon_{s} (v) \exp \left( -\frac{\varepsilon_{s} (v)}{\mathrm{k_B} T_F} \right). \label{eq:devib_mt_orig}
\end{equation}

The ground-electronic-state $\mathrm{N_2}$ diatomic potential of the Minnesota PESs accommodates nearly 60 discrete vibrational levels, whereas for $\mathrm{O_2}$ it is about 45 (see supplemental information of Ref.~\cite{torres24b} for detailed listings). The exact formulation given by Eq.~(\ref{eq:devib_mt_orig}) requires one to perform two separate Boltzmann-weighted sums spanning all the diatom's vibrational energy levels $\varepsilon_{s}(v)$ at pseudotemperature $T_F (T,T_\mathrm{v})$, one for the numerator and another one for the vibrational partition function:
\begin{equation}
 Q_{\mathrm{v},s} (T_F) = \sum_{\mathrm{v}=0}^{n_{v,s}-1} \exp \left( - \frac{\varepsilon_{s} (v)}{\mathrm{k_B} T_F} \right) \label{eq:exact_partfunc}
\end{equation}
in the denominator. However, Knab et al.~\cite{KnabFM1995} avoid this costly calculation by approximating the exact state sums with simpler expressions derived from sums over simple harmonic oscillator (SHO) levels $[\varepsilon_{s} (v)]^\mathrm{SHO} = v \, \mathrm{k_B} \Theta_{\mathrm{v},s} $ truncated below the dissociation threshold $D_{0,s}$. This is how they arrive at the approximate expression for the partition function given by Eq.~(\ref{eq:approx_partfunc}) and are able to approximate Eq.~(\ref{eq:devib_mt_orig}) with Eq.~(\ref{eq:Knab}), the form originally given as Eq.~(39) in Knab et al.~\cite{KnabFM1995}. 

\begin{figure}%[htb]
 \begin{subfigure}[t]{0.48\linewidth}
      \centering
      \includegraphics[width=\textwidth]{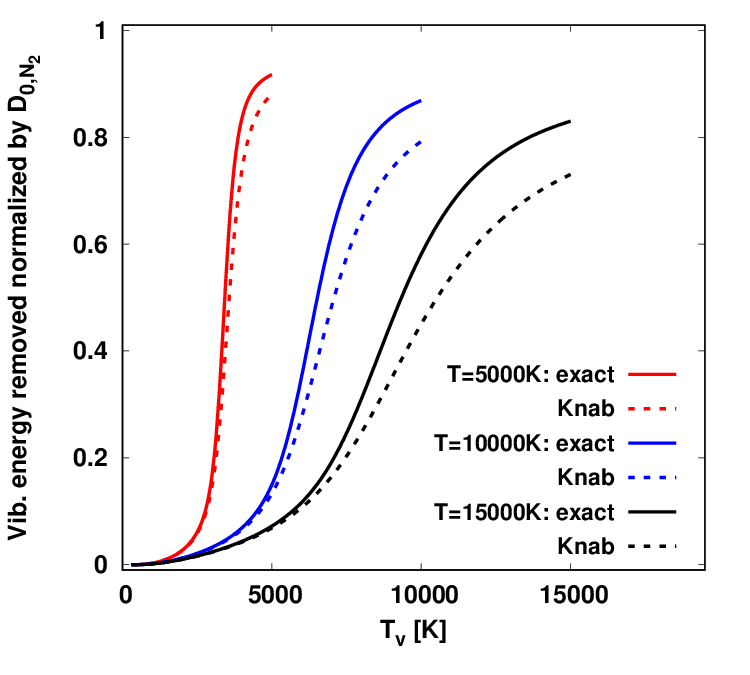}
      \caption{Vibrational energy loss in dissociation from exact expression [Eq.~(\ref{eq:devib_mt_orig})] vs. Knab's formula [Eq.~(\ref{eq:Knab})].}
      \label{fig:devib_comparison_1}
   \end{subfigure}~
   \begin{subfigure}[t]{0.04\linewidth}
    
   \end{subfigure}
   \begin{subfigure}[t]{0.48\linewidth}
      \centering
      \includegraphics[width=\textwidth]{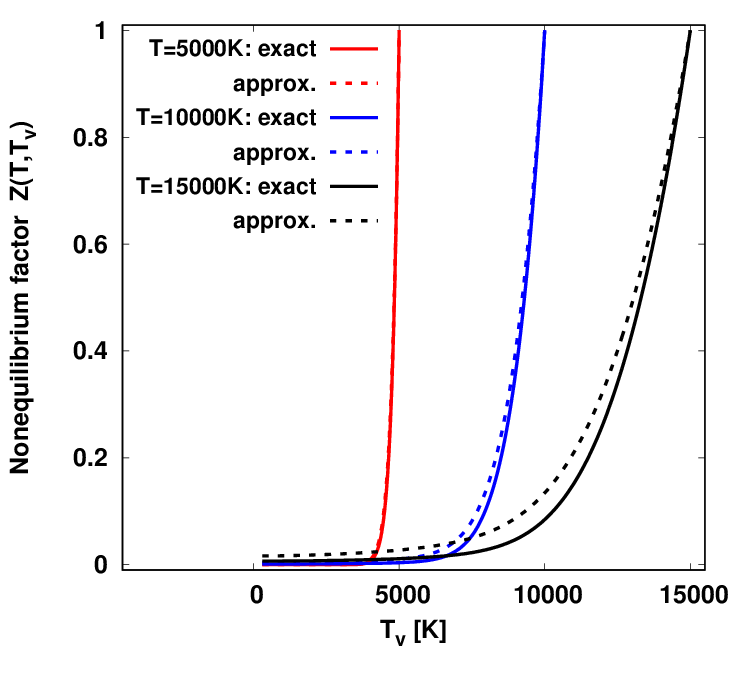}
      \caption{Nonequilibrium factor [Eq.~(\ref{eq:Zfact})] computed using exact expressions for $Q_\mathrm{v, N_2}$ [Eq.~(\ref{eq:exact_partfunc})] vs. approximate partition functions $\tilde{Q}_\mathrm{v, N_2}$ [Eq.~(\ref{eq:approx_partfunc})].}
      \label{fig:z_comparison_1}
   \end{subfigure}
   
 \caption{Exact (solid lines) vs. approximate (dotted lines) expressions for vibrational energy loss $\langle \varepsilon_\mathrm{v, N_2} \rangle_\mathrm{diss} (T,T_\mathrm{v})$ and nonequilibrium factor $Z (T,T_\mathrm{v})$. Parameters for reaction $\mathrm{2N_2 \rightarrow 2N + N_2}$ from Table~\ref{tab:td_variable}.}
 \label{fig:Knab_error_comparison}
\end{figure}

In Fig.~\ref{fig:devib_comparison_1} we compare predictions for $\langle \varepsilon_{\mathrm{v},s} \rangle_\mathrm{diss} \, (T,T_\mathrm{v})$ for reaction $\mathrm{2N_2 \rightarrow 2N + N_2}$ from the exact expression, i.e. Eq.(\ref{eq:devib_mt_orig}), against Knab's approximate formula, i.e. Eq.~(\ref{eq:Knab}). We conduct sweeps over $T_\mathrm{v} \le T$ for three distinct values of $T$. Knab's formula (dotted lines) consistently under-predicts the values obtained with the exact expression (solid lines). At all three translational temperatures shown, the gap between both curves grows with increasing vibrational temperature and becomes largest at thermal equilibrium. But even at $T = T_\mathrm{v} = 15\,000\, \mathrm{K}$ (black curves), the absolute energy difference only maxes out at about 10\% of $D_{0, \mathrm{N_2}}$. 

In Fig.~\ref{fig:z_comparison_1} we compare the result of evaluating Eq.~(\ref{eq:Zfact}) for the nonequilibrium factor $Z (T,T_\mathrm{v})$ using the exact vibrational partition functions according to Eq.~(\ref{eq:exact_partfunc}), vs. the truncated SHO ones from Eq.~(\ref{eq:approx_partfunc}). At $T_\mathrm{v} \ll T$ the approximate formula (dotted lines) predicts slightly larger values than the exact one (continuous lines). This is most clearly visible for $T = 15\,000 \, \mathrm{K}$ (black lines). However, in absolute terms the error in $Z$ remains small and vanishes at thermal equilibrium.

%-------------------------------------------------------------------------------

%-------------------------------------------------------------------------------

Despite the discrepancies, switching to the approximate expressions is still worthwhile, as it is extremely useful in reducing the computational cost of the MMT model in large-scale CFD computations, where Eqs.~(\ref{eq:approx_partfunc}) and (\ref{eq:Knab}) must be evaluated for every dissociation reaction multiple times per cell and per time step. Furthermore, by applying the SHO approximation in this manner we avoid the need for including detailed listings of $\varepsilon_s (v)$ for every dissociating species as part of the two-temperature model. Instead, we merely require their characteristic vibrational and dissociation temperatures.

In hindsight, the preceding discussion justifies fitting $\langle \varepsilon_{\mathrm{v},s} \rangle_\mathrm{diss} (T, T_\mathrm{v})$ to Eq.~(\ref{eq:Knab}) for the Marrone-Treanor model. We focus on results generated with the variable dissociation temperatures listed in Table~\ref{tab:td_variable}. An equivalent fit using the fixed $T_\mathrm{D,s}$ values from Table~\ref{tab:td_fixed} can be performed, but is not shown here. As mentioned in Sec.~\ref{sec:mmt_model_explanation}, the characteristic vibrational temperature $\Theta_{\mathrm{v},s}$ is defined using the energy gap between the two lowest-lying vibrational levels derived from the corresponding diatomic potential.

The results for oxygen dissociation with collision partner $\mathrm{N_2}$ are shown in Fig.~\ref{fig:O2_Noneq_Fits}(b). QCT data from the nonequilibrium test set are shown as black dots and the best fit to Eq.~(\ref{eq:Knab}) with $U = 21\,399 \, \mathrm{K}$ is shown in green. Equivalent plots for $\langle \varepsilon_{\mathrm{v},\mathrm{O_2}} \rangle_\mathrm{diss}$ in $\mathrm{O_2}$ dissociation with partners $\mathrm{O_2}$ and $\mathrm{O}$ can be found in Figs.~3.38(d) and 3.38(e) of Ref.~\cite{Chaudhry2018}. Just as for oxygen, we have analyzed $\langle \varepsilon_{\mathrm{v},\mathrm{N_2}} \rangle_\mathrm{diss}$ for nitrogen dissociation with the collision partners listed in Table~\ref{tab:mmt_params} in the same fashion (not shown).

Overall, we found the Marrone-Treanor model to accurately describe the data: the maximum error is $0.13 \, \mathrm{eV}$ for oxygen dissociation and $0.49 \, \mathrm{eV}$ for nitrogen dissociation. We found the best-fit model parameter $U$ to be similar when fitting to rate data and when fitting to change in energy data (within 7\% for oxygen dissociation and 15\% for nitrogen dissociation). For all cases, the fitted model parameter is between $\frac{1}{3.5}$ and $\frac{1}{2}$ of the dissociation temperature.

%-------------------------------------------------------------------------------
\subsection{Marrone-Treanor characteristic probability temperature} \label{sec:mmt_pseudotemp}

In the original derivation of their preferential dissociation model Marrone and Treanor~\cite{MarroneT1963} had to rely on shock tube data to infer $U$ (what they called the characteristic probability temperature) essentially by trial and error. Given the large uncertainty in the experimental data, this meant they could only obtain a single, rough approximate value for a given reaction instead of a generally valid functional dependence for $U$. By contrast, with the help of our extensive QCT database we have the ability to calculate $U$ from first principles and may examine its variation over a wide range of nonequilibrium conditions. Based on the results of Secs.~\ref{sec:neq_diss_rate_coeff}-\ref{sec:neq_devib}, we thus propose a modification to the Marrone-Treanor model meant to yield a good fit to both quantities in the equilibrium ($T\!=\!T_\mathrm{v}$) and nonequilibrium ($T\!\ne\!T_\mathrm{v}$) test sets and for all effective dissociation reactions considered. 

As was seen in Fig.~\ref{fig:O2_Noneq_Fits}, the nonequilibrium factor and vibrational energy change per dissociation are each well described by a particular value of $U$ at a fixed value of $T$ and the two lie fairly close to one another. Specifically, for $T = 10\,000 \, \mathrm{K}$ the value of $U$ that simultaneously fits both quantities well, lies somewhere between $20\,836 \, \mathrm{K}$ and $21\,399\, \mathrm{K}$. Similar behavior was observed at other temperatures and for the other reactions studied.

The goal is now to find a simple expression that allows us to estimate $U$ (and by extension $T_F$) with good enough accuracy for any $T,T_\mathrm{v}$ combination and for any reaction within the entire QCT data set. We face the choice of whether to base our analysis on fitting to data for $Z$, for $\langle \varepsilon_{\mathrm{v},s} \rangle_\mathrm{diss}$, or some combination thereof. Because we are primarily interested in minimizing the error in vibrational energy decrease per dissociation, we opt for fitting to this quantity and proceed as follows. By construction the value of $U = 21\,399\, \mathrm{K}$ in Fig.~\ref{fig:O2_Noneq_Fits}(b) also fits $\langle \varepsilon_{\mathrm{v},s} \rangle_\mathrm{diss}$ at thermal equilibrium, i.e. the single point where $T_\mathrm{v}\!=\!T=\!10\,000 \, \mathrm{K}$. Thus, for a given temperature $T$ we may find the value of $U$ that describes the entire nonequilibrium behavior by only having to match the value for $\langle \varepsilon_{\mathrm{v},s} \rangle_\mathrm{diss}$ from QCT at the corresponding equilibrium condition.

\begin{figure}%[t]
   \begin{subfigure}[t]{0.48\textwidth}
      \centering
      \includegraphics[width=\textwidth]{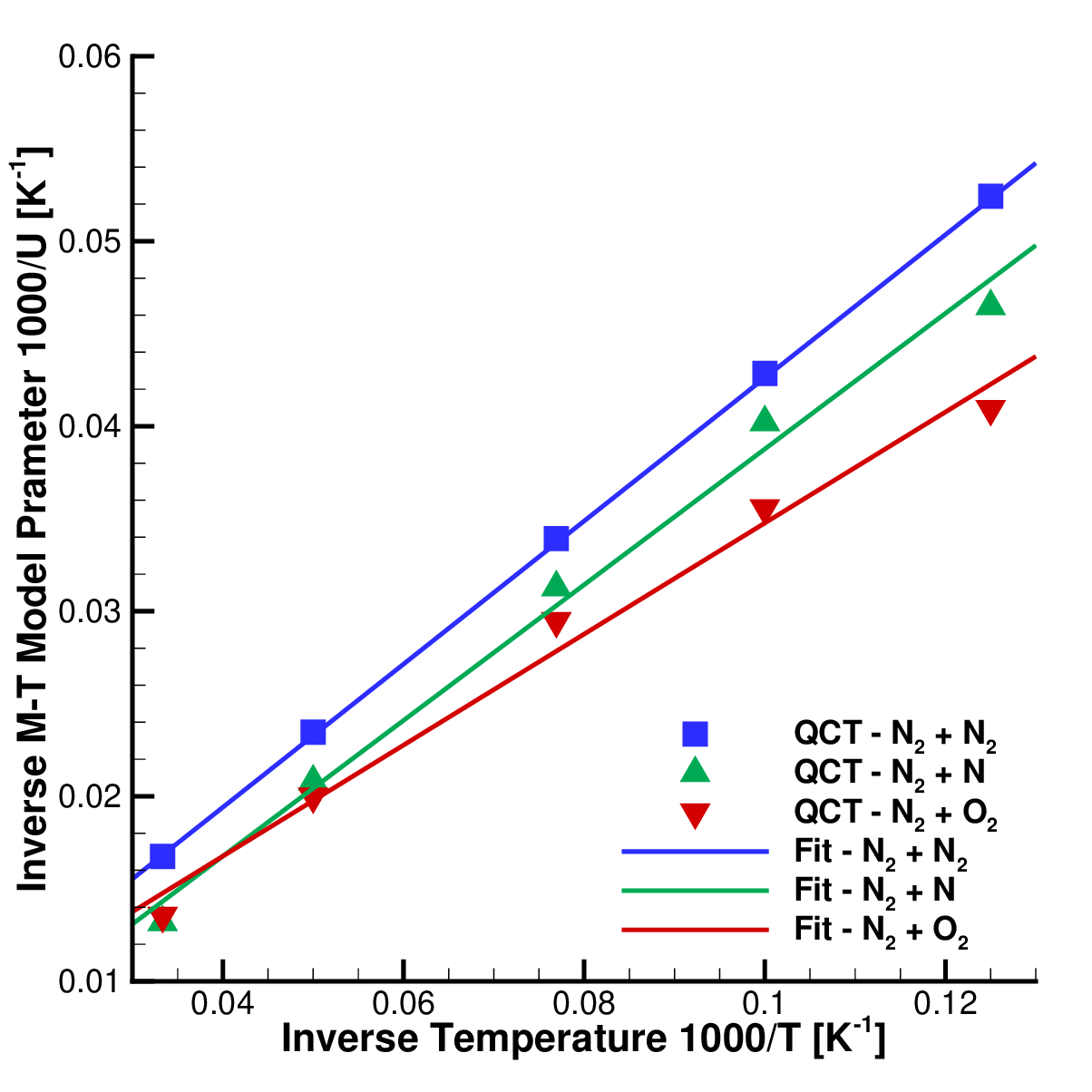}
      \caption{Nitrogen dissociation}
   \end{subfigure}
   \begin{subfigure}[t]{0.48\textwidth}
      \centering
      \includegraphics[width=\textwidth]{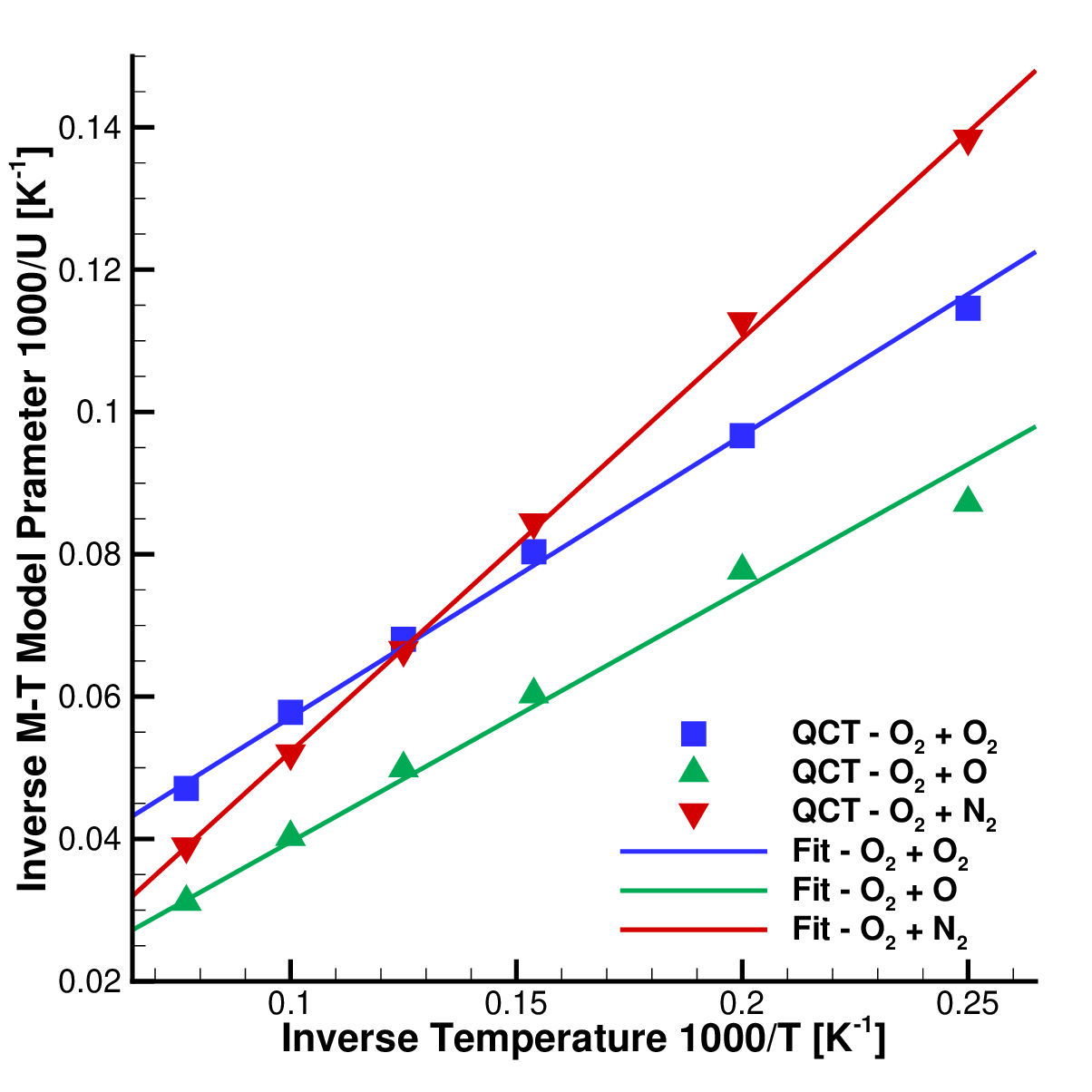}
      \caption{Oxygen dissociation}
   \end{subfigure}
   
   \caption{Dependence of Marrone-Treanor parameter $1/U$ on $1/T$, determined from vibrational energy loss per dissociation in equilibrium test set (symbols) and linear fit to Eq.~(\ref{eq:NewFit_U}) (lines).}
   \label{fig:MT_U}
\end{figure}

For each temperature condition in the equilibrium QCT dataset, we therefore determine the value of $U$ that best matches $\langle \varepsilon_{\mathrm{v},s} \rangle_\mathrm{diss} (T,T_\mathrm{v}\!=\!T)$, via a Newton iteration on Eq.~(\ref{eq:Knab}). We now have the dependence of $U$ on $T$, which we observe to be fit well by the linear relationship in inverse temperature space that is Eq.~(\ref{eq:NewFit_U}). We confirm this by plotting the aggregate results of the fitting procedure in Fig.~\ref{fig:MT_U} for the six reactions of Table~\ref{tab:mmt_params}. Nitrogen dissociation data in Fig.~\ref{fig:MT_U}(a) were fit within the range $8\,000 \, \mathrm{K} \le T \le 30\,000\, \mathrm{K}$, oxygen dissociation data in Fig.~\ref{fig:MT_U}(b) were fit within the range $4\,000 \, \mathrm{K} \le T \le 13\,000\, \mathrm{K}$.

By substituting Eq.~(\ref{eq:NewFit_U}) into Eq.~(\ref{eq:Knab}) we may end up determining the model parameters $a_U$ and $U^*$ by simply fitting to $\langle \varepsilon_{\mathrm{v},s} \rangle_\mathrm{diss} \, (T,T_\mathrm{v}=T)$ directly. In the end this means that the dependence of $U$ on $T$ is extracted entirely from the equilibrium vibrational energy change due to dissociation. The resulting values for $a_U$ and $U^*$ have been listed in Table~\ref{tab:mmt_params}. Note that these values for a given reaction differ between sub-tables~\ref{tab:td_variable} and \ref{tab:td_fixed} due to the precise values for $T_{\mathrm{D},s}$ employed.

%-------------------------------------------------------------------------------

%-------------------------------------------------------------------------------
\subsection{Comparison of MMT model with QCT data} \label{sec:comparison_mmt_qct}

\begin{figure}%[p]
   \centering
   \begin{subfigure}[t]{0.48\textwidth}
      \centering
      \includegraphics[width=\textwidth]{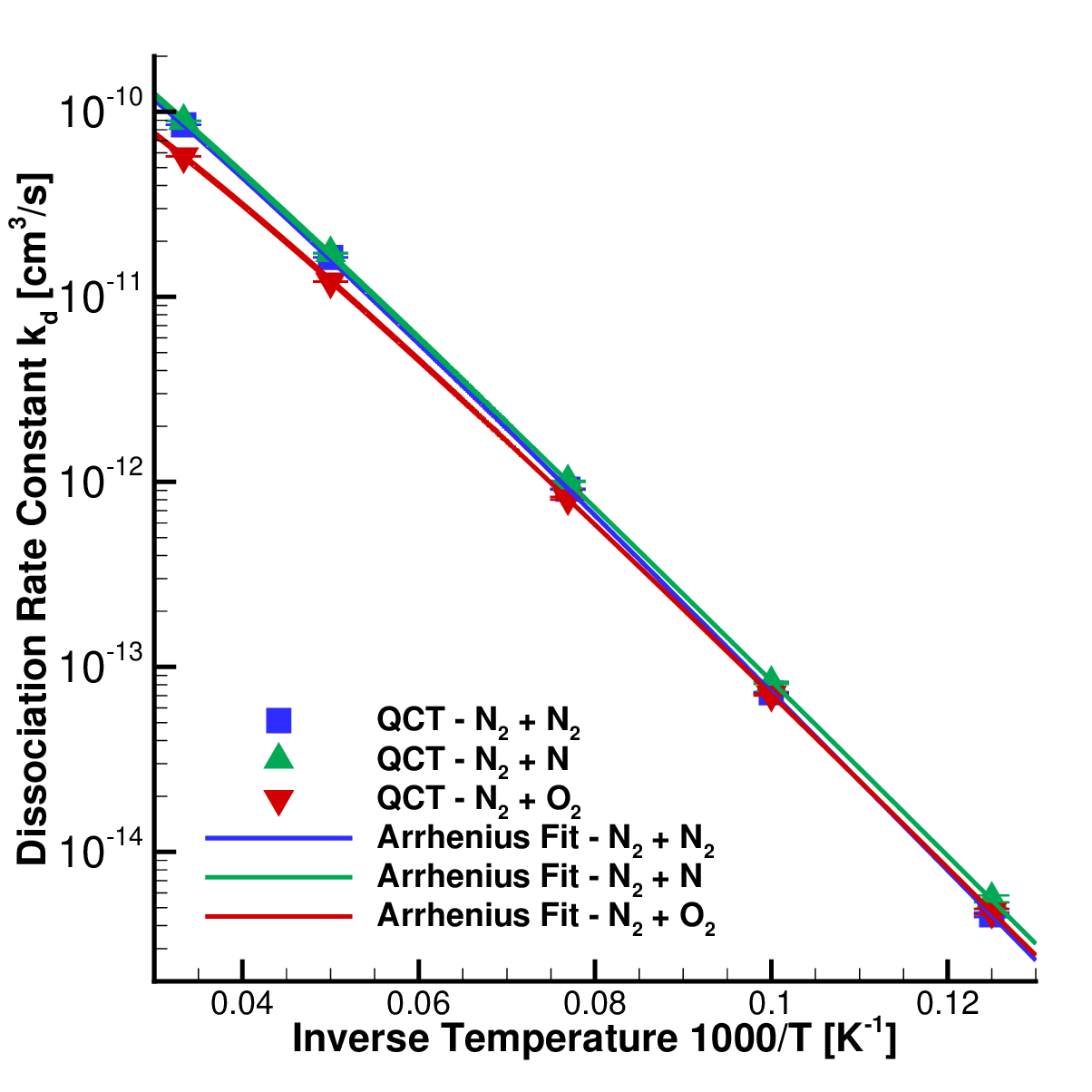}
      \caption{Dissociation rate coefficient, $T = T_\mathrm{v}$}
      \label{subfig:N2-dissoc_Arr}
   \end{subfigure}
   \begin{subfigure}[t]{0.48\textwidth}
      \centering
      \includegraphics[width=\textwidth]{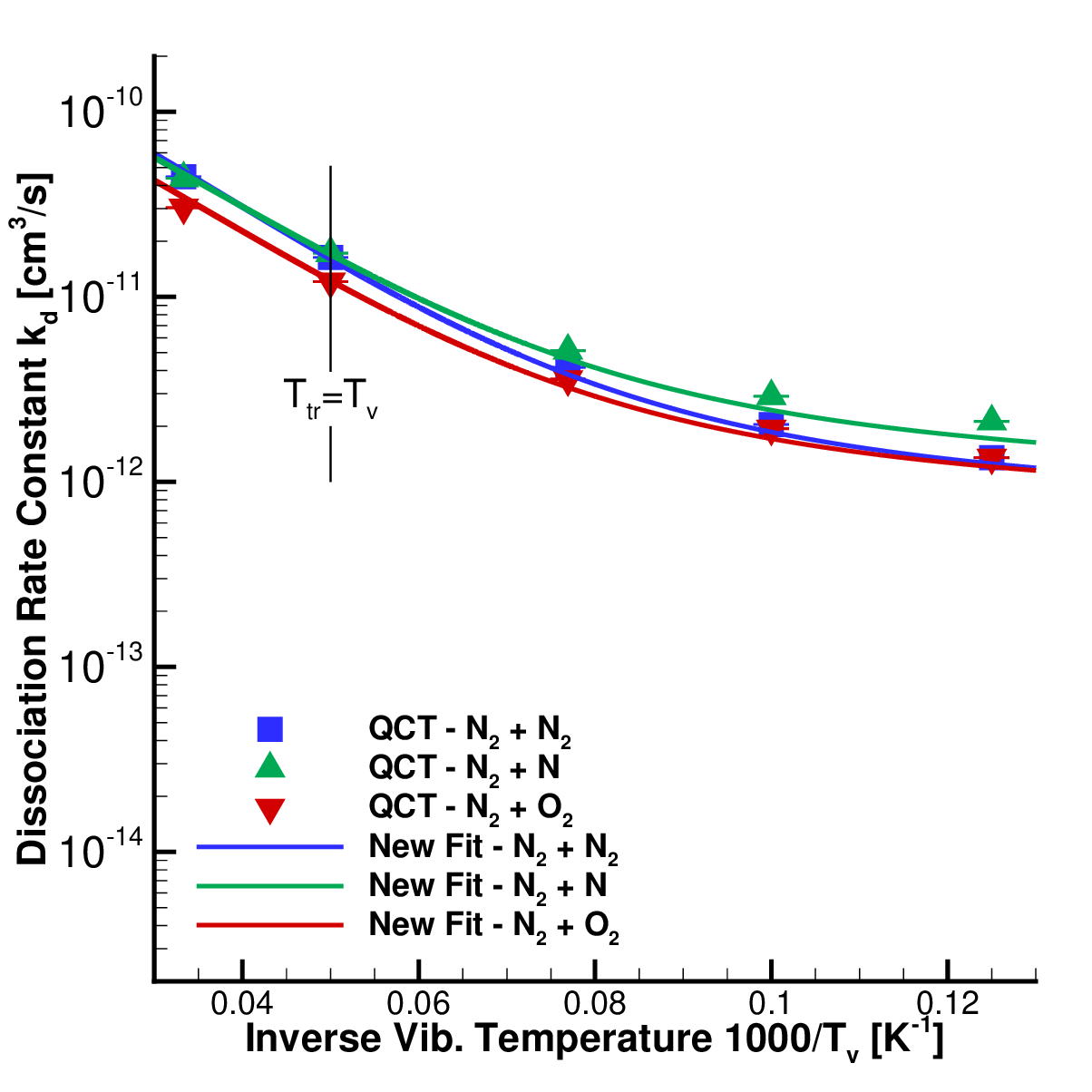}
      \caption{Dissociation rate coefficient, $T = 20\,000 \, \mathrm{K}$}
   \end{subfigure}

   \begin{subfigure}[t]{0.48\textwidth}
      \centering
      \includegraphics[width=\textwidth]{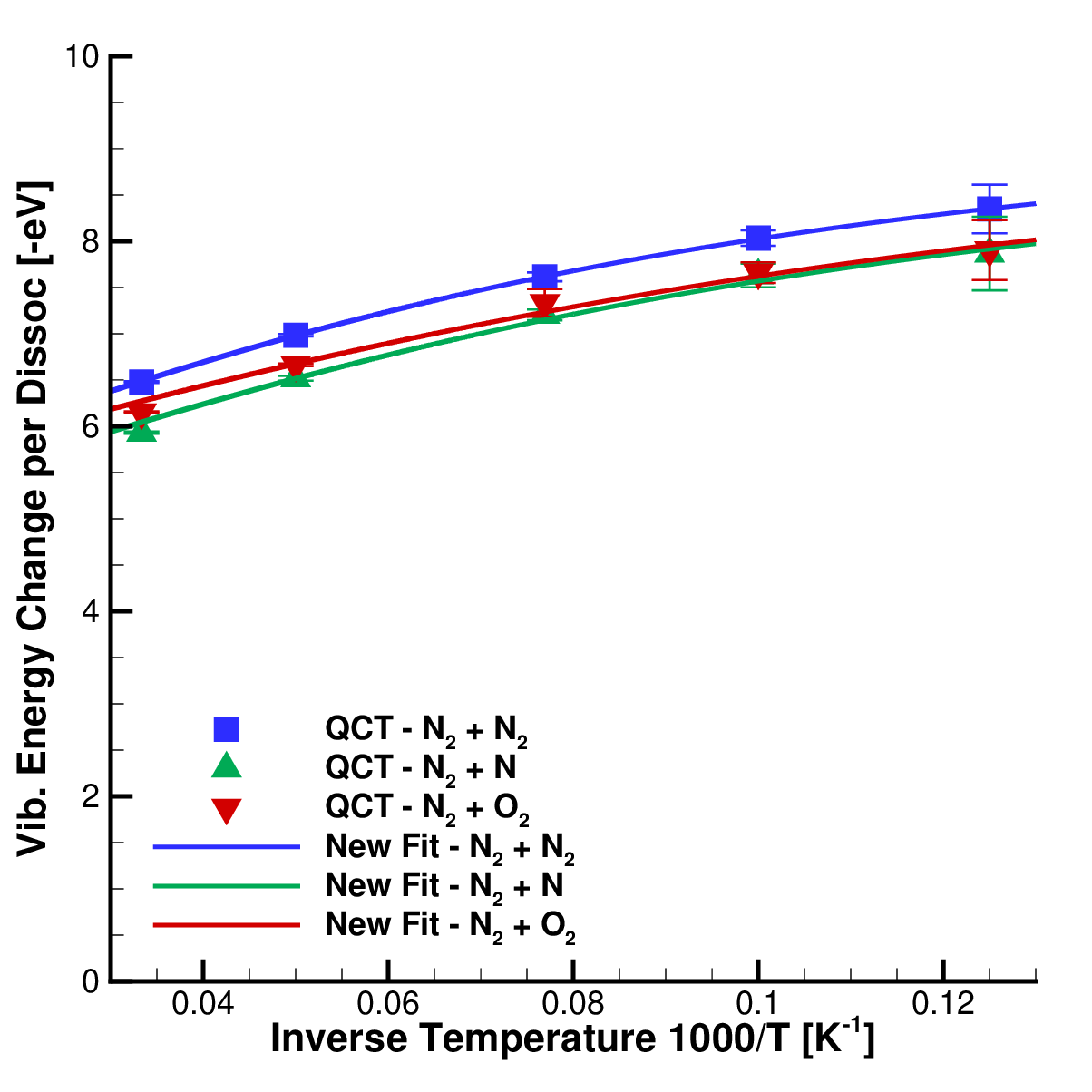}
      \caption{Change in vibrational energy, $T=T_\mathrm{v}$}
   \end{subfigure}
   \begin{subfigure}[t]{0.48\textwidth}
      \centering
      \includegraphics[width=\textwidth]{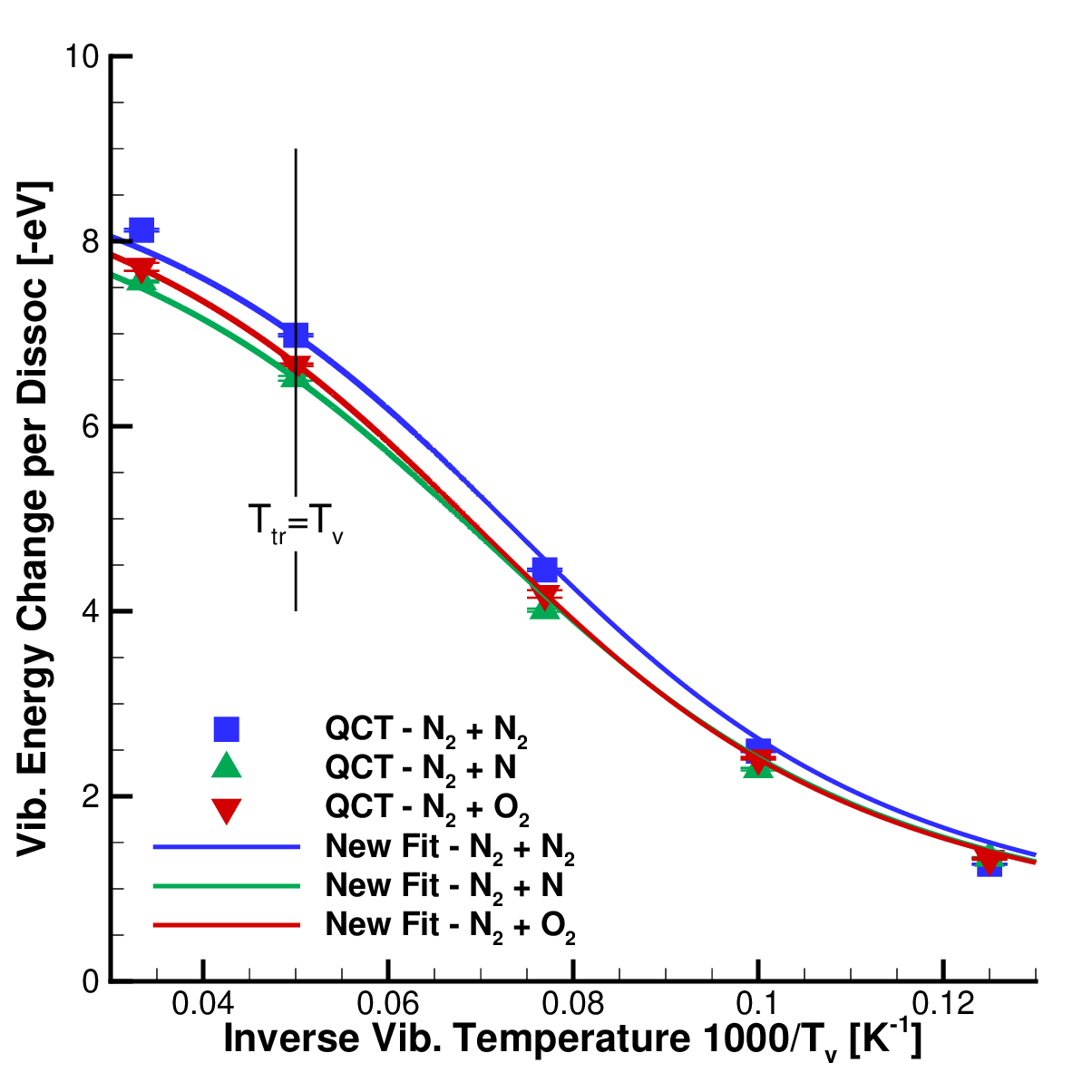}
      \caption{Change in vibrational energy, $T = 20\,000 \, \mathrm{K}$}
   \end{subfigure}
   
   \caption{Fitting to nitrogen dissociation QCT data. Maximum relative error in rate coefficient is 22\% and maximum error in vibrational energy change is $0.24\,\mathrm{eV}$.}
   \label{fig:N2_New_Fit}
\end{figure}

\begin{figure}%[p]
   \centering
   \begin{subfigure}[t]{0.48\textwidth}
      \centering
      \includegraphics[width=\textwidth]{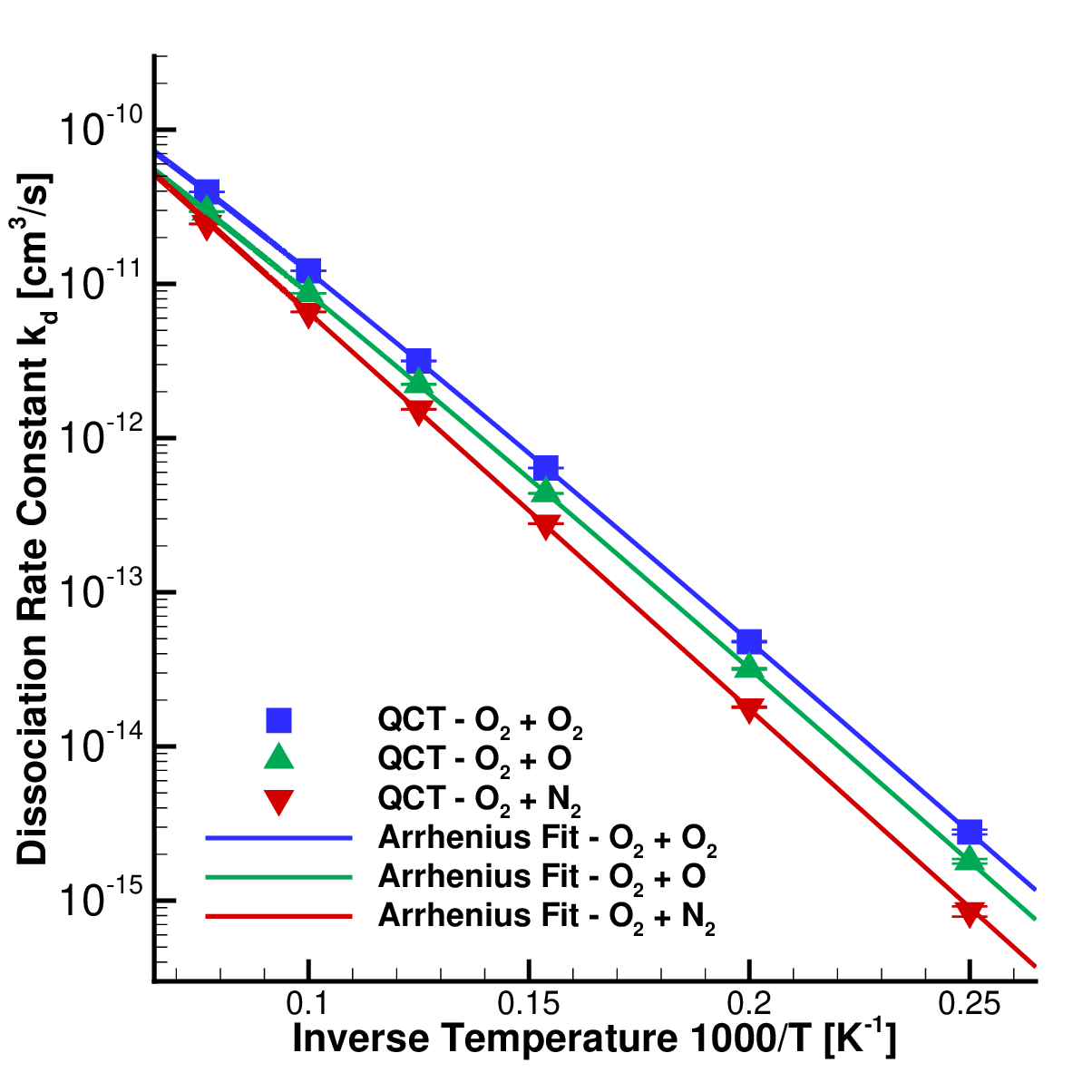}
      \caption{Dissociation rate coefficient, $T = T_\mathrm{v}$}
      \label{subfig:O2-dissoc_Arr}
   \end{subfigure}~
   \begin{subfigure}[t]{0.48\textwidth}
      \centering
      \includegraphics[width=\textwidth]{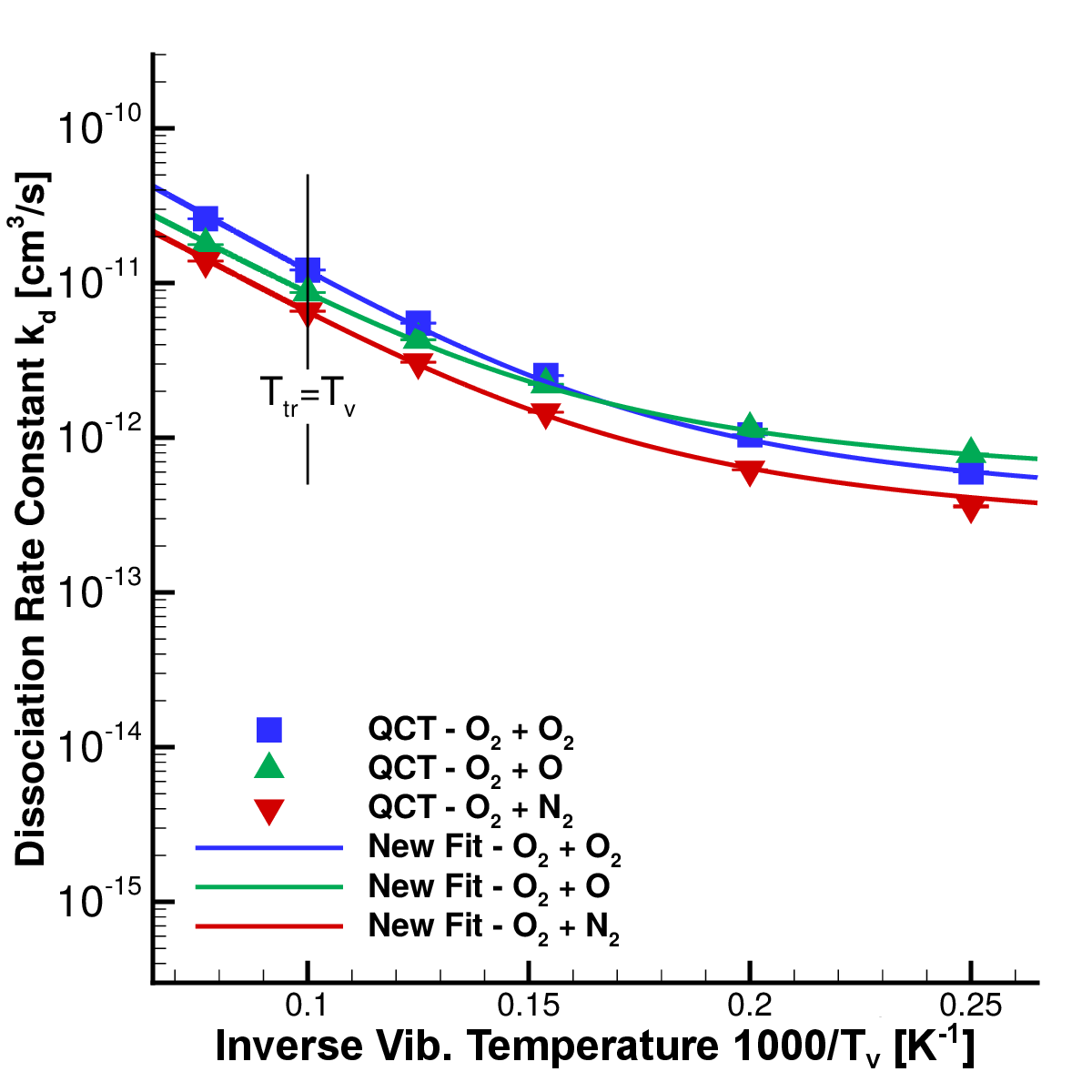}
      \caption{Dissociation rate coefficient, $T = 10\,000 \, \mathrm{K}$}
   \end{subfigure}

   \begin{subfigure}[t]{0.48\textwidth}
      \centering
      \includegraphics[width=\textwidth]{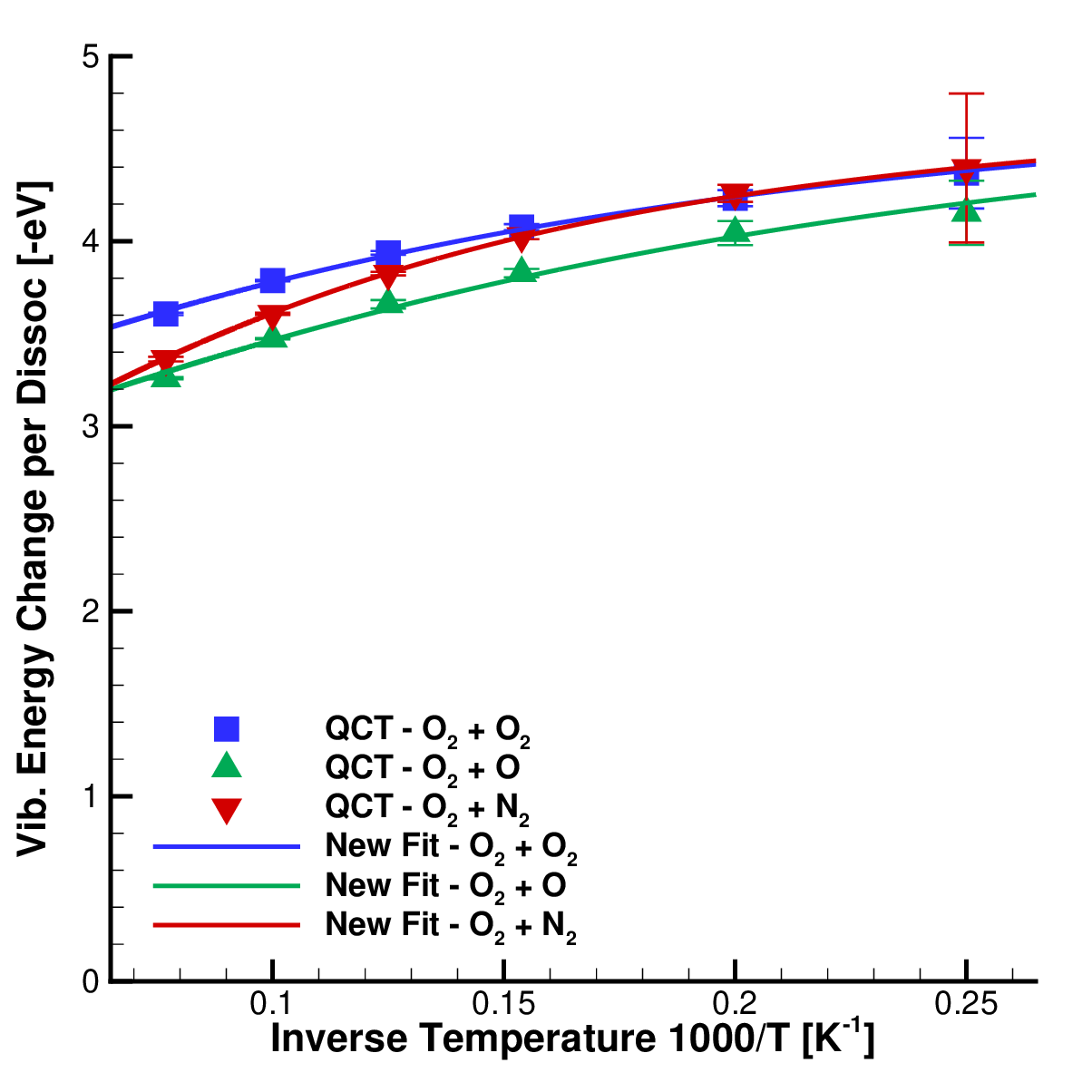}
      \caption{Change in vibrational energy, $T = T_\mathrm{v}$}
   \end{subfigure}~
   \begin{subfigure}[t]{0.48\textwidth}
      \centering
      \includegraphics[width=\textwidth]{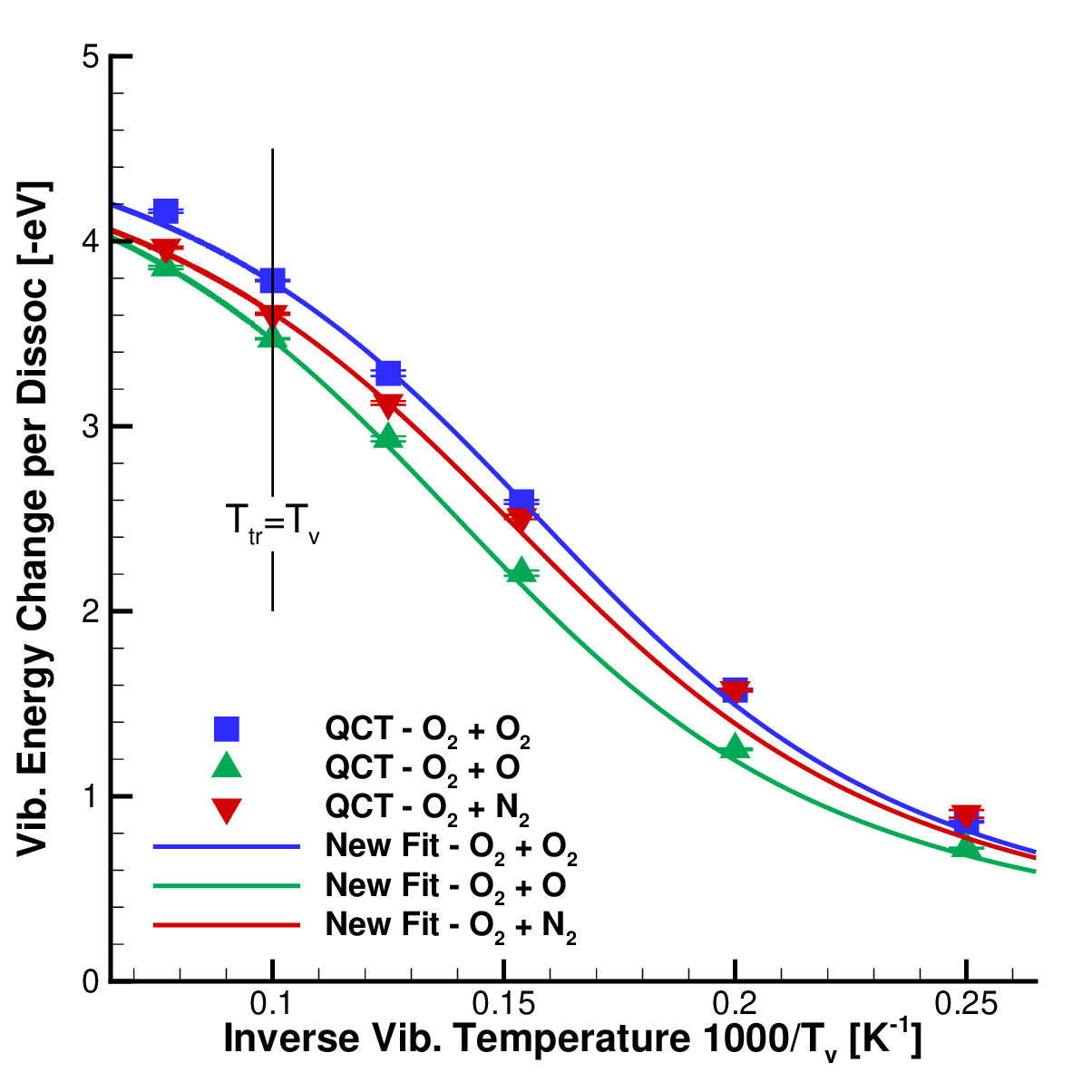}
      \caption{Change in vibrational energy, $T = 10\,000 \, \mathrm{K}$}
   \end{subfigure}
   
   \caption{Fitting to oxygen dissociation QCT data. Maximum relative error in rate coefficient is 15\% and maximum error in energy change is $0.18 \, \mathrm{eV}$.}
   \label{fig:O2_New_Fit}
   %; effective dissociation with fixed $T_{\mathrm{D},s}$
\end{figure}

We now compare the original QCT data to the model prediction: in Fig.~\ref{fig:N2_New_Fit} for nitrogen dissociation and in Fig.~\ref{fig:O2_New_Fit} for oxygen dissociation. We find this modified Marrone-Treanor model to accurately fit all quantities of interest and for all dissociation reactions.

The dissociation rate coefficient in the equilibrium test set, described using the modified Arrhenius form of Eq.~(\ref{eq:karr}), has a mean deviation at or below 1\% for most cases; oxygen dissociation with partner $\mathrm{N_2}$ is an outlier here with a maximum deviation of 5.8\%. For all dissociation rate coefficients in the nonequilibrium test set, the model has a maximum deviation of 22\% and the average deviation is below 5\% for all oxygen dissociation cases. Vibrational energy per dissociation in the equilibrium test set is fit extremely well; the maximum error is $0.12 \, \mathrm{eV}$ for nitrogen dissociation and $0.05 \, \mathrm{eV}$ for oxygen dissociation. Finally, the vibrational energy change in the nonequilibrium test set has a maximum error of $0.24 \, \mathrm{eV}$. For reference, the dissociation energy of nitrogen is $9.8 \, \mathrm{eV}$ and oxygen is $5.1 \, \mathrm{eV}$; these energy changes are therefore accurate to within 4\% of the corresponding dissociation energy.

It is worth recalling again that we have determined all model parameters using data in the equilibrium test set only: dissociation rate coefficients to determine the modified Arrhenius form, and average energy change per dissociation to determine the novel two-parameter fit to $U$. This means that the goodness of fit demonstrated for the nonequilibrium test set is predictive. Predictive capability of this accuracy is a strong indicator that we are capturing the dominant physical mechanisms, instead of simply providing sufficient degrees of freedom to intersect the data. The Marrone-Treanor model is old and relatively simple, but it continues to be recommended for use by chemical kinetics authors~\cite{KnabFM1995, LosevKMPS1996}. It has also been extended to exchange reactions by Knab et al.~\cite{KnabFM1995}.

The modification we propose is consistent with qualitative and quantitative observations from the QCT data. Thus, the proposed modified Marrone-Treanor model is a simple, self-consistent description of dissociation at conditions characterized by $T$ and $T_\mathrm{v}$.

%-------------------------------------------------------------------------------

%-------------------------------------------------------------------------------

%-------------------------------------------------------------------------------
%\clearpage

\section{Non-Boltzmann correction factors} \label{sec:nb_effects}

In Sec.~\ref{sec:mmt_model_explanation} we introduced all analytical expressions needed for implementing the MMT model, which at its core consists of evaluating the dissociation rate coefficient with Eq.~(\ref{eq:mmt_rate_coefficient}) and associated vibrational energy removed per dissociation according to Eq.~(\ref{eq:devib_nb_factor}). Thus the rate coefficient is evaluated as the product of the two-temperature expression $k^\mathrm{Arr} (T) \, Z (T,T_\mathrm{v})$ and a non-Boltzmann correction factor $f_k^\mathrm{NB}$. In a similar fashion, the dissociating species' energy removed per dissociation is evaluated as the product of Knab's two-temperature formula for $\langle \varepsilon_{\mathrm{v},s} \rangle_\mathrm{diss}^\mathrm{Knab} (T, T_\mathrm{v})$ and its own corresponding non-Boltzmann factor $f_\varepsilon^\mathrm{NB}$. Whereas the model's dependence on $T$ and $T_\mathrm{v}$ is justified by the analysis in Secs.~\ref{sec:qct_analysis} and \ref{sec:fitting_qct_data}, a more in-depth justification for including these correction factors was deferred to the present section.

In Sec.~\ref{sec:introduction}, we referenced DMS~\cite{ValentiniSBNC2015, ValentiniSBC2016, grover19a, grover19b, torres24a} and state-to-state master equation studies~\cite{PanesiJSM2013, KimB2013,JaffeSP2015,macdonald20b} of coupled rovibrational excitation and dissociation of suddenly heated air species in isothermal heat baths. Regardless of the mixture in question, all these studies showed that after early transients the vibrational level populations of $\mathrm{N_2}$ and $\mathrm{O_2}$ adopted nearly time-invariant non-Boltzmann distributions as dissociation takes place with the respective $T_\mathrm{v}$ adopting a near-constant value slightly below the heat bath temperature $T$. During this phase low-lying energy levels remain Boltzmann-populated at $T_\mathrm{v}$, but the high-energy tail will be noticeably depleted~\cite{singh20a, SinghS2018} below the expected Boltzmann populations. This depletion is a consequence of a dynamic balance established between preferential dissociation from high-$v$ levels and their comparatively slow replenishment through vibrational excitation, all the while recombination reaction rates which may also replenish the high-$v$ tail, are still negligible. Such conditions have sometimes been referred to as the QSS dissociation regime~\cite{Park1990} and the associated non-Boltzmann internal energy populations as QSS distributions. A direct consequence of molecules adopting such depleted-tail non-Boltzmann distributions is that the macroscopic dissociation and vibrational energy removal rates during the QSS phase drop noticeably below their thermal equilibrium counterparts. The analysis in Secs.~\ref{sec:qct_analysis} and \ref{sec:fitting_qct_data} could not capture these non-Boltzmann depletion effects, since all pre-collision vibrational populations were sampled from fixed Boltzmann distributions at a given $T_\mathrm{v}$. Only dynamic calculation methods involving state-to-state master equations, or DMS can accurately establish the detailed dynamic balance between preferential dissociation and vibrational excitation that ultimately produces such QSS distributions. 

Methodologies for numerically estimating QSS dissociation rate coefficients can be found in literature dating back several decades. Even before the availability of the current ab initio-based QCT databases for air species, Gonzales and Varghese~\cite{gonzales93a, gonzales94a} simulated nonequilibrium dissociation of various diatomic species in highly diluted isothermal heat baths using linearized master equations. They observed the gradual evolution from a vibrationally cold initial state toward the gas adopting its characteristic depleted-tail non-Boltzmann distribution and estimated the dissociation rate coefficient during the QSS phase. In subsequent work~\cite{gonzales95a}, they further showed that by including recombination reactions via detailed balance relations the gas eventually evolves past the QSS regime toward thermodynamic equilibrium characterized by the expected Boltzmann distribution at the heat bath temperature $T$. The more recent rovibrational-specific master equation and DMS studies~\cite{PanesiJSM2013, KimB2013, JaffeSP2015, macdonald20b} for the $\mathrm{N_2}(v,j) - \mathrm{N}$ system all confirm the presence of these features.

Beyond this, vibrational population distributions with depleted high-energy tails are not exclusively a product of dissociation in isothermal heat baths. Similar non-Boltzmann distributions have also been observed in DMS calculations under adiabatic conditions~\cite{torres20b} and downstream of normal shocks~\cite{torres22b}. This implies that depletion effects are generally present in thermo-chemical nonequilibrium flow fields and must be taken into account as part of the MMT model formulation. From a macroscopic modeling viewpoint, when the diatomic species in a gas mixture adopt QSS vibrational distributions, this manifests in a reduction of dissociation rates (by factors $2-5$), as well as in a decrease of roughly 10-20\% in average vibrational energy removed per dissociation relative to values sampled from equivalent-temperature Boltzmann distributions.

\begin{figure}%[htb]

  \begin{subfigure}[t]{0.48\textwidth}
   \centering
   \includegraphics[width=\textwidth]{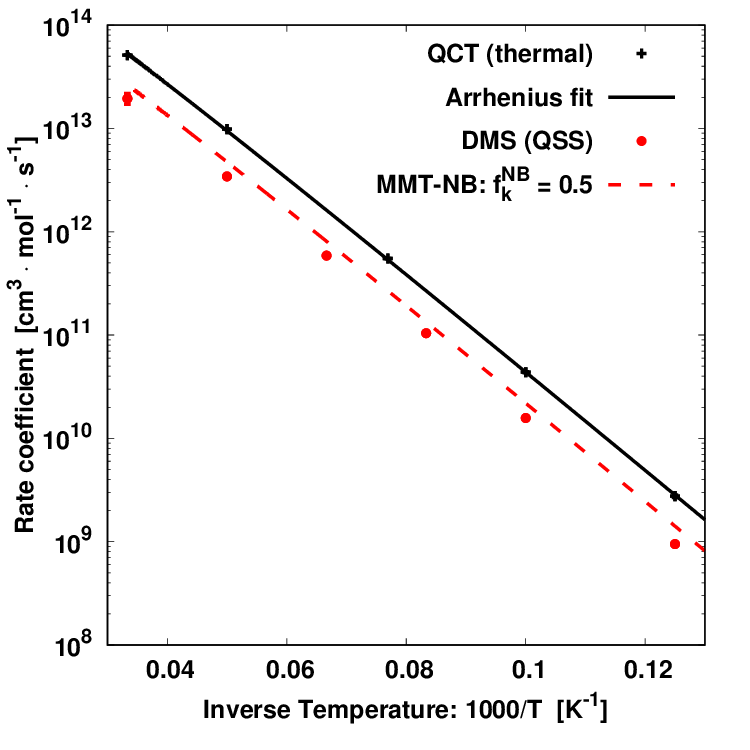}
   \caption{Rate coefficient for $\mathrm{2N_2 \rightarrow 2N + N_2}$}
   \label{fig:N2-N2_diss_arrhenius_fits}
  \end{subfigure}~
  \begin{subfigure}[t]{0.48\textwidth}
   \centering
   \includegraphics[width=\textwidth]{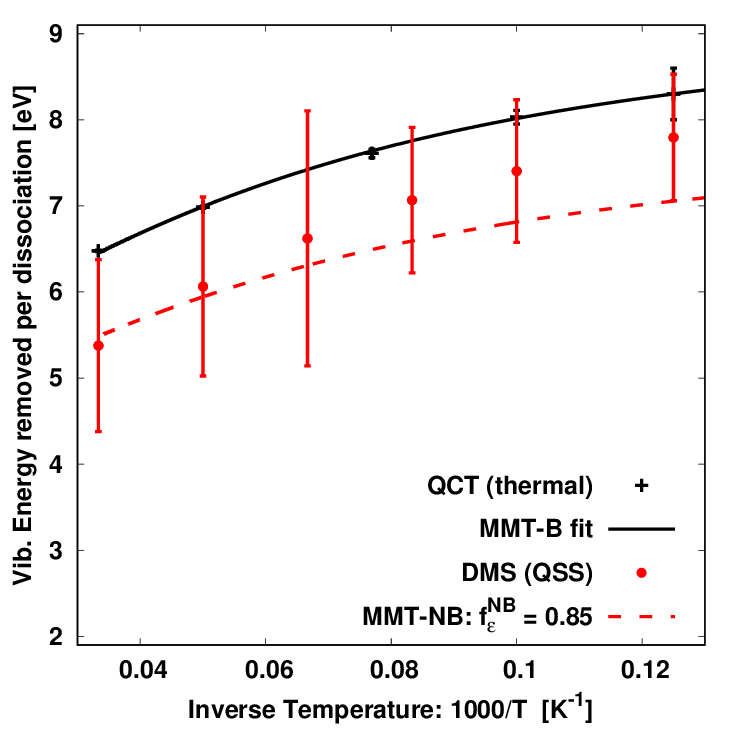}
   \caption{Vibrational energy removed per dissociation for $\mathrm{2N_2 \rightarrow 2N + N_2}$}
   \label{fig:mmt_evib_fit_N2-N2_diss.eps}
  \end{subfigure}
  
  \begin{subfigure}[t]{0.48\textwidth}
   \centering
   \includegraphics[width=\textwidth]{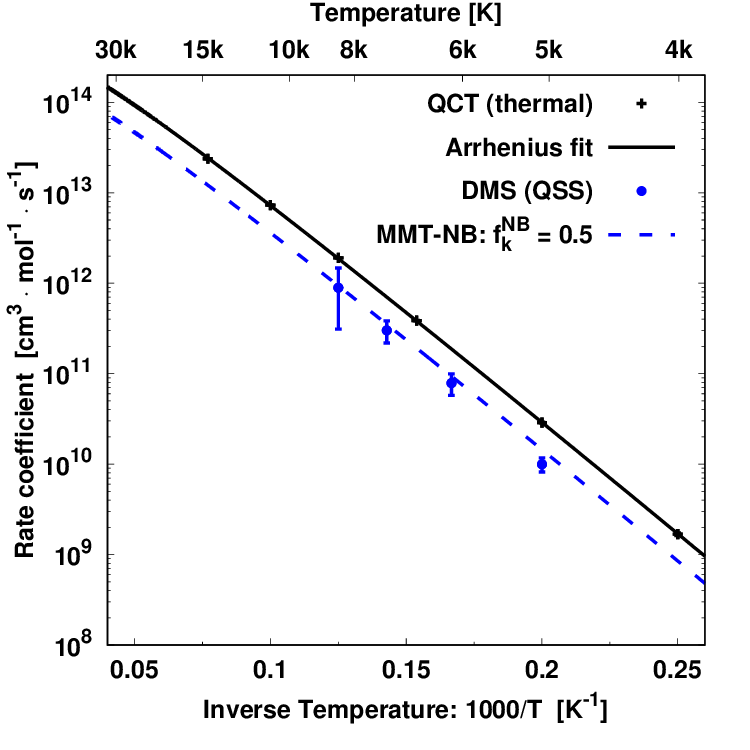}
   \caption{Rate coefficient for $\mathrm{2O_2 \rightarrow 2O + O_2}$}
   \label{fig:O2-O2_diss_arrhenius_fits}
  \end{subfigure}~
  \begin{subfigure}[t]{0.48\textwidth}
   \centering
   \includegraphics[width=\textwidth]{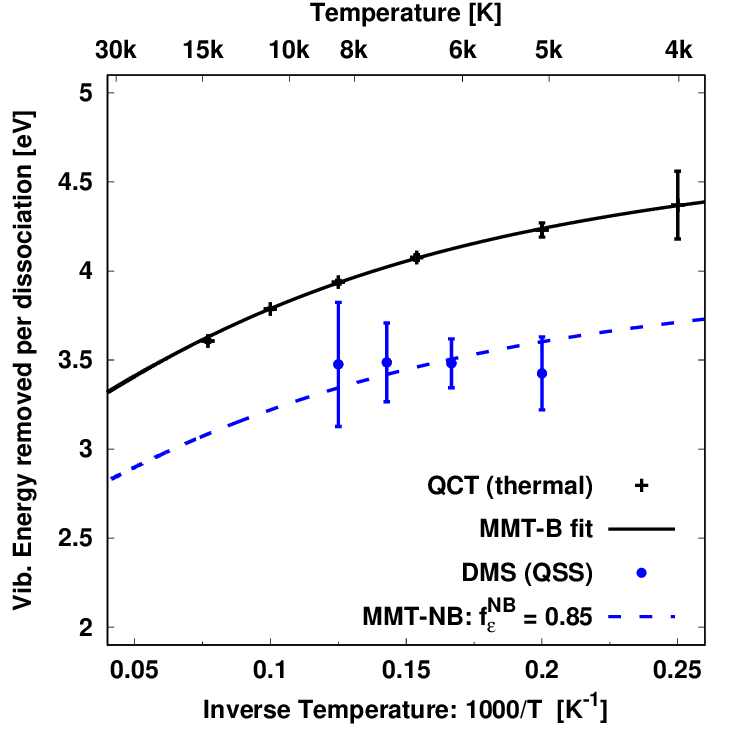}
   \caption{Vibrational energy removed per dissociation for $\mathrm{2O_2 \rightarrow 2O + O_2}$}
   \label{fig:mmt_evib_fit_O2-O2_diss}
  \end{subfigure}

  \caption{Rate coefficient and vibrational energy removed in dissociation reactions with non-Boltzmann correction factors (for molecule-molecule collisions)}
  \label{fig:non_boltzmann_factor_comparison_1}
\end{figure}

\begin{figure}%[htb]

  \begin{subfigure}[t]{0.48\textwidth}
   \centering
   \includegraphics[width=\textwidth]{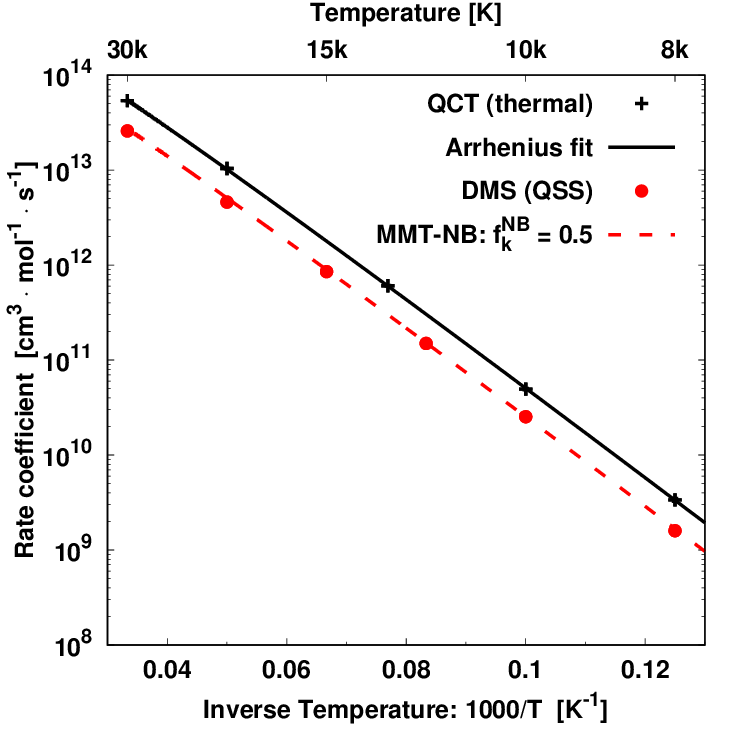}
   \caption{Rate coefficient for $\mathrm{N_2 + N \rightarrow 3N}$}
   \label{fig:N2-N_diss_arrhenius_fits}
  \end{subfigure}~
  \begin{subfigure}[t]{0.48\textwidth}
   \centering
   \includegraphics[width=\textwidth]{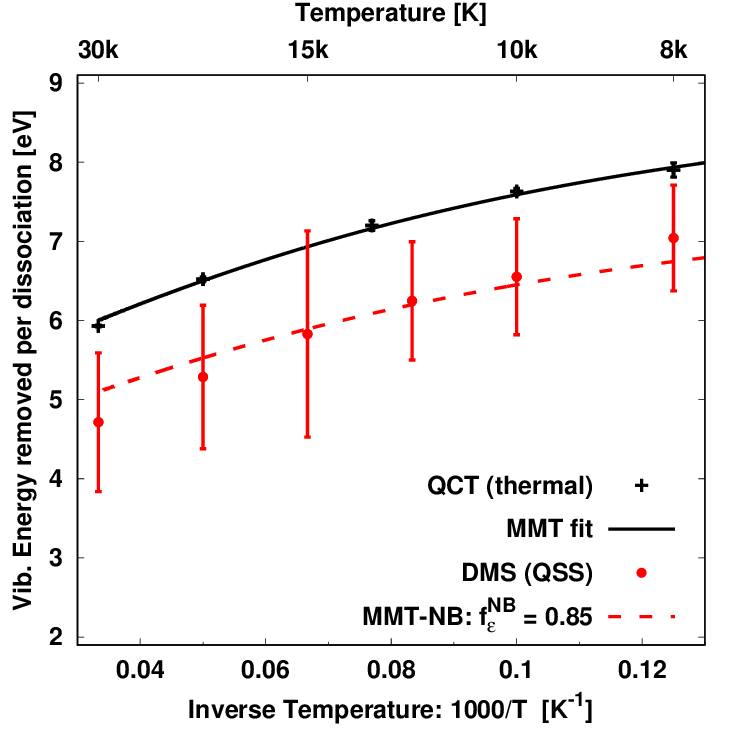}
   \caption{Vibrational energy removed per dissociation for $\mathrm{N_2 + N \rightarrow 3N}$}
   \label{fig:mmt_evib_fit_N2-N_diss.eps}
  \end{subfigure}
  
  \begin{subfigure}[t]{0.48\textwidth}
   \centering
   \includegraphics[width=\textwidth]{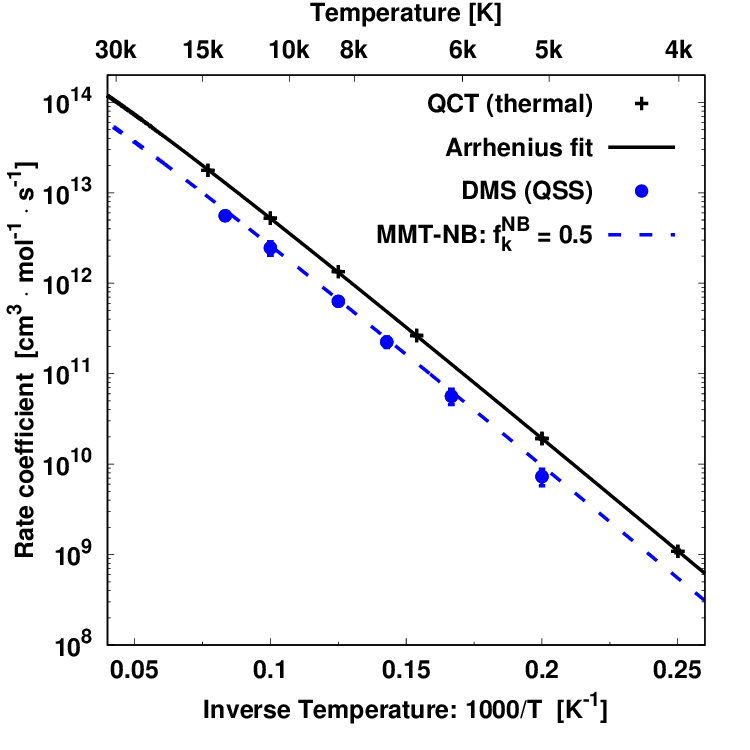}
   \caption{Rate coefficient for $\mathrm{O_2 + O \rightarrow 3O}$}
   \label{fig:O2-O_diss_arrhenius_fits}
  \end{subfigure}~
  \begin{subfigure}[t]{0.48\textwidth}
   \centering
   \includegraphics[width=\textwidth]{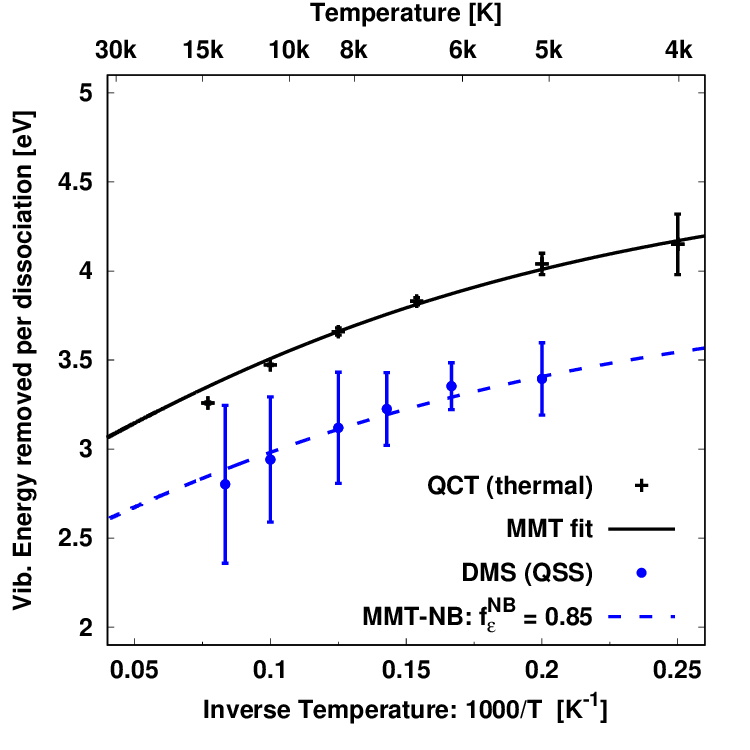}
   \caption{Vibrational energy removed per dissociation for $\mathrm{O_2 + O \rightarrow 3O}$}
   \label{fig:mmt_evib_fit_O2-O_diss}
  \end{subfigure}

  \caption{Rate coefficient and vibrational energy removed in dissociation reactions with non-Boltzmann correction factors (for molecule-atom collisions)}
  \label{fig:non_boltzmann_factor_comparison_2}
\end{figure}

Figures~\ref{fig:non_boltzmann_factor_comparison_1} and \ref{fig:non_boltzmann_factor_comparison_2} illustrate the need to correct for such non-Boltzmann effects. Starting in Fig.~\ref{fig:N2-N2_diss_arrhenius_fits} we plot the dissociation rate coefficient for reaction $\mathrm{2N_2 \rightarrow 2N + N_2}$. Black symbols represent the QCT-derived \emph{thermal equilibrium} rate coefficient (where $T = T_\mathrm{v}$), the continuous black line shows the corresponding Arrhenius fit (using Eq.~(\ref{eq:karr}) and parameters from Table~\ref{tab:mmt_params}). Red symbols represent the equivalent dissociation rate coefficient for quasi-steady-state (QSS) dissociation conditions, extracted from a series of isothermal heat bath DMS calculations~\cite{torres24a} (where $T \gtrapprox T_\mathrm{v}$). Within the temperature range shown, these QSS rate coefficients consistently remain below their corresponding thermal equilibrium counterparts by a near-constant factor. When the Arrhenius rate coefficient is re-scaled by $f_k^\mathrm{NB} = 0.5$ (equivalent to Eq.~(\ref{eq:mmt_rate_coefficient}) with $Z(T, T_\mathrm{v}{=}T) {=} 1$) the dashed red line is obtained. The corresponding plot for $\mathrm{N_2}$ vibrational energy removed by dissociation is shown in Fig.~\ref{fig:mmt_evib_fit_N2-N2_diss.eps}. Black symbols again represent the QCT-derived values at thermal equilibrium, while the black continuous line shows the ``Boltzmann'' portion $\langle \varepsilon_{\mathrm{v},\mathrm{N_2}} \rangle_\mathrm{diss}^\mathrm{Knab} (T,  T_\mathrm{v} {=} T)$ evaluated from Knab's formula with the MMT parameters $a_U$ and $U^*$ from Table~\ref{tab:mmt_params}. Red symbols again represent the corresponding DMS-derived values extracted under QSS dissociation conditions. Analogous to what was observed for the rate coefficient, the $\mathrm{N_2}$ vibrational energy removed by dissociation at QSS conditions remains slightly below the corresponding thermal equilibrium QCT value over the entire temperature range shown. This behavior again suggests there exists a simple way of bringing the ``Boltzmann'' estimate of Eq.~(\ref{eq:Knab}) closer in line with the DMS predictions. The dashed red line is obtained with Eq.~(\ref{eq:devib_nb_factor}), when $\langle \varepsilon_{\mathrm{v},\mathrm{N_2}} \rangle_\mathrm{diss}^\mathrm{Knab} (T,  T_\mathrm{v} {=} T)$ is multiplied by $f_\varepsilon^\mathrm{NB} = 0.85$.

Figures~\ref{fig:O2-O2_diss_arrhenius_fits} and \ref{fig:mmt_evib_fit_O2-O2_diss} present equivalent plots for reaction $\mathrm{2O_2 \rightarrow 2O + O_2}$, while analogous data for reactions $\mathrm{N_2 + N \rightarrow 3N}$ and $\mathrm{O_2 + O \rightarrow 3O}$ are compared in Fig.~\ref{fig:non_boltzmann_factor_comparison_2}. All three reactions exhibit similar features to the first one: a dissociation rate coefficient about 50\% smaller and roughly 15\% less average vibrational energy removed per reaction in QSS than at thermal equilibrium. This consistent behavior makes it possible to capture the complex non-Boltzmann depletion effects in our two-temperature CFD model by employing the two simple correction factors. Of course, for every reaction the precise non-Boltzmann factors yielding the best possible fit between QCT and DMS data differ slightly and also vary somewhat with temperature. However, we choose to apply the same factors $f_k^\mathrm{NB} = 0.5$ and $f_\varepsilon^\mathrm{NB} = 0.85$ to all four reactions so as to not introduce unnecessary complexity to the MMT model.

%-------------------------------------------------------------------------------

%-------------------------------------------------------------------------------
%-------------------------------------------------------------------------------
%\clearpage

\section{Approach to chemical equilibrium} \label{sec:approach_to_equilibrium}

In Sec.~\ref{sec:nb_effects} we motivated the use of non-Boltzmann correction factors as part of the MMT model by comparisons between QCT and DMS data. We found that values $f_{k}^\mathrm{NB} = 0.5$ and $f_{\varepsilon}^\mathrm{NB} = 0.85$ produce very good $k_\mathrm{diss}$ and $\langle \varepsilon_{\mathrm{v},s} \rangle_\mathrm{diss}$ estimates for the major nitrogen and oxygen dissociation reactions during the QSS dissociation phase. Thus, the MMT model is able to accurately reproduce the DMS-predicted rates of dissociation and vibrational energy removal experienced by the gas immediately downstream of a strong shock. However, blindly employing that same pair of constant non-Boltzmann factors becomes non-physical as the dissociating mixture approaches chemical equilibrium.

For illustration, consider the generic diatom dissociation-recombination reaction $\mathrm{AB} + \mathrm{M} \rightleftharpoons \mathrm{A} + \mathrm{B} + \mathrm{M}$. The forward and backward rates are evaluated according to Eq.~(\ref{eq:net_rate}) to be $\mathcal{R}_\mathrm{diss} = k_\mathrm{diss} \, \left( \rho_\mathrm{AB} / M_\mathrm{AB} \right) \left( \rho_\mathrm{M} / M_\mathrm{M} \right)$ and $\mathcal{R}_\mathrm{rec} = k_\mathrm{rec} \, \left( \rho_\mathrm{A} / M_\mathrm{A} \right) \left( \rho_\mathrm{B} / M_\mathrm{B} \right) \left( \rho_\mathrm{M} / M_\mathrm{M} \right)$ respectively. In a two-temperature model the forward (dissociation) rate coefficient at an arbitrary nonequilibrium condition $T \ne T_\mathrm{v}$ becomes $k_\mathrm{diss} = k_\mathrm{diss} (T,T_\mathrm{v})$ and in MMT specifically it is computed from Eq~(\ref{eq:mmt_rate_coefficient}). But calculating the backward (recombination) rate coefficient $k_\mathrm{rec}$ is not as straightforward, especially in a nonequilibrium setting. Despite recent QCT studies~\cite{geistfeld23a} on the available ab initio PESs, at present we are unaware of any general and reliable method to calculate the recombination rates directly from first principles. The main impediments, as we see them, are the added complexity to sampling initial parameters for the recombining three-body collisions (as opposed to dissociating two-body collisions), as well as the lack of a fully developed kinetic theory to calculate a three-body collision rate.

The most commonly used approach is to instead tie the recombination rate to its corresponding dissociation rate using detailed balance relations. Such relations would be expressed at the kinetic scale by linking together cross sections for forward and backward reactions, but a reasonable simplification at the fluid scale is to assume Maxwellian velocity distributions at translational temperature $T$ for all collision partners. In this limit it becomes possible to formulate detailed balance relations in terms of rate coefficients and the equilibrium constant of a given reaction. This approach has for instance been used in master equation studies~\cite{PanesiJSM2013, KimB2013, JaffeSP2015} to enforce detailed balance relations in terms of state-specific dissociation-recombination rate coefficient pairs and drive the simulated mixture to thermodynamic equilibrium. Beyond this, in two-temperature models such as Park~\cite{Park1990} or MMT, it is implicitly assumed that the low-lying vibrational levels of diatomic species are populated according to Boltzmann distributions at temperature $T_\mathrm{v}$. Therefore, the rate coefficients of forward and backward reactions at thermal equilibrium must naturally be evaluated at the single equilibrium temperature $T_\mathrm{v} = T$.

In the present discussion we proceed as follows. First, recall the detailed balance relation that holds at thermodynamic equilibrium:
\begin{equation}
 K_\mathrm{c}^\mathrm{eq} (T) = \left[ \frac{\left( \rho_\mathrm{A} / M_\mathrm{A} \right) \left( \rho_\mathrm{B} / M_\mathrm{B} \right)}{\left( \rho_\mathrm{AB} / M_\mathrm{AB} \right)} \right]^{\star} = \frac{k_\mathrm{diss}^\star (T)}{k_\mathrm{rec}^\star (T)}. \label{eq:detailed_balance}
\end{equation}

Here $K_\mathrm{c}^\mathrm{eq} (T)$ is the equilibrium constant for the generic dissociation-recombination reaction and the ``$\star$'' superscripts indicate that all quantities take on their respective thermodynamic equilibrium values. As implied by the first equality in Eq.~(\ref{eq:detailed_balance}), the equilibrium constant at temperature $T$ is equivalent to the equilibrium product-to-reactant molar concentration ratio, which is the one that minimizes the Gibbs free energy. In practice, the $K_\mathrm{c}^\mathrm{eq}$ are typically calculated from thermodynamic databases, such as the NASA Lewis polynomial fits~\cite{mcbride93a} and are themselves independent of the reaction rates. But at thermodynamic equilibrium the second equality in Eq.~(\ref{eq:detailed_balance}) must also hold. It stems from of the fact that at this specific condition the net reaction rate must be zero:
\begin{equation}
 [\mathcal{R}_\mathrm{diss} - \mathcal{R}_\mathrm{rec}]^{\star} = k_\mathrm{diss}^\star (T) \left( \frac{\rho_\mathrm{AB}^{\star}}{M_\mathrm{AB}} \right) \left( \frac{\rho_\mathrm{M}}{M_\mathrm{M}} \right) - k_\mathrm{rec}^\star (T) \left( \frac{\rho_\mathrm{A}^{\star}}{M_\mathrm{A}} \right) \left( \frac{\rho_\mathrm{B}^{\star}}{M_\mathrm{B}} \right) \left( \frac{\rho_\mathrm{M}}{M_\mathrm{M}} \right) \overset{!}{=} 0
\end{equation}

Thus, from the second equality in Eq.~(\ref{eq:detailed_balance}) one derives that $k_\mathrm{rec}^\star (T) = k_\mathrm{diss}^\star (T) / K_\mathrm{c}^\mathrm{eq} (T)$, where $k_\mathrm{diss}^\star (T)$ represents the \emph{thermal} dissociation rate coefficient. As discussed in Sec.~\ref{sec:equil_diss_rate_coeff}, we have fit the thermal rate coefficients of all dissociation reactions in Table~\ref{tab:mmt_params} to the modified Arrhenius form. Thus, in our notation, the equality $k_\mathrm{rec}^\star (T) = k^\mathrm{Arr} (T) / K_\mathrm{c}^\mathrm{eq} (T)$ must hold. 

This indirect way of calculating the recombination rate coefficient is only strictly valid at thermodynamic equilibrium, a state characterized by Maxwell-Boltzmann energy distributions for all species at an unequivocal single temperature $T$. However, to our knowledge all current CFD implementations of two-temperature models invoke the detailed balance relation even when the gas is locally out of thermal equilibrium. One thus assumes approximately that $k_\mathrm{rec} = k_\mathrm{diss} / K_\mathrm{c}^\mathrm{eq} (T)$, even when locally nonequilibrium effects may be present.

In the current implementation we also follow this path. We calculate the forward rate by evaluating the dissociation rate coefficient from Eq.~(\ref{eq:mmt_rate_coefficient}), yielding $\mathcal{R}_\mathrm{diss}^\mathrm{MMT} = f_k^\mathrm{NB} \, Z(T,T_\mathrm{v}) \, k^\mathrm{Arr} \left(\rho_\mathrm{AB} / M_\mathrm{AB} \right) \left(\rho_\mathrm{M} / M_\mathrm{M} \right)$. But when we determine the recombination rate coefficient using the approximate detailed balance relation, we explicitly set the nonequilibrium factor $Z$ in Eq.~(\ref{eq:mmt_rate_coefficient}) to unity. We do this, because for the recombination rate coefficient a dependence on vibrational temperature is not justified.
Recall that the factor $Z(T,T_\mathrm{v})$ was introduced in the Marrone-Treanor model to account for vibrational bias found in dissociation and is thus only meaningful for the diatomic reactants possessing vibrational energy modes. Introducing such a bias makes no sense for the recombining atoms. At the same time, the non-Boltzmann factor remains part of the backward rate coefficient expression. We thus obtain $k_\mathrm{rec} = f_k^\mathrm{NB} \, k^\mathrm{Arr} (T) / K_\mathrm{c}^\mathrm{eq} (T)$ and the recombination rate then becomes $\mathcal{R}_\mathrm{rec}^\mathrm{MMT} = f_k^\mathrm{NB} \, k^\mathrm{Arr} (T) \left( \rho_A / M_A \right) \left( \rho_B / M_B \right) \left( \rho_\mathrm{M} / M_\mathrm{M} \right)$. When put together, these two expressions yield for the net rate:
\begin{equation}
 \mathcal{R}_\mathrm{diss} - \mathcal{R}_\mathrm{rec} = f_k^\mathrm{NB} \, k^\mathrm{Arr} (T) \, \left( \frac{\rho_\mathrm{M}}{M_\mathrm{M}} \right) \left[ Z(T, T_\mathrm{v}) \left( \frac{\rho_\mathrm{AB}}{M_\mathrm{AB}} \right) - \frac{1}{K_\mathrm{c}^\mathrm{eq} (T)} \, \left( \frac{\rho_\mathrm{A}}{M_\mathrm{A}} \right) \left( \frac{\rho_\mathrm{B}}{M_\mathrm{B}} \right) \right]. \label{eq:net_rate_ab_diss}
\end{equation}

This expression makes it clear that, according to the MMT model as formulated in Sec.~\ref{sec:mmt_model_explanation}, the non-Boltzmann factor re-scales both forward and backward rates simultaneously. This does not constitute a major problem at the location where fully recombined freestream gas initially traverses a shock front. Under such conditions the precise value of the recombination rate will be negligible anyway and only the forward reaction will contribute to any significant degree to Eq.~(\ref{eq:net_rate_ab_diss}). During this early QSS dissociation phase the MMT model will thus reproduce the dissociation rates predicted from an equivalent DMS calculation fairly well.

However, as the gas then proceeds downstream of the shock front, vibrational and translation-rotational modes should gradually relax to a common temperature and eventually the mixture should also approach chemical equilibrium. Forward and backward reactions should then approach their respective thermal rates, i.e. $\mathcal{R}_\mathrm{diss}^\star = k^\mathrm{Arr} (T)$ $\left( \rho_\mathrm{AB}^{\star} / M_\mathrm{AB} \right)$ $\left( \rho_\mathrm{M} / M_\mathrm{M} \right)$ and $\mathcal{R}_\mathrm{rec}^\star = k^\mathrm{Arr} (T) / K_\mathrm{c}^\mathrm{eq} \left( \rho_\mathrm{A}^{\star}/M_\mathrm{A} \right) \left( \rho_\mathrm{B}^{\star}/M_\mathrm{B} \right) \left( \rho_\mathrm{M} / M_\mathrm{M} \right)$ respectively. However, the non-Boltzmann factor originally introduced to help adjust the value of the dissociation rate coefficient for the QSS regime, remains part of the rate coefficient formulation and, if kept at its original value of $f_k^\mathrm{NB} = 0.5$, now prevents the forward and backward processes from reaching their correct thermal rates.

This issue also affects the vibrational energy-chemistry coupling term. Evaluating Eq.~(\ref{eq:evib_chem_source_eff}) and (\ref{eq:devib_nb_factor}) together with Eq.~(\ref{eq:net_rate_ab_diss}) for the present reaction yields:
\begin{equation}
 w_{\mathrm{v}, \mathrm{diss-rec}, \mathrm{AB}}^{\mathrm{chem}} = - f_k^\mathrm{NB} \, f_\varepsilon^\mathrm{NB} \, k^\mathrm{Arr} (T) \, \left( \frac{\rho_\mathrm{M}}{M_\mathrm{M}} \right) \left[ Z (T, T_\mathrm{v}) \left( \frac{\rho_\mathrm{AB}}{M_\mathrm{AB}} \right) - \frac{1}{K_\mathrm{c}^\mathrm{eq} (T)} \, \left( \frac{\rho_\mathrm{A}}{M_\mathrm{A}} \right) \left( \frac{\rho_\mathrm{B}}{M_\mathrm{B}} \right) \right] \, \langle \varepsilon_{\mathrm{v},\mathrm{AB}} \rangle_r^\mathrm{Knab} (T,T_\mathrm{v}), \label{eq:net_evib_rate_ab_diss}
\end{equation}
with both non-Boltzmann factors seen to be re-scaling the coupling source term. As was the case for the reaction rates themselves, this does not constitute a problem during the early QSS dissociation phase, where only the term for the forward reaction contributes meaningfully to $w_{\mathrm{v}, \mathrm{diss-rec}, \mathrm{AB}}^{\mathrm{chem}}$. However, when the same constant non-Boltzmann factors are employed regardless of the local conditions, they will continue to incorrectly re-scale both forward and reverse vibrational energy-chemistry removal/replenishment rates even as the gas is meant to approach thermal and chemical equilibrium.

To address this problem, we propose \emph{variable} non-Boltzmann factors $f_{k}^\mathrm{VNB}$ and $f_{\varepsilon}^\mathrm{VNB}$, whose values change dynamically with the local state of the gas. Following the approach of Singh and Schwartzentruber~\cite{singh22a} we define the reaction's nonequilibrium concentration ratio:
\begin{equation}
 \zeta = \frac{(\rho_\mathrm{A} / M_\mathrm{A}) \, (\rho_\mathrm{B} / M_\mathrm{B})}{( \rho_\mathrm{AB} / M_\mathrm{AB} ) \, K_\mathrm{c}^\mathrm{eq}(T)}, \label{eq:concentration_ratio}
\end{equation}
which will depend on the local mixture composition and the equilibrium constant for the reaction at temperature $T$. Three limiting cases are of particular interest. (1) When the gas is close to ``fully recombined'', i.e. $\rho_\mathrm{A} \rightarrow 0$ and $\rho_\mathrm{B} \rightarrow 0$, so $\zeta \rightarrow 0$. (2) When the gas takes on its equilibrium composition $\rho_\mathrm{A} = \rho_\mathrm{A}^{\star}$, $\rho_\mathrm{B} = \rho_\mathrm{B}^{\star}$ and $\rho_\mathrm{AB} = \rho_\mathrm{AB}^{\star}$, and since $K_\mathrm{c}^\mathrm{eq} = (\rho_\mathrm{A}^{\star} / M_\mathrm{A}) (\rho_\mathrm{B}^{\star} / M_\mathrm{B}) / (\rho_\mathrm{AB}^{\star} / M_\mathrm{AB})$, one ends up with $\zeta = 1$. (3) Finally, when the gas is close to ``fully dissociated'', $\rho_\mathrm{AB} \rightarrow 0$, therefore $\zeta \rightarrow \infty$. 

Downstream of a strong shock, where fully recombined freestream gas experiences sudden heating, this concentration ratio will start out at $\zeta = 0$. At such conditions the non-Boltzmann factors should take on the previously found values $f_{k}^\mathrm{NB} = 0.5$ and $f_{\varepsilon}^\mathrm{NB} = 0.85$ to accurately reproduce the QSS dissociation and vibrational energy removal rates. As the gas is advected downstream and relaxes toward post-shock thermal and chemical equilibrium, $\zeta$ will gradually approach unity. At this point the variable non-Boltzmann factors should both approach unity themselves, so as to reproduce the correct equilibrium reaction and vibrational energy removal rates. Finally, in boundary layers forming near cold surfaces, or in nozzle expansion flows of partially dissociated air, approach to chemical equilibrium will occur from the opposite direction, starting out from some concentration ratio $\zeta > 1$ and gradually approaching $\zeta = 1$ as atoms recombine. At present we do not know which precise values the two non-Boltzmann factors should take on under such conditions, but there are grounds to expect them both to be greater than unity in a recombination-dominated gas. 

The reasoning is as follows. In net-recombining flows where dissociation is negligible, preferential formation of molecules possessing internal energies close to the dissociation threshold $D_0$ (a.k.a. preferential recombination) may cause temporary over-population in the high-energy tail of the diatom's vibrational energy distribution. Such a sudden over-abundance of highly excited diatoms, themselves prime candidates for dissociation, in turn may lead to non-Boltzmann correction factors greater than unity. Note that preferential recombination, the counterpart to preferential dissociation, has been observed in recent three-body QCT~\cite{geistfeld23a} and Molecular Dynamics-based~\cite{pahlani23a} studies, as well as a recent state-resolved master equation study~\cite{macdonald24a}. Indeed, in state-specific master equation modeling, preferential recombination into rovibrational levels close to $D_0$ is a necessary consequence of enforcing detailed balance relations for state-specific dissociation-recombination reaction pairs (e.g. see Eq.~(11) in Ref.~\cite{KimB2013} or the equivalent Eq.~(5) in Ref.~\cite{PanesiJSM2013}). 

At the present time though, we do not possess any suitable data to systematically confirm or reject the aforementioned speculations. Studying net-recombining flows from first principles would require us to perform DMS calculations with recombination reactions included, which in turn would presuppose the availability of robust algorithms to simulate on-the-fly three-body trajectories as part of our DMS collision routines. Absent these prerequisites, we are currently unable to estimate the non-Boltzmann factors for concentration rations $\zeta > 1$. And as a consequence, we \emph{choose} to cap both of them at unity for the recombining regime and propose a piece-wise linear interpolation function that spans all three limiting cases:
\begin{equation}
 f_{k/\varepsilon}^\mathrm{VNB} (\zeta) = \left\lbrace \begin{array}{l l}
                                  (1 - f_{k/\varepsilon, \, 0}^\mathrm{NB}) \, \zeta + f_{k/\varepsilon, \, 0}^\mathrm{NB}  & \text{if} \quad 0 < \zeta < 1 \\
                                  1 & \text{if} \quad \zeta \ge 1
                                 \end{array} \right. \label{eq:vnb_factor}
\end{equation}

\begin{figure}%[htb]
 \centering
 
 \includegraphics[width=0.5\textwidth]{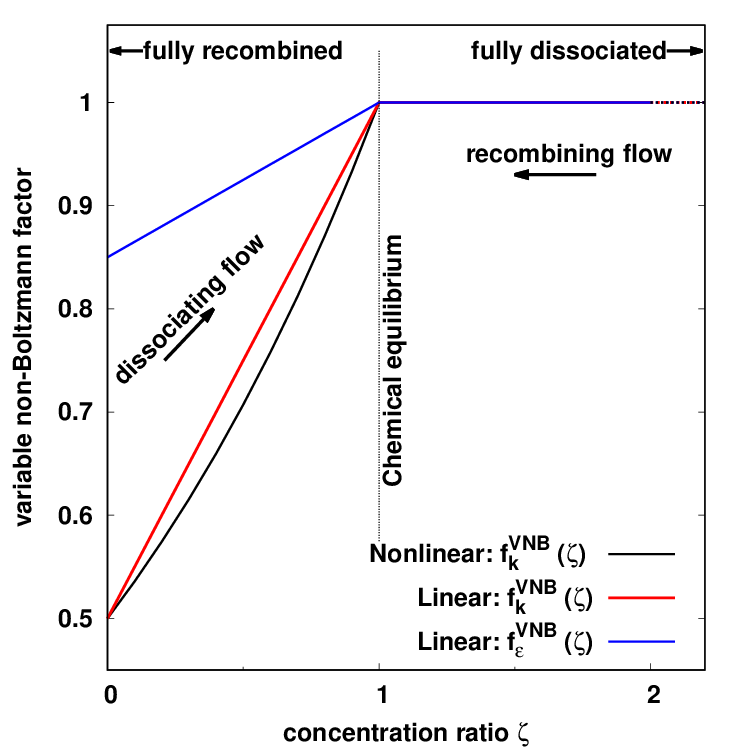}
 
 \caption{Variable non-Boltzmann factors as function of concentration ratio, for use in Eqs.~(\ref{eq:mmt_rate_coefficient}) and (\ref{eq:devib_nb_factor}) respectively.}
 \label{fig:vnb_factor}
\end{figure}

This functional form is applied to both non-Boltzmann factors, with $f_{k,0}^\mathrm{NB} = 0.5$ for the rate coefficient and $f_{\varepsilon, 0}^\mathrm{NB} = 0.85$ for vibrational energy removed by dissociation. Figure~\ref{fig:vnb_factor} illustrates the behavior just described. Solid red and blue lines represent the variation of $f_k^\mathrm{VNB}(\zeta)$ and $f_\varepsilon^\mathrm{VNB}(\zeta)$ respectively. It should be emphasized that Eq.~(\ref{eq:vnb_factor}) is ultimately a modeling choice. Linear interpolation was chosen for no other reason than its simplicity. Indeed, an alternative non-linear function has been proposed for $f_{k}^\mathrm{VNB} (\zeta)$ (see Eq.~(12) of Ref.~\cite{chaudhry20b}) and is shown as the black solid line in Fig.~\ref{fig:vnb_factor}. Ultimately, the non-Boltzmann factors account for the net effect on rate coefficients and internal energy change in dissociation reactions due to either depleted, or overpopulated high-energy vibrational levels. As discussed in Ref.~\cite{singh22a} the degree of overpopulation (when $f_{k/\varepsilon}^\mathrm{VNB}(\zeta > 1) > 1$) can be estimated using depletion statistics combined with state-resolved analysis. However, such estimates have so far not been thoroughly confirmed by first-principles methods, which is why in the MMT model we restrict $f_{k/\varepsilon}^\mathrm{VNB}(\zeta) \le 1$. Therefore, MMT does not currently model overpopulation effects, although it could be extended to include them in the future.

%-------------------------------------------------------------------------------

%-------------------------------------------------------------------------------

%-------------------------------------------------------------------------------
%\clearpage
\section{Benchmarking against Direct Molecular Simulations} \label{sec:verif_vs_dms}

Our overarching goal for the Modified Marrone-Treanor model has been to construct a \emph{predictive} dissociation model with a functional form and parametrization based as much as possible on detailed ab initio results, in contrast to existing empirical models that have primarily been fit to experimental data. In the preceding sections we have presented the MMT model's analytical formulation and listed the parameters required to evaluate dissociation rate coefficients, vibrational energy removal rates, as well as characteristic vibrational relaxation times, all derived from detailed ab initio calculations for nitrogen and oxygen~\cite{BenderVNPVTSC2015, ValentiniSBC2016, ChaudhryBSC2018, Chaudhry2018, torres24b}.

A complete validation of MMT for real-world applications will require large-scale CFD simulations of nonequilibrium reacting flows produced at hypersonic flight conditions, or in ground test facilities. However, before achieving this milestone, we must assess to what degree our CFD implementation of the MMT model agrees with DMS results that already describe at the microscopic scale the most relevant thermochemical nonequilibrium processes in $\mathrm{N_2/N}$ and $\mathrm{O_2/O}$) mixtures. In these comparison calculations we focus on the dissociation chemistry coupled to vibrational relaxation, which in the CFD model is fully driven by the source terms of Eqs.~(\ref{eq:species_continuity}) and (\ref{eq:vibconservation}). It is most convenient to study these phenomena isolated from other fluid dynamics effects, such as convection and diffusion, by simulating simple space-homogeneous heat baths (also referred to as 0D reactors), either under isothermal (constant translational temperature), or adiabatic (constant total energy) conditions.

In order to achieve the most meaningful comparison, we must take care to only include physical processes in the CFD-MMT calculations that are also captured by DMS. Thus, the chemistry parameters taken from Table~\ref{tab:mmt_params} (either sub-tables) and vibrational relaxation from Table~\ref{tab:tau_v_fit_reduced} were derived on the same PES set as the one employed in the DMS reference calculations. This implies that in both solution methods the dissociation and vibrational relaxation processes are based exclusively on electronically adiabatic trajectories between ground-state atoms and molecules. It further means that the thermodynamic properties (i.e. enthalpies, heat capacities, equilibrium constants, etc.) employed in our CFD-MMT benchmarking calculations explicitly \emph{do not} account for contributions of molecular or atomic species' electronic excited states.

Refer to Tables~\ref{tab:mmt_params} and \ref{tab:tau_v_fit_reduced} for the parameters used to generate the MMT results presented in this section. Further note that constant non-Boltzmann factors of $f_k^\mathrm{NB} = 0.5$ and $f_\varepsilon^\mathrm{NB} = 0.85$ are employed in Eqs.~(\ref{eq:mmt_rate_coefficient}) and (\ref{eq:devib_nb_factor}) respectively  throughout all MMT calculations presented in this section.

%-------------------------------------------------------------------------------
\subsection{Vibrational excitation and dissociation of nitrogen and oxygen in isothermal heat baths} \label{sec:nitro_oxy_iso}

\begin{figure}%[htb]
 \centering
 
 \begin{subfigure}[t]{0.5\textwidth}
  \centering
  \includegraphics[width=\textwidth]{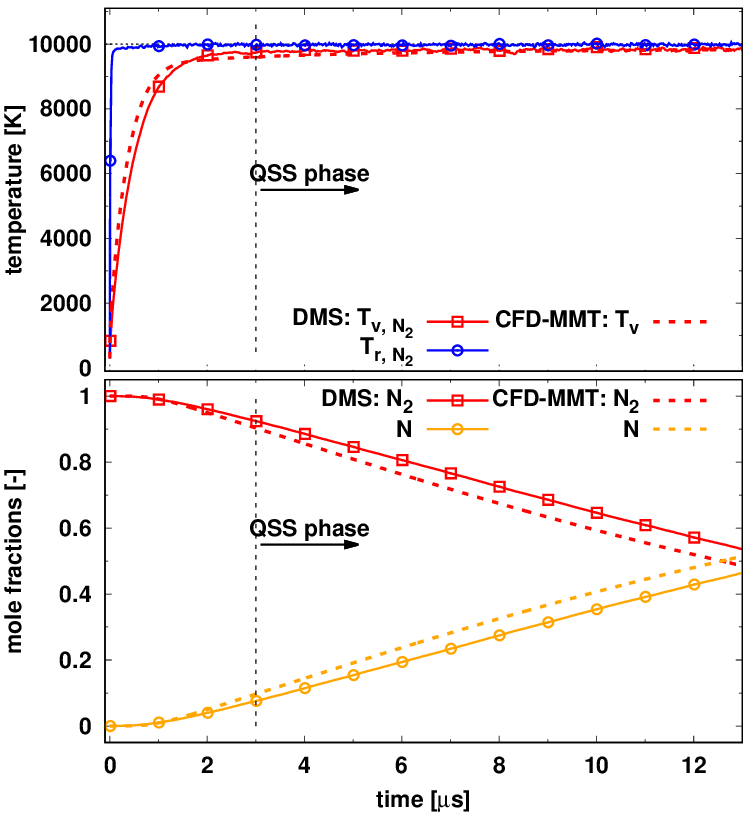}
  \caption{$\mathrm{N_2/N}$ isothermal heat bath at $T = 10\,000 \, \mathrm{K}$}
  \label{fig:iso_n2_10k_combined_vs_dms}
 \end{subfigure}~
 \begin{subfigure}[t]{0.5\textwidth}
  \centering
  \includegraphics[width=\textwidth]{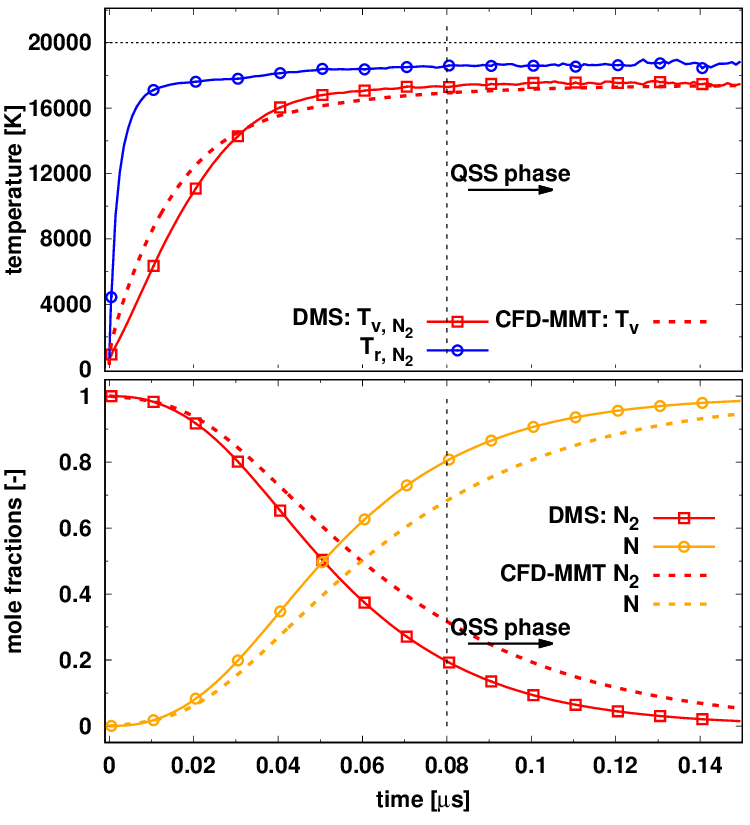}
  \caption{$\mathrm{N_2/N}$ isothermal heat bath at $T = 20\,000 \, \mathrm{K}$}
  \label{fig:iso_n2_20k_combined_vs_dms}
 \end{subfigure}
 
 \caption{Benchmarking of CFD-MMT model against DMS for nitrogen dissociation in isothermal heat bath}
 \label{fig:iso_n2_vs_dms}
\end{figure}

We benchmark the MMT model in scenarios where the gas initially consists only of a single diatomic species, either pure $\mathrm{N_2}$, or pure $\mathrm{O_2}$. Figure~\ref{fig:iso_n2_vs_dms} shows two cases of nitrogen dissociation in isothermal, isochoric heat baths. In both these CFD-MMT calculations the gas is initially composed of pure $\mathrm{N_2}$ with an initial number density of $n_0 = 10^{24} \, \mathrm{m^{-3}}$ at initial vibrational temperature $T_\mathrm{v} \, (t\!\!=\!\!0) = 300 \, \mathrm{K}$, while the combined translation-rotational mode is kept artificially constant at the heat bath temperature. Macroscopic conditions imposed in the DMS reference calculations are similar to those of the CFD-MMT runs, with one major exception. The initial internal energies of all DMS particles representing the $\mathrm{N_2}$-gas are sampled from a Maxwell-Boltzmann distribution with equal rotational and vibrational temperatures, i.e. $T_\mathrm{r, N_2} \, (t\!\!=\!\!0) = T_\mathrm{v, N_2} \, (t\!\!=\!\!0) = 300 \, \mathrm{K}$, while their center-of-mass velocities are sampled from a Maxwellian distribution at the reservoir temperature $T$. In order to maintain isothermal conditions throughout the remainder of the DMS calculation, center-of-mass velocities of the $\mathrm{N_2}$ molecules, as well as of $\mathrm{N}$-atoms produced by dissociation, are re-sampled from their respective Maxwellians at reservoir temperature $T$ after every time step. This means that in the DMS calculations both $T_\mathrm{r, N_2}$ and $T_\mathrm{v, N_2}$ will begin to rise from their initial values and gradually approach the reservoir temperature. By contrast, in the CFD-MMT calculations it is implicitly assumed that the rotational mode starts out fully excited at the reservoir temperature and only the vibrational mode is initialized out of equilibrium.

In Figs.~\ref{fig:iso_n2_10k_combined_vs_dms} and \ref{fig:iso_n2_20k_combined_vs_dms} we compare CFD-MMT (dotted lines) and DMS (continuous lines with symbols) results at reservoir temperatures $T = 10\,000 \, \mathrm{K}$ and $T = 20\,000 \, \mathrm{K}$ respectively. At both conditions, the CFD-MMT and DMS-derived $T_\mathrm{v}$ profiles rise together at almost equal rates from the initial value at $300 \, \mathrm{K}$. In the $10\,000 \, \mathrm{K}$-case this phase is seen to last about $2 \, \mathrm{\mu s}$, while in the $20\,000 \, \mathrm{K}$-case it is roughly $0.04 \, \mathrm{\mu s}$ long. The vibrational temperatures then level off at near-constant values somewhat below the respective heat bath temperatures. For the first case this $T_\mathrm{v}$-plateau lies only about $150 \, \mathrm{K}$ below $T = 10\,000 \, \mathrm{K}$, whereas for the second case the gap between $T = 20\,000 \, \mathrm{K}$ and $T_\mathrm{v}$ is closer to $2600 \, \mathrm{K}$, about ten times greater in relative terms. The observed gaps imply that during the simulated time neither of the two cases manages to attain thermal equilibrium. Instead, after the rapid early rise in $T_\mathrm{v}$, the mixture settles into a QSS dissociation regime with slight thermal nonequilibrium for the remainder of simulated time.

As mentioned in Sec.~\ref{sec:nb_effects}, this phenomenon is the result of a temporary balance established between two processes acting in opposite directions. First, vibrational excitation drives $e_{\mathrm{v}, \mathrm{N_2}} (T_\mathrm{v})$ up toward matching its corresponding local equilibrium value $e_{\mathrm{v}, \mathrm{N_2}} (T)$. Simultaneously, with every dissociation reaction a certain amount of vibrational energy is removed from the gas, pushing $e_{\mathrm{v}, \mathrm{N_2} (T_\mathrm{v})}$ in the opposite direction. As discussed for instance in Refs.~\cite{macdonald20b, torres24a}, in the DMS solutions this quasi-steady-state dissociation phase is naturally captured by the method itself and vertical dashed lines in Figs.~\ref{fig:iso_n2_10k_combined_vs_dms} and \ref{fig:iso_n2_20k_combined_vs_dms} mark its approximate beginning. The $T - T_\mathrm{v}$ gap in DMS occurs as a macroscopic manifestation of a temporary balance at the microscopic level between vibrational energy gain in inelastic collisions and nearly equal energy loss due to reactive collisions respectively. In the CFD-MMT solutions the same process is captured macroscopically by the difference between the two vibrational energy source terms of Eq.~(\ref{eq:vib_relax_equation}). Given that $T_\mathrm{v} < T$ throughout the entire simulated time, the vibrational relaxation term $w_v^\mathrm{relax}$ remains positive and is greatest in magnitude early on. During the vibrational excitation phase this is the dominant source term, only being counteracted by the vibrational energy-chemistry coupling term $w_v^\mathrm{chem}$ once nitrogen dissociation begins to gain prominence. After the dissociation rate has fully ramped up, $w_v^\mathrm{relax}$ and $w_v^\mathrm{chem}$ become almost equal in magnitude, but opposite in sign. Thus, according to Eq.~(\ref{eq:vib_relax_equation}) vibrational energy density, and with it $T_\mathrm{v}$, in the heat bath do not change appreciably during the QSS phase.

Although both methods produce largely similar vibrational excitation profiles, there remain slight differences. At both heat bath conditions the CFD-derived vibrational temperature (dotted red) profile leads its DMS-derived (continuous red) counterpart during the early excitation phase. The lag in the DMS $T_\mathrm{v}$-profiles can be attributed to competition between rotational and vibrational relaxation taking place simultaneously in the DMS calculations. Here part of the energy provided by the heat bath is channeled toward raising the gas' rotational energy, delaying transfer to the vibrational mode in the process. These differences are especially pronounced at the higher-temperature heat bath condition, where rotational relaxation occurs at a time scale comparable to that of vibrational energy. Indeed, the $T_{\mathrm{r}, \mathrm{N_2}}$ profile (continuous blue line) in Fig.~\ref{fig:iso_n2_20k_combined_vs_dms} gradually flattens during the QSS dissociation phase, implying that in the DMS solution the rotational mode never fully equilibrates with the heat bath temperature. By contrast, in the CFD-MMT solutions the rotational mode starts out and remains fully excited at the heat bath temperature the entire time. %This is one major difference between to keep in mind when comparing DMS and CFD-MMT solutions thro
At the lower-temperature condition the ``interference'' of rotational relaxation is less severe. At $T = 10\,000 \, \mathrm{K}$ the characteristic time scale for rotational relaxation is much shorter than that of the vibrational mode and, as seen by the rapid approach of the DMS-derived $T_{\mathrm{r}, \mathrm{N_2}}$ profile to match $T$ in Fig.~\ref{fig:iso_n2_10k_combined_vs_dms} (continuous blue line), is complete within the first microsecond. Thus, the implicit assumption of a fully excited rotational mode in the CFD-MMT solution is satisfied to a much greater degree in the DMS reference solution at this lower temperature.

The corresponding CFD-MMT and DMS profiles for mixture composition are shown in the lower halves of Figs.~\ref{fig:iso_n2_10k_combined_vs_dms} and \ref{fig:iso_n2_20k_combined_vs_dms}. At the lower-temperature condition the $\mathrm{N_2}$ and $\mathrm{N}$ mole fraction profiles of the CFD-MMT solution (red and orange dotted lines) lead their DMS counterparts (solid lines) by a small amount. In the higher-temperature heat bath this relationship is inverted and the CFD-MMT solution lags behind its DMS counterpart by a small, equally noticeable amount. Such isothermal heat bath simulations involve extreme conditions as energy is continually being added to the system, and exact agreement between a two-temperature CFD model and DMS is not expected.

%-------------------------------------------------------------------------------

%-------------------------------------------------------------------------------

\begin{figure}%[htb]
 \centering
 
 \begin{subfigure}[t]{0.5\textwidth}
  \centering
  \includegraphics[width=\textwidth]{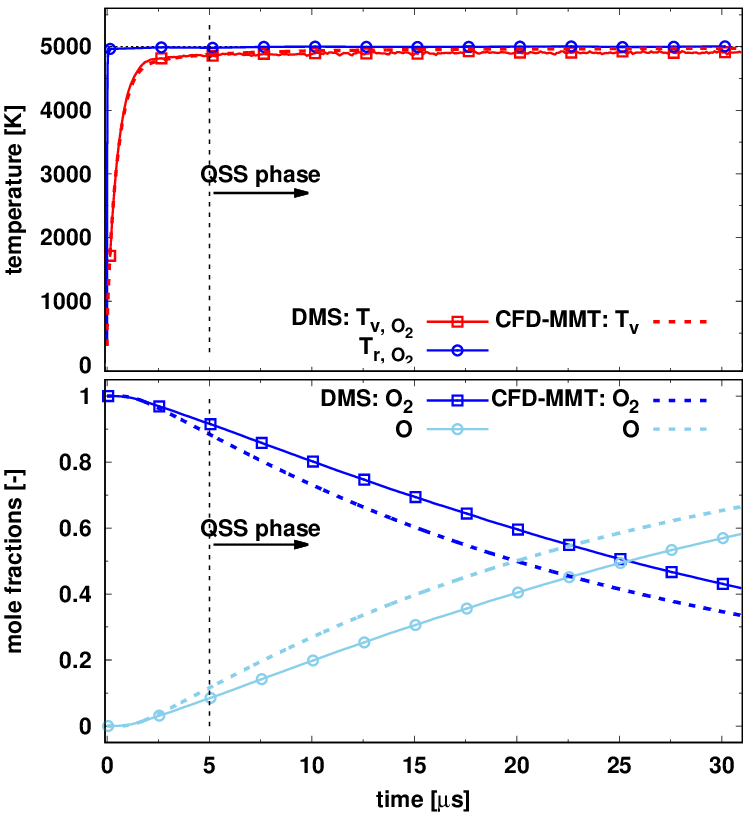}
  \caption{$\mathrm{O_2/O}$ isothermal heat bath at $T = 5\,000 \, \mathrm{K}$}
  \label{fig:iso_o2_5k_combined_vs_dms}
 \end{subfigure}~
 \begin{subfigure}[t]{0.5\textwidth}
  \centering
  \includegraphics[width=\textwidth]{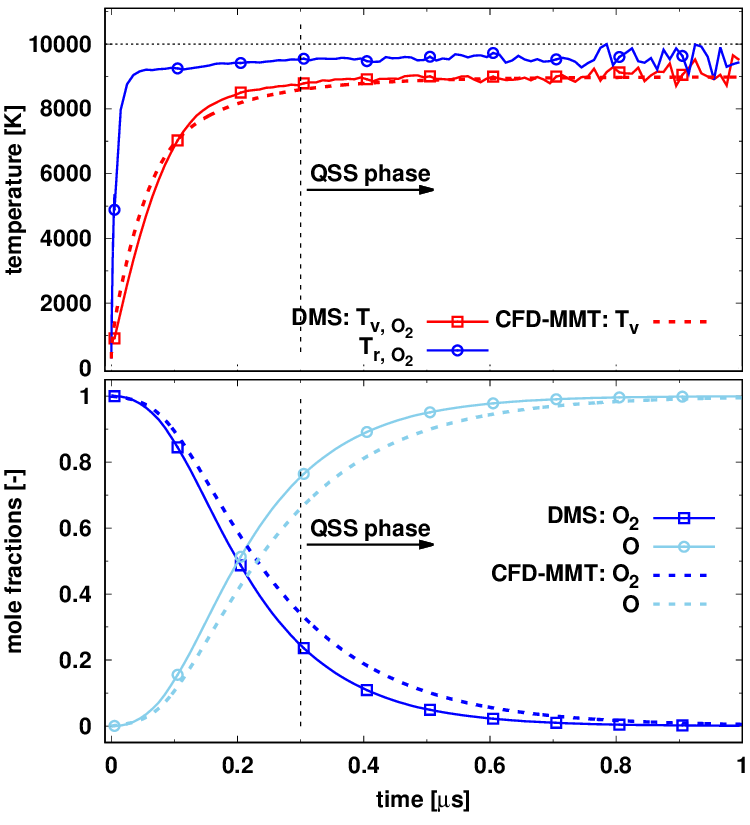}
  \caption{$\mathrm{O_2/O}$ isothermal heat bath at $T = 10\,000 \, \mathrm{K}$}
  \label{fig:iso_o2_10k_combined_vs_dms}
 \end{subfigure} 
 
 \caption{Benchmarking of CFD-MMT model against DMS for oxygen dissociation in isothermal heat bath}
 \label{fig:iso_o2_vs_dms}
\end{figure}

In Fig.~\ref{fig:iso_o2_vs_dms} we report results for CFD/DMS comparisons in an $\mathrm{O_2}/\mathrm{O}$ mixture at conditions similar to the preceding ones for nitrogen. Since the dissociation threshold for oxygen is about half that for nitrogen ($T_\mathrm{D} = 59\,300 \, \mathrm{K}$ for $\mathrm{O_2}$ vs. $113\,200 \, \mathrm{K}$ for $\mathrm{N_2}$), it readily dissociates at temperatures several thousand kelvin lower. Therefore, oxygen comparisons are carried out in isothermal heat baths at $T = 5\,000 \, \mathrm{K}$ and $10\,000 \, \mathrm{K}$ respectively. 

As seen in Figs.~\ref{fig:iso_o2_5k_combined_vs_dms} and \ref{fig:iso_o2_10k_combined_vs_dms}, temperature and mixture composition profiles for the $T = 5\,000 \, \mathrm{K}$-oxygen case qualitatively resemble those of the $10\,000 \, \mathrm{K}$-nitrogen case, and a similar observation can be made for the $10\,000 \, \mathrm{K}$-oxygen vs. $20\,000 \, \mathrm{K}$-nitrogen heat baths. Focusing on the lower-temperature condition first, in Fig.~\ref{fig:iso_o2_5k_combined_vs_dms} one sees a rapid rise in $\mathrm{O_2}$ vibrational temperature and approach to a QSS-phase plateau only a few hundred kelvin below the heat bath's constant $T = 5\,000 \, \mathrm{K}$. The CFD-derived (dotted red) and DMS (solid red) $T_\mathrm{v}$ profiles lie almost on top of each other during vibrational excitation, confirming that the values for calculating $\tau_\mathrm{O_2 - O_2}^\mathrm{v}$ and $\tau_\mathrm{O_2 - O}^\mathrm{v}$ from Table~\ref{tab:tau_v_fit_reduced} do an excellent job of reproducing the DMS behavior. Just as was seen for nitrogen in Fig.~\ref{fig:iso_n2_10k_combined_vs_dms}, the DMS-derived $\mathrm{O_2}$ rotational temperature in Fig.~\ref{fig:iso_o2_5k_combined_vs_dms} (solid blue line) almost instantly jumps to match the heat bath temperature. This is a consequence of the characteristic time scale for rotational relaxation being so much shorter than that for vibration at these conditions.

By contrast, in the $10\,000 \, \mathrm{K}$-oxygen heat bath, DMS-derived $\mathrm{O_2}$ rotational and vibrational temperatures rise over similar time scales (solid blue and red lines in Fig.~\ref{fig:iso_o2_10k_combined_vs_dms}). At this condition the rotational temperature of $\mathrm{O_2}$ levels off at its own QSS plateau and does not reach the heat bath temperature before the simulation ends. Thus, as was seen for the $20\,000 \, \mathrm{K}$-nitrogen case in Fig.~\ref{fig:iso_n2_20k_combined_vs_dms}, in the $10\,000 \, \mathrm{K}$-oxygen DMS calculation a certain degree of translation-rotational nonequilibrium persists for the entire simulated time. Figure~\ref{fig:iso_o2_10k_combined_vs_dms} also shows that the CFD-derived $T_\mathrm{v}$ profile (dotted red) agrees well with the DMS-derived one (solid red), throughout the excitation and QSS phases. Both vibrational temperature profiles level off at nearly equal QSS plateaus.

Finally, the $\mathrm{O_2/O}$ mole fraction profiles in the bottom half of Figs.~\ref{fig:iso_o2_5k_combined_vs_dms} and \ref{fig:iso_o2_10k_combined_vs_dms} exhibit behavior qualitatively very similar to the $\mathrm{N_2/N}$ equivalents in the bottom half of Figs.~\ref{fig:iso_n2_10k_combined_vs_dms} and \ref{fig:iso_n2_20k_combined_vs_dms} respectively. At $T = 5\,000 \, \mathrm{K}$ the CFD-derived $\mathrm{O_2}$ and $\mathrm{O}$ mole fraction profiles (dotted blue lines in bottom half of Fig.~\ref{fig:iso_o2_5k_combined_vs_dms}) lead their DMS counterparts (solid lines) by a small amount. At $T = 10\,000 \, \mathrm{K}$ this relationship is reversed, with the CFD-derived $\mathrm{O_2}$ and $\mathrm{O}$ mole fraction profiles slightly lagging behind their DMS counterparts. We consider deviations of this magnitude acceptable given how much the choice of a single, temperature-independent non-Boltzmann factor allows us to simplify the MMT model.

%-------------------------------------------------------------------------------

%-------------------------------------------------------------------------------

\subsection{Vibrational excitation and dissociation of nitrogen and oxygen in adiabatic heat baths} \label{sec:nitro_oxy_adia}

After the isothermal heat bath comparisons of Sec.~\ref{sec:nitro_oxy_iso}, we perform a similar analysis for adiabatic conditions. In this scenario the heat bath's total energy remains constant and translational temperature freely adjusts in response to internal energy redistribution processes and chemical reactions, which gradually drive the gas closer to equilibrium. These conditions allow us to better assess the MMT model's behavior over a wide temperature range within a single calculation and it is arguably more representative of post-shock conditions.

Four separate cases, two for nitrogen and two for oxygen are simulated using the MMT model and compared to corresponding DMS reference solutions. In the following discussion each case is identified by the heat bath's specific enthalpy (constant throughout the simulation). The DMS results used for reference  were taken from previous work (nitrogen cases 3 and 4 in Ref.~\cite{torres20b} and ``low-enthalpy'' and ``high-enthalpy'' oxygen cases in Ref.~\cite{torres21a}). All initial conditions are summarized in Table~\ref{tab:nitro_oxy_adia_conditions} and, as was the case for the isothermal heat baths of Sec.~\ref{sec:nitro_oxy_iso}, they differ slightly between CFD and the equivalent DMS calculations. In the DMS calculations the translational temperature $T_\mathrm{t}^\mathrm{DMS}(t=0)$ is initialized at a higher value than the common rotational and vibrational temperatures $T_\mathrm{r}^\mathrm{DMS}(t=0) = T_\mathrm{v}^\mathrm{DMS}(t=0)$, causing the gas to undergo rapid early translation-rotational relaxation, in parallel with a slower vibrational excitation phase. By construction, in the two-temperature CFD model translational and rotational modes remain in equilibrium at all times. Thus, we must initialize the combined translation-rotational temperature to the weighted average $T_\mathrm{t-r}^\mathrm{CFD}(t=0) = \tfrac{3}{5} T_\mathrm{t}^\mathrm{DMS}(t=0) + \tfrac{2}{5} T_\mathrm{r, A_2}^\mathrm{DMS}(t=0)$ in order to match the enthalpy of the DMS case. Note that the weights $c_v^\mathrm{t} / (c_v^\mathrm{t} + c_v^\mathrm{r}) = 3/5$ and $c_v^\mathrm{r} / (c_v^\mathrm{t} + c_v^\mathrm{r}) = 2/5$ correspond to the relative energy contributions to heat capacity at constant volume of the translational and rotational modes in a fully excited diatomic gas. In each of the four cases the initial translation-rotational temperature of the CFD calculations will be lower than the DMS's initial translational one, because the two modes in CFD begin ``pre-equilibrated''. As seen in Table~\ref{tab:nitro_oxy_adia_conditions}, this also has repercussions for the initial pressures in the CFD and DMS calculations.

%-------------------------------------------------------------------------------

\begin{table}%[htb]
 \centering
 \caption{DMS and CFD initial conditions for adiabatic heat bath calculations}
 \label{tab:nitro_oxy_adia_conditions}
 
 \begin{tabular}{c c c c c c c c c c}
           &                  &        & initial & \multicolumn{3}{c}{DMS initial conditions} & \multicolumn{3}{c}{CFD initial conditions} \\
           & $h$              & $\rho \times 10^3$ & moles & $T_\mathrm{t}$ & $T_{\mathrm{r}, \mathrm{A_2}} = T_{\mathrm{v}, \mathrm{A_2}}$ & $p$ & $T$ & $T_\mathrm{v}$      & $p$ \\
  mixture  & $\mathrm{[MJ/kg]}$ & $\mathrm{[kg/m^3]}$ & $\mathrm{[-]}$ & $\mathrm{[K]}$ & $\mathrm{[K]}$ & $\mathrm{[kPa]}$ & $\mathrm{[K]}$ & $\mathrm{[K]}$      & $\mathrm{[kPa]}$ \\ \hline \hline
  nitrogen & $12.3$           & $2.11$ & $100\% \, \mathrm{N_2}$ & $24\,000$ & $300$ & $15.0$ & $14\,520$ & $300$ & $9.08$ \\
  nitrogen & $20.1$           & $2.65$ & $100\% \, \mathrm{N_2}$ & $42\,000$ & $300$ & $33.0$ & $25\,320$ & $300$ & $25.3$ \\
  oxygen   & $4.67$           & $34.5$ & $100\% \, \mathrm{O_2}$ & $10\,300$ & $298$ & $92.0$ & $6\,280$  & $298$ & $56.3$ \\
  oxygen   & $9.81$           & $20.5$ & $100\% \, \mathrm{O_2}$ & $22\,500$ & $298$ & $120$  & $13\,600$ & $298$ & $72.7$ \\
 \end{tabular}
 
\end{table}

%-------------------------------------------------------------------------------

\begin{figure}%[htb]
 \centering
 
 \begin{subfigure}[t]{0.5\textwidth}
  \centering
  \includegraphics[width=\textwidth]{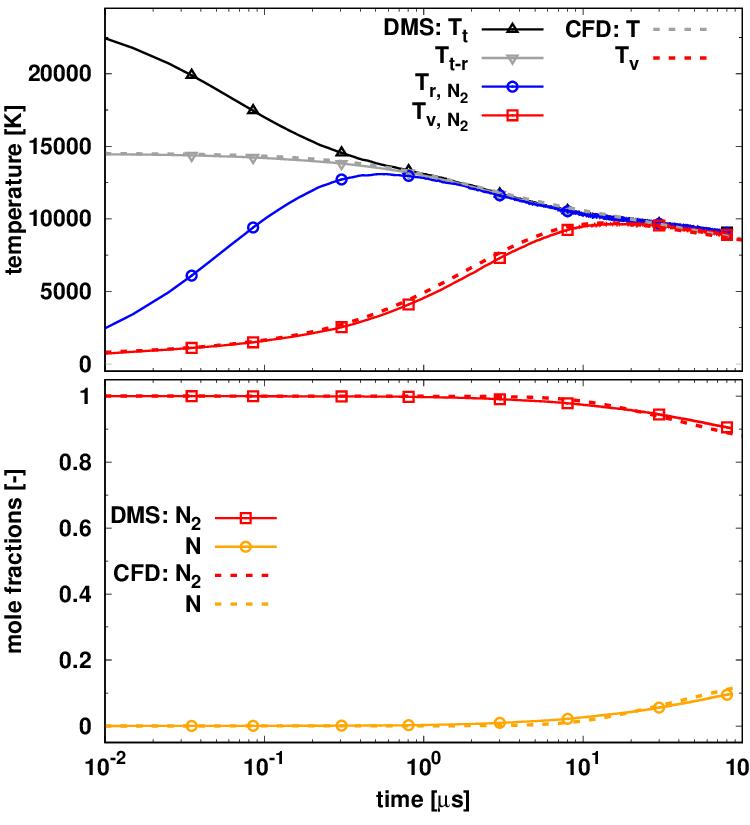}
  \caption{$\mathrm{N_2/N}$ adiabatic heat bath at $h = 12.3 \, \mathrm{MJ/kg}$}
  \label{fig:adia_n2_h12p3_combined_vs_dms}
 \end{subfigure}~
 \begin{subfigure}[t]{0.5\textwidth}
  \centering
  \includegraphics[width=\textwidth]{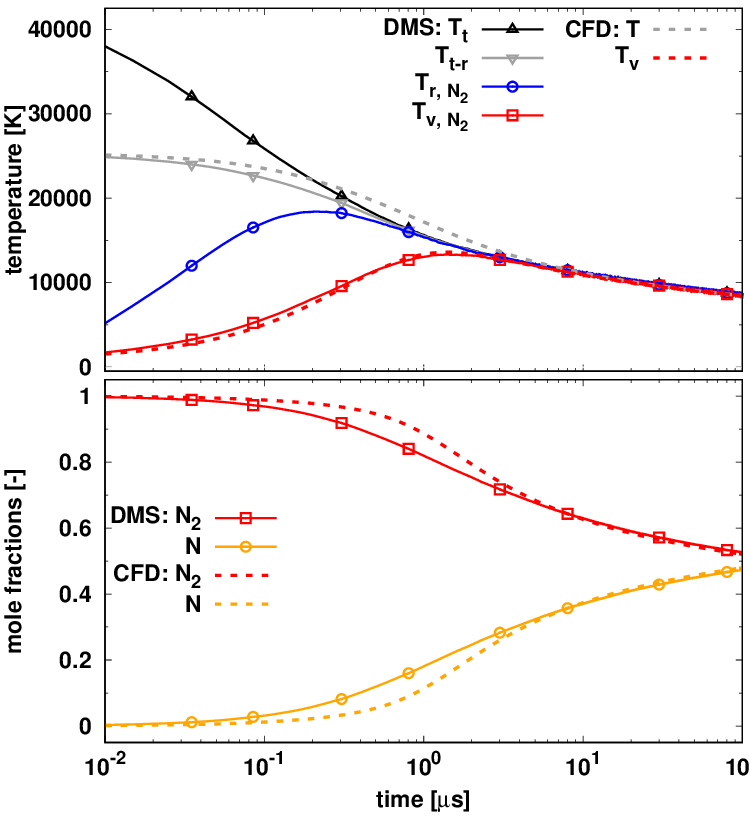}
  \caption{$\mathrm{N_2/N}$ adiabatic heat bath at $h = 20.1 \, \mathrm{MJ/kg}$}
  \label{fig:adia_n2_h21p0_combined_vs_dms}
 \end{subfigure}

 \caption{Benchmarking of CFD-MMT model against DMS for nitrogen dissociation in adiabatic heat bath}
 \label{fig:adia_n2_vs_dms}
\end{figure}

Figure~\ref{fig:adia_n2_vs_dms} summarizes the comparisons for both nitrogen cases, the lower-enthalpy case in Fig.~\ref{fig:adia_n2_h12p3_combined_vs_dms} and the higher-enthalpy one in Fig.~\ref{fig:adia_n2_h21p0_combined_vs_dms}. As in each of the previous figures, the upper half shows temperatures and the lower half the corresponding mole fraction profiles. Most labeling conventions, line patterns and colors are carried over from Fig.~\ref{fig:iso_n2_vs_dms}, with two minor changes. Instead of a linear time axis, Figs.~\ref{fig:adia_n2_h12p3_combined_vs_dms} and \ref{fig:adia_n2_h21p0_combined_vs_dms} both use a logarithmic scale to better highlight the early rotational and vibrational relaxation phases in the DMS calculations, while also allowing one to observe the slow decrease in temperature due to $\mathrm{N_2}$-dissociation later on. In addition to vibrational temperature profiles (red lines) from both methods, separate translational (solid black) and rotational (solid blue) temperatures are plotted for the DMS solution. A dotted gray curve represents the CFD-derived trans-rotational temperature $T$ and an equivalent one for $T_\mathrm{t-r}^\mathrm{DMS}$ is shown as a solid gray line. The latter is calculated, as just discussed, by weighing the DMS translational and rotational mode contributions according to $T_\mathrm{t-r}^\mathrm{DMS} = \tfrac{3}{5} \, T_\mathrm{t}^\mathrm{DMS} + \tfrac{2}{5} \, T_\mathrm{r, N_2}^\mathrm{DMS}$. During early translation-rotational relaxation captured by the DMS calculations this profile represents a fictitious average temperature, which nevertheless makes direct comparison between CFD and DMS results more straightforward. 

Focusing on the lower-enthalpy case first, the CFD-derived trans-rotational temperature profile agrees remarkably well with the corresponding DMS one. It is evident that the CFD-MMT model closely matches the vibrational relaxation behavior of the DMS reference calculations. With a specific enthalpy of $12.3 \, \mathrm{MJ/kg}$ this case represents the milder of the two conditions, where dissociation of nitrogen is less intense and the time scales for vibrational relaxation and chemistry are disparate enough for the gas to be nearly fully vibrationally excited, before significant dissociation begins to occur.

At the higher-enthalpy ($h = 20.1 \, \mathrm{MJ/kg}$) conditions  shown in Fig.~\ref{fig:adia_n2_h21p0_combined_vs_dms}, dissociation now takes place at a time scale comparable to that of vibrational relaxation. Agreement between the CFD-MMT and DMS solutions is very good overall, but discrepancies in translation-rotational temperature and early mixture composition profiles are more noticeable here than they were at the lower-enthalpy condition. However, these discrepancies become smaller as time progresses and past the first $10$ microseconds there is almost exact agreement between the CFD and DMS solutions.

%$T_\mathrm{t-r}$

%-------------------------------------------------------------------------------

%-------------------------------------------------------------------------------

\begin{figure}%[htb]
 \centering
 
 \begin{subfigure}[t]{0.5\textwidth}
  \centering
  \includegraphics[width=\textwidth]{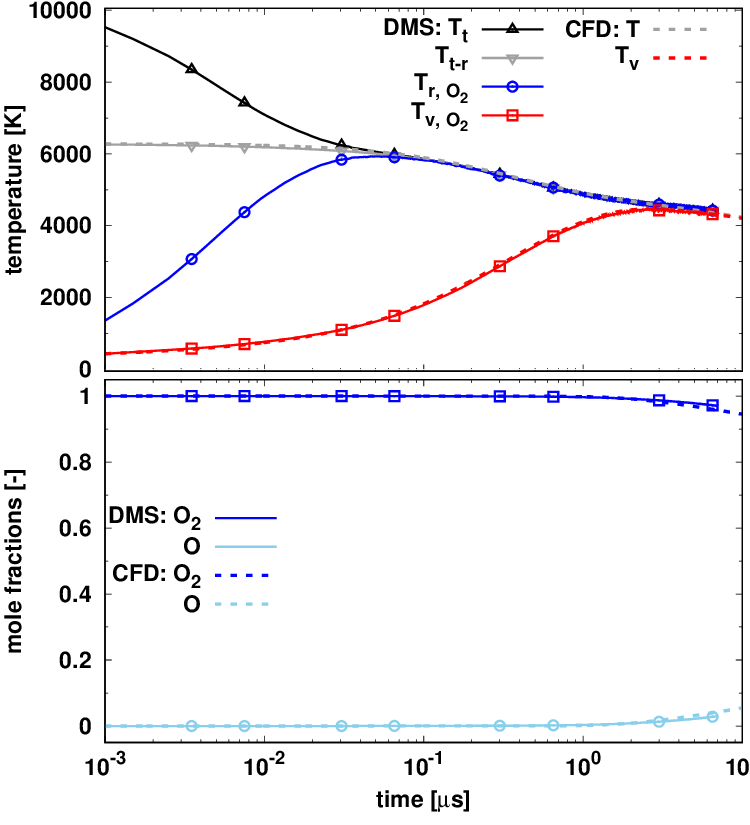}
  \caption{$\mathrm{O_2/O}$ adiabatic heat bath at $h = 4.67 \, \mathrm{MJ/kg}$}
  \label{fig:adia_o2_h4p665_combined_vs_dms}
 \end{subfigure}~
 \begin{subfigure}[t]{0.5\textwidth}
  \centering
  \includegraphics[width=\textwidth]{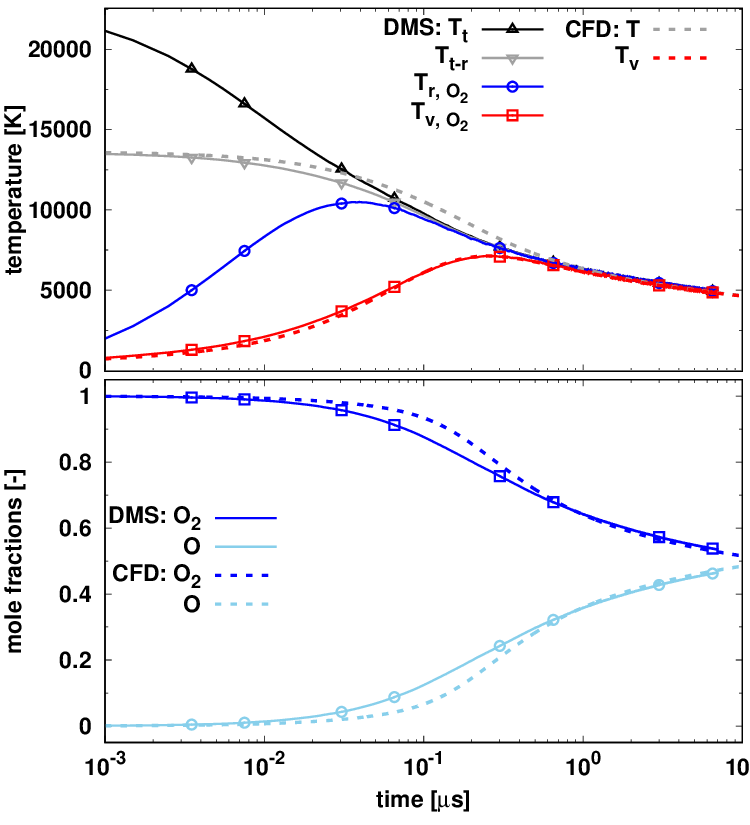}
  \caption{$\mathrm{O_2/O}$ adiabatic heat bath at $h = 9.81 \, \mathrm{MJ/kg}$}
  \label{fig:adia_o2_h9p813_combined_vs_dms}
 \end{subfigure}
 
 \caption{Benchmarking of CFD-MMT model against DMS for oxygen dissociation in adiabatic heat bath}
 \label{fig:adia_o2_vs_dms}
\end{figure}

The two adiabatic reservoir cases with pure oxygen are shown in Fig.~\ref{fig:adia_o2_vs_dms}. Much like for the preceding adiabatic nitrogen cases, at the lower-enthalpy condition shown in Fig.~\ref{fig:adia_o2_h4p665_combined_vs_dms} excellent agreement between CFD and DMS profiles is observed. More significant discrepancies between CFD and DMS solutions can be observed for the higher-enthalpy condition in Fig.~\ref{fig:adia_o2_h9p813_combined_vs_dms}. Here a time lag between $T$ from CFD and $T_\mathrm{t-r}^\mathrm{DMS}$, as well as both methods' species mole fraction profiles, qualitatively similar to the high-enthalpy nitrogen case of Fig.~\ref{fig:adia_n2_h21p0_combined_vs_dms}, can be observed during the first $0.7$ microseconds. The CFD solution (dotted lines) slightly trails behind the DMS one (solid lines) early on, but eventually catches up. Past the first microsecond both methods are in excellent agreement up until the point where the DMS solution cuts off.

%-------------------------------------------------------------------------------

%-------------------------------------------------------------------------------

%-------------------------------------------------------------------------------
%\clearpage
\section{Conclusion} \label{sec:conclusion}

In this paper we have presented a new two-temperature model for air species dissociation. The reaction set considered in this derivation encompasses $\mathrm{N_2}$ dissociation with collision partners $\mathrm{M = \{N_2, N, O_2 \}}$ and $\mathrm{O_2}$ dissociation with partners $\mathrm{M = \{O_2, O, N_2\}}$. The new model's outstanding feature is that all its  parameters were derived exclusively by fitting to rate data derived from QCT calculations on highly accurate ab initio potential energy surfaces. The entire PES set was provided by the Computational Chemistry group at the University of Minnesota. 

Our analysis of QCT datasets for reactant molecule internal energy states sampled from Maxwell-Boltzmann distributions over wide $(T,T_\mathrm{v})$ ranges revealed that vibrational energy change per dissociation, as well as the vibration-level-specific dissociation probability density (referred to as support factor) collapse to nearly unique curves when normalized by dissociation energy $D_0$ in a suitable way. The analysis also highlighted that a reactant diatom's vibrational energy has a much greater influence on its dissociation probability than that stored in its rotational motion.

In aggregate, the QCT results suggest that a rather simple two-temperature model is able to capture the major relevant physics of shock-heated dissociating diatomic species and that our proposed modified form of the classical Marrone-Treanor model (abbreviated as MMT) fits all the QCT data based on Boltzmann internal energy distributions reasonably well. We find that simple correction factors can account for non-Boltzmann depletion effects observed in DMS, or state-resolved master equation calculations employing the same ab initio PESs. Furthermore, a simple functional form proposed for these correction factors ensures that such depletion effects vanish as the mixture approaches chemical equilibrium.

We benchmarked our CFD implementation of the MMT model against DMS reference solutions in space-homogeneous isothermal and adiabatic heat baths representative of post-shock conditions in hypersonic flows. Simulations of $\mathrm{N_2/N}$ and $\mathrm{O_2/O}$ mixtures show that, with our appropriately calibrated kinetic rate parameters, the two-temperature formulation of the MMT model closely matches DMS results even when dissociation takes place under vibrational nonequilibrium conditions. The MMT model's mutually consistent expressions for nonequilibrium dissociation rate coefficient and associated vibrational energy removal term ensure that the correct amount of energy is transferred between the vibrational and trans-rotational modes at all times. This feature is particularly important for achieving close agreement between CFD-MMT and DMS predictions under adiabatic conditions, where $T$ and $T_\mathrm{v}$ both vary dynamically in response to dissociation coupled to vibrational relaxation.

The proposed MMT model accurately reproduces the major features observed in DMS calculations, but remains computationally inexpensive enough to be employed in large-scale CFD simulations. In a follow-up article we will parameterize the MMT model for 5-species air, including {new QCT data for the Zeldovich exchange reactions, and benchmark} its predictions against DMS reference results for partially dissociated air.

%-------------------------------------------------------------------------------

\section*{Acknowledgments}

This work was sponsored by the Air Force Office of Scientific Research under grants FA9550-16-1-0161, FA9550-19-1-0219 and FA9550-23-1-0446. The views and conclusions contained herein are those of the authors and should not be interpreted as necessarily representing the official policies or endorsements, either expressed or implied, of the funding agencies or the U.S.~Government.

\bibliography{Bibliography} 

\begin{thebibliography}{67}
\newcommand{\enquote}[1]{``#1''}
\providecommand{\natexlab}[1]{#1}
\providecommand{\url}[1]{\texttt{#1}}
\providecommand{\urlprefix}{URL }
\expandafter\ifx\csname urlstyle\endcsname\relax
  \providecommand{\doi}[1]{doi:\discretionary{}{}{}#1}\else
  \providecommand{\doi}{doi:\discretionary{}{}{}\begingroup
  \urlstyle{rm}\Url}\fi

\bibitem[{Gnoffo(1999)}]{Gnoffo1999}
Gnoffo, P.~A., \enquote{Planetary-Entry Gas Dynamics,} \emph{Annual Review of
  Fluid Mechanics}, Vol.~31, No.~1, 1999, pp. 459--494.
\newblock \doi{10.1146/annurev.fluid.31.1.459}.

\bibitem[{Candler(2019)}]{Candler2019}
Candler, G.~V., \enquote{Rate Effects in Hypersonic Flows,} \emph{Annual Review
  of Fluid Mechanics}, Vol.~51, No.~1, 2019, pp. 379 -- 402.
\newblock \doi{10.1146/annurev-fluid-010518-040258}.

\bibitem[{Hammerling et~al.(1959)Hammerling, Teare, and
  Kivel}]{HammerlingTK1959}
Hammerling, P., Teare, J.~D., and Kivel, B., \enquote{Theory of Radiation from
  Luminous Shock Waves in Nitrogen,} \emph{The Physics of Fluids}, Vol.~2,
  No.~4, 1959, pp. 422--426.
\newblock \doi{10.1063/1.1724413}.

\bibitem[{Treanor and Marrone(1962)}]{TreanorM1962}
Treanor, C.~E., and Marrone, P.~V., \enquote{Effect of Dissociation on the Rate
  of Vibrational Relaxation,} \emph{The Physics of Fluids}, Vol.~5, No.~9,
  1962, pp. 1022--1026.
\newblock \doi{10.1063/1.1724467}.

\bibitem[{Marrone and Treanor(1963)}]{MarroneT1963}
Marrone, P.~V., and Treanor, C.~E., \enquote{Chemical Relaxation with
  Preferential Dissociation from Excited Vibrational Levels,} \emph{The Physics
  of Fluids}, Vol.~6, No.~9, 1963, pp. 1215--1221.
\newblock \doi{10.1063/1.1706888}.

\bibitem[{Gnoffo et~al.(1989)Gnoffo, Gupta, and Shinn}]{GnoffoGS1989}
Gnoffo, P.~A., Gupta, R.~N., and Shinn, J.~L., \enquote{Conservation Equations
  and Physical Models for Hypersonic Air Flows in Thermal and Chemical
  Nonequilibrium,} Tech. Rep. TP-2867, NASA, February 1989.

\bibitem[{Losev and Generalov(1962)}]{LosevG1962}
Losev, S.~A., and Generalov, N.~A., \enquote{A Study of the Excitation of
  Vibrations and Dissociation of Oxygen Molecules at High Temperatures,}
  \emph{Soviet Physics Doklady}, Vol.~6, 1962, p. 1081.

\bibitem[{Park(1988{\natexlab{a}})}]{Park1988}
Park, C., \enquote{Assessment of a Two-Temperature Kinetic Model for
  Dissociating and Weakly Ionizing Nitrogen,} \emph{Journal of Thermophysics
  and Heat Transfer}, Vol.~2, No.~1, 1988{\natexlab{a}}, pp. 8--16.
\newblock \doi{10.2514/3.55}.

\bibitem[{Knab et~al.(1995)Knab, Fruehauf, and Messerschmid}]{KnabFM1995}
Knab, O., Fruehauf, H.~H., and Messerschmid, E.~W., \enquote{Theory and
  Validation of the Physically Consistent Coupled Vibration-Chemistry-Vibration
  Model,} \emph{Journal of Thermophysics and Heat Transfer}, Vol.~9, No.~2,
  1995, pp. 219--226.
\newblock \doi{10.2514/3.649}.

\bibitem[{Luo et~al.(2018)Luo, Alexeenko, and Macheret}]{LuoAM2018}
Luo, H., Alexeenko, A.~A., and Macheret, S.~O., \enquote{Assessment of
  Classical Impulsive Models of Dissociation in Thermochemical Nonequilibrium,}
  \emph{Journal of Thermophysics and Heat Transfer}, Vol.~32, No.~4, 2018, pp.
  861--868.
\newblock \doi{10.2514/1.T5375}.

\bibitem[{Millikan and White(1963)}]{MillikanW1963}
Millikan, R.~C., and White, D.~R., \enquote{Systematics of Vibrational
  Relaxation,} \emph{The Journal of Chemical Physics}, Vol.~39, No.~12, 1963,
  pp. 3209--3213.
\newblock \doi{10.1063/1.1734182}.

\bibitem[{Truhlar and Muckerman(1979)}]{truhlar79a}
Truhlar, D., and Muckerman, J., \enquote{{R}eactive scattering cross sections
  {III}: {Q}uasiclassical and semiclassical methods,} \emph{Atom-Molecule
  Collision Theory}, Springer, 1979, pp. 505--566.

\bibitem[{Paukku et~al.(2013)Paukku, Yang, Varga, and Truhlar}]{PaukkuYVT2013}
Paukku, Y., Yang, K.~R., Varga, Z., and Truhlar, D.~G., \enquote{Global
  \emph{ab initio} Ground-State Potential Energy Surface of N$_4$,} \emph{The
  Journal of Chemical Physics}, Vol. 139, No.~4, 2013, p. 044309.
\newblock \doi{10.1063/1.4811653}.

\bibitem[{Paukku et~al.(2014)Paukku, Yang, Varga, and Truhlar}]{PaukkuYVT2014}
Paukku, Y., Yang, K.~R., Varga, Z., and Truhlar, D.~G., \enquote{Erratum:
  ``Global \emph{ab initio} Ground-State Potential Energy Surface of N$_4$''
  [J. Chem. Phys. 139, 044309 (2013)],} \emph{The Journal of Chemical Physics},
  Vol. 140, No.~1, 2014, p. 019903.
\newblock \doi{10.1063/1.4861562}.

\bibitem[{Bender et~al.(2015)Bender, Valentini, Nompelis, Paukku, Varga,
  Truhlar, Schwartzentruber, and Candler}]{BenderVNPVTSC2015}
Bender, J.~D., Valentini, P., Nompelis, I., Paukku, Y., Varga, Z., Truhlar,
  D.~G., Schwartzentruber, T., and Candler, G.~V., \enquote{An Improved
  Potential Energy Surface and Multi-Temperature Quasiclassical Trajectory
  Calculations of N$_2$ + N$_2$ Dissociation Reactions,} \emph{The Journal of
  Chemical Physics}, Vol. 143, No.~5, 2015, p. 054304.
\newblock \doi{10.1063/1.4927571}.

\bibitem[{Varga and Truhlar(2021)}]{varga21b}
Varga, Z., and Truhlar, D., \enquote{Potential energy surface for high-energy
  $\mathrm{N} + \mathrm{N_2}$ collisions,} \emph{Phys. Chem. Chem. Phys.},
  Vol.~23, 2021, pp. 26273--26284.
\newblock \doi{10.1039/D1CP04373K}.

\bibitem[{Paukku et~al.(2017)Paukku, Yang, Varga, Song, Bender, and
  Truhlar}]{PaukkuYVSBT2017}
Paukku, Y., Yang, K.~R., Varga, Z., Song, G., Bender, J.~D., and Truhlar,
  D.~G., \enquote{Potential Energy Surfaces of Quintet and Singlet O$_4$,}
  \emph{The Journal of Chemical Physics}, Vol. 147, No.~3, 2017, p. 034301.
\newblock \doi{10.1063/1.4993624}.

\bibitem[{Paukku et~al.(2018)Paukku, Varga, and Truhlar}]{PaukkuVT2018}
Paukku, Y., Varga, Z., and Truhlar, D.~G., \enquote{Potential Energy Surface of
  Triplet O$_4$,} \emph{The Journal of Chemical Physics}, Vol. 148, No.~12,
  2018, p. 124314.
\newblock \doi{10.1063/1.5017489}.

\bibitem[{Varga et~al.(2017)Varga, Paukku, and Truhlar}]{VargaPT2017}
Varga, Z., Paukku, Y., and Truhlar, D.~G., \enquote{Potential Energy Surfaces
  for O + O$_2$ Collisions,} \emph{The Journal of Chemical Physics}, Vol. 147,
  No.~15, 2017, p. 154312.
\newblock \doi{10.1063/1.4997169}.

\bibitem[{Varga et~al.(2016)Varga, Meana-Pañeda, Song, Paukku, and
  Truhlar}]{VargaMSPT2016}
Varga, Z., Meana-Pañeda, R., Song, G., Paukku, Y., and Truhlar, D.~G.,
  \enquote{Potential Energy Surface of Triplet N$_2$O$_2$,} \emph{The Journal
  of Chemical Physics}, Vol. 144, No.~2, 2016, p. 024310.
\newblock \doi{10.1063/1.4939008}.

\bibitem[{Valentini et~al.(2016)Valentini, Schwartzentruber, Bender, and
  Candler}]{ValentiniSBC2016}
Valentini, P., Schwartzentruber, T.~E., Bender, J.~D., and Candler, G.~V.,
  \enquote{Dynamics of Nitrogen Dissociation from Direct Molecular Simulation,}
  \emph{Physical Review Fluids}, Vol.~1, No.~4, 2016, p. 043402.
\newblock \doi{10.1103/PhysRevFluids.1.043402}.

\bibitem[{Chaudhry et~al.(2018{\natexlab{a}})Chaudhry, Bender,
  Schwartzentruber, and Candler}]{ChaudhryBSC2018}
Chaudhry, R.~S., Bender, J.~D., Schwartzentruber, T.~E., and Candler, G.~V.,
  \enquote{Quasiclassical Trajectory Analysis of Nitrogen for High-Temperature
  Chemical Kinetics,} \emph{Journal of Thermophysics and Heat Transfer},
  Vol.~32, No.~4, 2018{\natexlab{a}}, pp. 833--845.
\newblock \doi{10.2514/1.T5484}.

\bibitem[{Chaudhry et~al.(2016)Chaudhry, Bender, Valentini, Schwartzentruber,
  and Candler}]{ChaudhryBVSC2016}
Chaudhry, R.~S., Bender, J.~D., Valentini, P., Schwartzentruber, T.~E., and
  Candler, G.~V., \enquote{Analysis of Dissociation and Internal Energy
  Transfer in High-Energy N$_{2}$+O$_{2}$ Collisions using the Quasiclassical
  Trajectory Method,} \emph{46th AIAA Thermophysics Conference}, AIAA Paper
  2016-4319, 2016.
\newblock \doi{10.2514/6.2016-4319}.

\bibitem[{Chaudhry et~al.(2018{\natexlab{b}})Chaudhry, Grover, Bender,
  Schwartzentruber, and Candler}]{ChaudhryGBSC2018}
Chaudhry, R.~S., Grover, M.~S., Bender, J.~D., Schwartzentruber, T.~E., and
  Candler, G.~V., \enquote{Quasiclassical Trajectory Analysis of Oxygen
  Dissociation via O$_2$, O, and N$_2$,} \emph{2018 AIAA Aerospace Sciences
  Meeting}, AIAA Paper 2018-0237, 2018{\natexlab{b}}.
\newblock \doi{10.2514/6.2018-0237}.

\bibitem[{Chaudhry(2018)}]{Chaudhry2018}
Chaudhry, R.~S., \enquote{Modeling and Analysis of Chemical Kinetics for
  Hypersonic Flows in Air,} Ph.D. thesis, University of Minnesota, Minneapolis,
  MN, November 2018.
\newblock \urlprefix\url{https://hdl.handle.net/11299/201709}.

\bibitem[{Panesi et~al.(2013)Panesi, Jaffe, Schwenke, and
  Magin}]{PanesiJSM2013}
Panesi, M., Jaffe, R.~L., Schwenke, D.~W., and Magin, T.~E.,
  \enquote{Rovibrational Internal Energy Transfer and Dissociation of
  $N_2(^1\Sigma_g^+)$ -- $N(^4S_u)$ System in Hypersonic Flows,} \emph{The
  Journal of Chemical Physics}, Vol. 138, No.~4, 2013, p. 044312.
\newblock \doi{10.1063/1.4774412}.

\bibitem[{Kim and Boyd(2013)}]{KimB2013}
Kim, J.~G., and Boyd, I.~D., \enquote{State-Resolved Master Equation Analysis
  of Thermochemical Nonequilibrium of Nitrogen,} \emph{Chemical Physics}, Vol.
  415, 2013, pp. 237--246.
\newblock \doi{10.1016/j.chemphys.2013.01.027}.

\bibitem[{Richard L.~Jaffe(2015)}]{JaffeSP2015}
Richard L.~Jaffe, M.~P., David W.~Schwenke, \enquote{First Principles
  Calculation of Heavy Particle Rate Coefficients,} \emph{Hypersonic
  Nonequilibrium Flows: Fundamentals and Recent Advances}, Vol. 247, edited by
  E.~Josyula, Progress in Aerospace Sciences, AIAA, New York, 2015, pp.
  103--158.
\newblock \doi{10.2514/5.9781624103292.0103.0158}.

\bibitem[{Schwartzentruber et~al.(2017)Schwartzentruber, Grover, and
  Valentini}]{SchwartzentruberGV2018}
Schwartzentruber, T.~E., Grover, M.~S., and Valentini, P., \enquote{Direct
  Molecular Simulation of Nonequilibrium Dilute Gases,} \emph{Journal of
  Thermophysics and Heat Transfer}, Vol.~32, No.~4, 2017, pp. 892--903.
\newblock \doi{10.2514/1.T5188}.

\bibitem[{Valentini et~al.(2015)Valentini, Schwartzentruber, Bender, Nompelis,
  and Candler}]{ValentiniSBNC2015}
Valentini, P., Schwartzentruber, T.~E., Bender, J.~D., Nompelis, I., and
  Candler, G.~V., \enquote{Direct Molecular Simulation of Nitrogen Dissociation
  Based on an \emph{ab initio} Potential Energy Surface,} \emph{Physics of
  Fluids}, Vol.~27, No.~8, 2015, p. 086102.
\newblock \doi{10.1063/1.4929394}.

\bibitem[{Grover et~al.(2019{\natexlab{a}})Grover, Schwartzentruber, Varga, and
  Truhlar}]{grover19a}
Grover, M., Schwartzentruber, T., Varga, Z., and Truhlar, D.,
  \enquote{Vibrational Energy Transfer and Collision-Induced Dissociation in
  $\mathrm{O + O_2}$ Collisions,} \emph{Journal of Thermophysics and Heat
  Transfer}, Vol.~33, No.~3, 2019{\natexlab{a}}, pp. 797--807.
\newblock \doi{10.2514/1.T5551}.

\bibitem[{Grover et~al.(2019{\natexlab{b}})Grover, Torres, and
  Schwartzentruber}]{grover19b}
Grover, M., Torres, E., and Schwartzentruber, T., \enquote{Direct molecular
  simulation of internal energy relaxation and dissociation in oxygen,}
  \emph{Physics of Fluids}, Vol.~31, 2019{\natexlab{b}}, p. 076107.
\newblock \doi{10.1063/1.5108666}.

\bibitem[{Torres and Schwartzentruber(2020)}]{torres20b}
Torres, E., and Schwartzentruber, T., \enquote{Direct molecular simulation of
  nitrogen dissociation under adiabatic post-shock conditions,} \emph{Journal
  of Thermophysics and Heat transfer}, Vol.~34, No.~4, 2020, pp. 801--815.
\newblock \doi{10.2514/1.T5970}.

\bibitem[{Torres et~al.(2024)Torres, Geistfeld, and
  Schwartzentruber}]{torres24a}
Torres, E., Geistfeld, E.~C., and Schwartzentruber, T.~E.,
  \enquote{High-Temperature Nonequilibrium Air Chemistry from First
  Principles,} \emph{Journal of Thermophysics and Heat Transfer}, Vol.~38,
  No.~2, 2024, pp. 260--291.
\newblock \doi{10.2514/1.T6863}.

\bibitem[{Singh and Schwartzentruber(2020{\natexlab{a}})}]{singh20b}
Singh, N., and Schwartzentruber, T., \enquote{Consistent kinetic--continuum
  dissociation model {I}. {K}inetic formulation,} \emph{The Journal of Chemical
  Physics}, Vol. 152, No.~22, 2020{\natexlab{a}}, p. 224302.
\newblock \doi{10.1063/1.5142752}.

\bibitem[{Singh and Schwartzentruber(2020{\natexlab{b}})}]{singh20c}
Singh, N., and Schwartzentruber, T., \enquote{Consistent kinetic-continuum
  dissociation model. {II}. {C}ontinuum formulation and verification,}
  \emph{The Journal of Chemical Physics}, Vol. 152, No.~22, 2020{\natexlab{b}},
  p. 224303.
\newblock \doi{10.1063/1.5142754}.

\bibitem[{Torres and Schwartzentruber(2025)}]{torres24b}
Torres, E., and Schwartzentruber, T.~E., \enquote{Characteristic vibrational
  and rotational relaxation times for air species from first-principles
  calculations,} \emph{Journal of Thermophysics and Heat Transfer}, Vol.~39,
  No.~2, 2025, pp. 223--249.
\newblock \doi{10.2514/1.T7042}.

\bibitem[{Varga et~al.(2021)Varga, Meana-Pa\~neda, Song, Paukku, and
  Truhlar}]{potlib21}
Varga, Z., Meana-Pa\~neda, R., Song, G., Paukku, Y., and Truhlar, D.~G.,
  \enquote{POTLIB: An Online Library of Potential Energy Surfaces,} , 2021.
\newblock \urlprefix\url{comp.chem.umn.edu/potlib/}.

\bibitem[{Bender(2016)}]{Bender2016}
Bender, J.~D., \enquote{Multiscale Computational Analysis of Nitrogen and
  Oxygen Gas-Phase Thermochemistry in Hypersonic Flows,} Ph.D. thesis,
  University of Minnesota, Minneapolis, MN, February 2016.
\newblock \urlprefix\url{https://hdl.handle.net/11299/178972}.

\bibitem[{Ibraguimova et~al.(2013)Ibraguimova, Sergievskaya, Levashov,
  Shatalov, Tunik, and Zabelinskii}]{IbraguimovaSLSTZ2013}
Ibraguimova, L.~B., Sergievskaya, A.~L., Levashov, V.~Y., Shatalov, O.~P.,
  Tunik, Y.~V., and Zabelinskii, I.~E., \enquote{Investigation of Oxygen
  Dissociation and Vibrational Relaxation at Temperatures 4000--10800 K,}
  \emph{The Journal of Chemical Physics}, Vol. 139, No.~3, 2013, p. 034317.
\newblock \doi{10.1063/1.4813070}.

\bibitem[{Streicher et~al.(2020)Streicher, Krish, and Hanson}]{streicher20c}
Streicher, J.~W., Krish, A., and Hanson, R.~K., \enquote{{Vibrational
  relaxation time measurements in shock-heated oxygen and air from 2000 K to
  9000 K using ultraviolet laser absorption},} \emph{Physics of Fluids},
  Vol.~32, No.~8, 2020, p. 086101.
\newblock \doi{10.1063/5.0015890}.

\bibitem[{Streicher et~al.(2021)Streicher, Krish, and Hanson}]{streicher21a}
Streicher, J., Krish, A., and Hanson, R., \enquote{Coupled
  vibration-dissociation time-histories and rate measurements in shock-heated,
  nondilute $\mathrm{O_2}$ and $\mathrm{O_2-Ar}$ mixtures from 6000 to 14000
  K,} \emph{Physics of Fluids}, Vol.~33, No.~5, 2021, p. 056107.
\newblock \doi{10.1063/5.0048059}.

\bibitem[{Streicher et~al.(2022{\natexlab{a}})Streicher, Krish, and
  Hanson}]{streicher22a}
Streicher, J.~W., Krish, A., and Hanson, R.~K., \enquote{High-temperature
  vibrational relaxation and decomposition of shock-heated nitric oxide. I.
  Argon dilution from 2200 to 8700 K,} \emph{Physics of Fluids}, Vol.~34,
  No.~11, 2022{\natexlab{a}}, p. 116122.
\newblock \doi{10.1063/5.0109109}.

\bibitem[{Streicher et~al.(2022{\natexlab{b}})Streicher, Krish, and
  Hanson}]{streicher22b}
Streicher, J.~W., Krish, A., and Hanson, R.~K., \enquote{High-temperature
  vibrational relaxation and decomposition of shock-heated nitric oxide: II.
  Nitrogen dilution from 1900 to 8200 K,} \emph{Physics of Fluids}, Vol.~34,
  No.~11, 2022{\natexlab{b}}, p. 116123.
\newblock \doi{10.1063/5.0122787}.

\bibitem[{Park(1993)}]{Park1993}
Park, C., \enquote{Review of Chemical-Kinetic Problems of Future NASA Missions,
  I: Earth Entries,} \emph{Journal of Thermophysics and Heat Transfer}, Vol.~7,
  No.~3, 1993, pp. 385--398.
\newblock \doi{10.2514/3.431}.

\bibitem[{Nikitin(1974)}]{Nikitin1974}
Nikitin, E., \emph{Theory of Elementary Atomic and Molecular Processes in
  Gases}, Clarendon Press, Oxford, 1974.

\bibitem[{Jaffe(1986)}]{Jaffe1986}
Jaffe, R.~L., \enquote{Rate Constants for Chemical Reactions in
  High-Temperature Nonequilibrium Air,} \emph{Thermophysical Aspects of
  Re-Entry Flows}, Vol. 103, edited by J.~N. Moss and C.~D. Scott, Progress in
  Aerospace Sciences, AIAA, New York, 1986, pp. 123--151.
\newblock \doi{10.2514/5.9781600865770.0123.0151}.

\bibitem[{Riley et~al.(2006)Riley, Hobson, and Bence}]{RileyHB2006}
Riley, K.~F., Hobson, M.~P., and Bence, S.~J., \emph{Mathematical Methods for
  Physics and Engineering}, 3\textsuperscript{rd} ed., Cambridge University
  Press, Cambridge, UK, 2006.

\bibitem[{Byron(1959)}]{Byron1959}
Byron, S.~R., \enquote{Measurement of the Rate of Dissociation of Oxygen,}
  \emph{The Journal of Chemical Physics}, Vol.~30, No.~6, 1959, pp. 1380--1392.
\newblock \doi{10.1063/1.1730209}.

\bibitem[{Park(1996)}]{Park1996}
Park, C., \enquote{Review of Finite-Rate Chemistry Models for Air Dissociation
  and Ionization,} \emph{Molecular Physics and Hypersonic Flows}, NATO ASI
  Series (Series C: Mathematical and Physical Sciences), Vol. 482, edited by
  M.~Capitelli, Kluwer Academic Publishers, Dordrecht, 1996, pp. 581--596.
\newblock \doi{10.1007/978-94-009-0267-1_39}.

\bibitem[{Losev et~al.(1996)Losev, Kovach, Makarov, Pogosbekjan, and
  Sergievskaya}]{LosevKMPS1996}
Losev, S., Kovach, E., Makarov, V., Pogosbekjan, M., and Sergievskaya, A.,
  \enquote{Chemistry Models for Air Dissociation,} \emph{Molecular Physics and
  Hypersonic Flows}, NATO ASI Series (Series C: Mathematical and Physical
  Sciences), Vol. 482, edited by M.~Capitelli, Kluwer Academic Publishers,
  Dordrecht, 1996, pp. 581--596.
\newblock \doi{10.1007/978-94-009-0267-1_40}.

\bibitem[{Park(1988{\natexlab{b}})}]{Park1988_Two-Temperature}
Park, C., \enquote{Two-Temperature Interpretation of Dissociation Rate Data for
  N$_2$ and O$_2$,} \emph{26th Aerospace Sciences Meeting}, AIAA Paper
  1988-458, 1988{\natexlab{b}}.
\newblock \doi{10.2514/6.1988-458}.

\bibitem[{Park(1990)}]{Park1990}
Park, C., \emph{Nonequilibrium Hypersonic Aerothermodynamics}, John Wiley \&
  Sons, Inc., New York, 1990.

\bibitem[{Macdonald et~al.(2020)Macdonald, Torres, Schwartzentruber, and
  Panesi}]{macdonald20b}
Macdonald, R., Torres, E., Schwartzentruber, T., and Panesi, M.,
  \enquote{State-to-State Master Equation and Direct Molecular Simulation Study
  of Energy Transfer and Dissociation for the $\mathrm{N_2 - N}$ System,}
  \emph{The Journal of Physical Chemistry A}, 2020.
\newblock \doi{10.1021/acs.jpca.0c04029}.

\bibitem[{Singh and Schwartzentruber(2020{\natexlab{c}})}]{singh20a}
Singh, N., and Schwartzentruber, T., \enquote{Non-Boltzmann vibrational energy
  distributions and coupling to dissociation rate,} \emph{The Journal of
  Chemical Physics}, Vol. 152, No.~22, 2020{\natexlab{c}}, p. 224301.
\newblock \doi{10.1063/1.5142732}.

\bibitem[{Singh and Schwartzentruber(2018)}]{SinghS2018}
Singh, N., and Schwartzentruber, T.~E., \enquote{Nonequilibrium Internal Energy
  Distributions during Dissociation,} \emph{Proceedings of the National Academy
  of Sciences}, Vol. 115, No.~1, 2018, pp. 47--52.
\newblock \doi{10.1073/pnas.1713840115}.

\bibitem[{Gonzales and Varghese(1993)}]{gonzales93a}
Gonzales, D.~A., and Varghese, P.~L., \enquote{A simple model for
  state-specific diatomic dissociation,} \emph{The Journal of Physical
  Chemistry}, Vol.~97, No.~29, 1993, pp. 7612--7622.
\newblock \doi{10.1021/j100131a034}.

\bibitem[{Gonzales and Varghese(1994)}]{gonzales94a}
Gonzales, D.~A., and Varghese, P.~L., \enquote{Evaluation of simple rate
  expressions for vibration-dissociation coupling,} \emph{Journal of
  Thermophysics and Heat Transfer}, Vol.~8, No.~2, 1994, pp. 236--243.
\newblock \doi{10.2514/3.529}.

\bibitem[{Gonzales and Varghese(1995)}]{gonzales95a}
Gonzales, D.~A., and Varghese, P.~L., \enquote{Vibrational relaxation models
  for dilute shock heated gases,} \emph{Chemical Physics}, Vol. 195, No.~1,
  1995, pp. 83--91.
\newblock \doi{10.1016/0301-0104(95)00078-3}.

\bibitem[{Torres and Schwartzentruber(2022)}]{torres22b}
Torres, E., and Schwartzentruber, T., \enquote{Direct molecular simulation of
  oxygen dissociation across normal shocks,} \emph{Theoretical and
  Computational Fluid Dynamics}, Vol.~36, No.~1, 2022, pp. 41--80.
\newblock \doi{10.1007/s00162-021-00596-6}.

\bibitem[{Geistfeld et~al.(2023)Geistfeld, Torres, and
  Schwartzentruber}]{geistfeld23a}
Geistfeld, E.~C., Torres, E., and Schwartzentruber, T.,
  \enquote{{Quasi-classical trajectory analysis of three-body collision induced
  recombination in neutral nitrogen and oxygen},} \emph{The Journal of Chemical
  Physics}, Vol. 159, No.~15, 2023, p. 154111.
\newblock \doi{10.1063/5.0163942}.

\bibitem[{{M}c{B}ride et~al.(1993){M}c{B}ride, {G}ordon, and
  {R}eno}]{mcbride93a}
{M}c{B}ride, B., {G}ordon, S., and {R}eno, M., \enquote{{C}oefficients for
  {C}alculating {T}hermodynamic and {T}ransport {P}roperties of {I}ndividual
  {S}pecies,} Tech. rep., NASA Technical Memorandum 4513, 1993.

\bibitem[{Singh and Schwartzentruber(2022)}]{singh22a}
Singh, N., and Schwartzentruber, T.~E., \enquote{Nonequilibrium Dissociation
  and Recombination Models for Hypersonic Flows,} \emph{AIAA Journal}, Vol.~60,
  No.~5, 2022, pp. 2810--2825.
\newblock \doi{10.2514/1.J061154}.

\bibitem[{Pahlani et~al.(2023)Pahlani, Torres, Schwartzentruber, and
  James}]{pahlani23a}
Pahlani, G., Torres, E., Schwartzentruber, T., and James, R.~D.,
  \enquote{{Objective molecular dynamics investigation of dissociation and
  recombination kinetics in high-temperature nitrogen},} \emph{Physics of
  Fluids}, Vol.~35, No.~6, 2023, p. 067111.
\newblock \doi{10.1063/5.0150492}.

\bibitem[{Macdonald(2024)}]{macdonald24a}
Macdonald, R.~L., \enquote{State-to-state study of non-equilibrium
  recombination of oxygen and nitrogen molecules,} \emph{The Journal of
  Chemical Physics}, Vol. 160, No.~13, 2024, p. 134307.
\newblock \doi{10.1063/5.0195238}.

\bibitem[{Chaudhry et~al.(2020)Chaudhry, Boyd, and Candler}]{chaudhry20b}
Chaudhry, R., Boyd, I., and Candler, G., \enquote{{V}ehicle-{S}cale
  {S}imulations of {H}ypersonic {F}lows using the {MMT} {C}hemical {K}inetics
  {M}odel,} \emph{AIAA Aviation 2020 Forum}, 2020.
\newblock \doi{10.2514/6.2020-3272}, {AIAA} 2020-3272.

\bibitem[{Torres and Schwartzentruber(2021)}]{torres21a}
Torres, E., and Schwartzentruber, T., \enquote{Direct molecular simulation of
  dissociating oxygen under adiabatic and normal shock wave conditions,}
  \emph{AIAA Scitech 2021 Forum}, 2021.
\newblock \doi{10.2514/6.2021-0318}, {AIAA} 2021-0318.

\end{thebibliography}

\end{document}